\documentclass[a4paper,11pt]{article}
\usepackage{jinstpub} 
\pdfoutput=1 

\usepackage[parfill]{parskip} 

\usepackage{cleveref} 
\usepackage[utf8]{inputenc} 
\usepackage{isotope} 
\usepackage{lineno} 
\usepackage{tablefootnote} 
\usepackage{textcomp} 
\usepackage[normalem]{ulem} 
\usepackage[usenames,dvipsnames]{xcolor} 
\usepackage{xfrac} 
\usepackage{xspace} 

\newboolean{showedits}
\setboolean{showedits}{true} 
\newboolean{showcomments}
\setboolean{showcomments}{true} 

\newcommand{\ee}[1]{$\,\times\,10^{#1}$} 
\newcommand{\dbd}[1]{$#1\nu\beta\beta$} 
\newcommand{\degC}{\,\textdegree{}C} 
\newcommand{\tubii}{TUB\lowercase\expandafter{\romannumeral2}\xspace} 

\ifthenelse{\boolean{showedits}}
{
    \newcommand{\del}[1]{\textcolor{red}{\sout{#1}}}   
}{
    \newcommand{\del}[1]{}   
}

\newcommand{\id}[1]{$-$ \textsc{\rversion} $-$}

\ifthenelse{\boolean{showcomments}}
{
    \newcommand{\ob}[2]{ 
        {\noindent\colorbox{Orange}
            {\bfseries\sffamily\scriptsize\textcolor{white}{#1}}}
        {\textcolor{RedOrange}
            {\sf\small$\blacktriangleright${#2}$\blacktriangleleft$}}
    }
    \newcommand{\bb}[2]{ 
        {\noindent\colorbox{MidnightBlue}
            {\bfseries\sffamily\scriptsize\textcolor{white}{#1}}}
        {\textcolor{MidnightBlue}
            {\sf\small$\blacktriangleright${#2}$\blacktriangleleft$}}
    }
    \newcommand{\rb}[2]{
        {\noindent\colorbox{BrickRed}
            {\bfseries\sffamily\scriptsize\textcolor{white}{#1}}}
        {\textcolor{BrickRed}
            {\sf\small$\blacktriangleright${#2}$\blacktriangleleft$}}
    }
    \newcommand{\gb}[2]{ 
        {\noindent\colorbox{Green}
            {\bfseries\sffamily\scriptsize\textcolor{white}{#1}}}
        {\textcolor{Green}
            {\sf\small$\blacktriangleright${#2}$\blacktriangleleft$}}
    }
    
}
{
    \newcommand{\ob}[2]{}
    \newcommand{\rb}[2]{}
    \newcommand{\bb}[2]{}
    \newcommand{\gb}[2]{}
    
}

\ExplSyntaxOn
\newcommand{\latinabbrev}[1]{
    \peek_meaning:NTF .
    {#1\xspace}
    {
        #1.\xspace
    }
}

\newcommand{\latinabbrevstyled}[1]{
    \peek_meaning:NTF .
    {\emph{#1}\xspace}
    {
        \emph{#1.}\xspace
    }
}
\ExplSyntaxOff

\newcommand{\ie}{\latinabbrev{i.e}}
\newcommand{\eg}{\latinabbrev{e.g}}

\title{\boldmath The SNO+ Experiment}
\collaboration{
    \includegraphics[height=17mm]{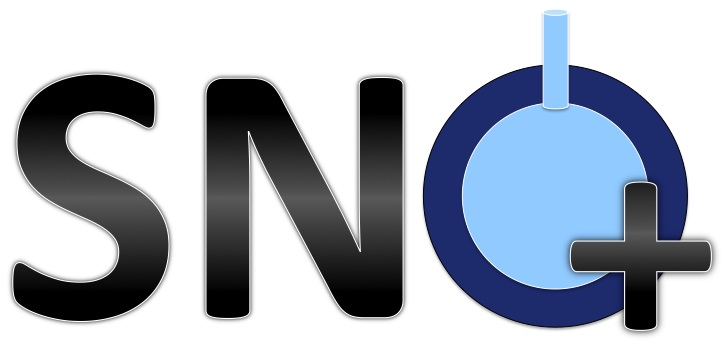}\\
    [6pt] SNO+ collaboration
}

\author[a]{V.\,Albanese,}
\author[b]{R.\,Alves,}
\author[c]{M.\,R.\,Anderson,}
\author[d]{S.\,Andringa,}
\author[e]{L.\,Anselmo,}
\author[f]{E.\,Arushanova,}
\author[c]{S.\,Asahi,}
\author[g,h,i]{M.\,Askins,}
\author[j]{D.\,J.\,Auty,}

\author[f,k]{A.\,R.\,Back,}
% ...

\affiliation[a]{\it Laurentian University, Department of Physics, \\ 935 Ramsey Lake Road, Sudbury, ON P3E 2C6, Canada}
\affiliation[b]{\it Laborat\'{o}rio de Instrumenta\c{c}\~{a}o e F\'{\i}sica Experimental de Part\'{\i}culas (LIP), \\ 3004-516, Coimbra, Portugal}
\affiliation[c]{\it Queen's University, Department of Physics, Engineering Physics \& Astronomy, \\ 64 Bader Lane, Kingston, ON K7L 3N6, Canada}
\affiliation[d]{\it Laborat\'{o}rio de Instrumenta\c{c}\~{a}o e  F\'{\i}sica Experimental de Part\'{\i}culas (LIP), \\ Av. Prof. Gama Pinto 2, 1649-003 Lisboa, Portugal}
\affiliation[e]{\it SNOLAB, Creighton Mine \#9,\\ 1039 Regional Road 24, Sudbury, ON P3Y 1N2, Canada}
\affiliation[f]{\it Queen Mary University of London, School of Physics and Astronomy, \\ 327 Mile End Road, London, E1 4NS, UK}
\affiliation[g]{\it University of California, Berkeley, Department of Physics, \\ 366 Physics North MC 7300, Berkeley, CA 94720-7300, USA}
\affiliation[h]{\it Lawrence Berkeley National Laboratory, \\ 1 Cyclotron Road, Berkeley, CA 94720-8153, USA}
\affiliation[i]{\it University of California, Davis, Department of Physics, \\ 1 Shields Avenue, Davis, CA 95616, USA}
\affiliation[j]{\it University of Alberta, Department of Physics, \\ 4-181 CCIS,  Edmonton, AB T6G 2E1, Canada}
\affiliation[k]{\it University of Sussex, Physics \& Astronomy, \\ Pevensey II, Falmer, Brighton, BN1 9QH, UK}
% ...

\emailAdd{erica.caden@snolab.ca}
\emailAdd{martti.nirkko@cern.ch}
\emailAdd{benjamin.tam@queensu.ca}

\abstract{
    The SNO+ experiment is located 2\,km underground at SNOLAB in Sudbury, Canada.
    A low background search for neutrinoless double beta (\dbd{0}) decay will be conducted using 780\,tonnes of liquid scintillator loaded with 3.9\,tonnes of natural tellurium, corresponding to 1.3\,tonnes of \isotope[130]{Te}.
    This paper provides a general overview of the SNO+ experiment, including detector design, construction of process plants, commissioning efforts, electronics upgrades, data acquisition systems, and calibration techniques.
    The SNO+ collaboration is reusing the acrylic vessel, PMT array, and electronics of the SNO detector, having made a number of experimental upgrades and essential adaptations for use with the liquid scintillator.
    With low backgrounds and a low energy threshold, the SNO+ collaboration will also pursue a rich physics program beyond the search for \dbd{0} decay, including studies of geo- and reactor antineutrinos, supernova and solar neutrinos, and exotic physics such as the search for invisible nucleon decay.
    The SNO+ approach to the search for \dbd{0} decay is scalable: a future phase with high \isotope[130]{Te}-loading is envisioned to probe an effective Majorana mass in the inverted mass ordering region.
}


\arxivnumber{2104.11687}

\begin{document}

    \pagenumbering{Alph}
    \maketitle
    \flushbottom 

    \section{Introduction}
\label{sec:introduction}

\pagestyle{myplain}\pagenumbering{arabic}
\setcounter{page}{1}

The Sudbury Neutrino Observatory (SNO) experiment~\cite{snonim} took data from 1999 to 2006, before returning its target heavy water ($^2$H$_2$O or D$_2$O) to the supplier.\footnote{Atomic Energy of Canada Limited (AECL). \href{https://www.aecl.ca/}{https://www.aecl.ca}}
Using much of the original SNO hardware, the SNO+ experiment will operate in three phases distinguished by target medium, each with a diverse and complementary physics program.
In the first operating phase of the experiment, or ``water phase'', the SNO+ detector was initially filled with light water (H$_2$O) in May 2017.
For the second operating phase, or ``scintillator phase'', the inner detector volume was replaced with 780\,tonnes of organic liquid scintillator.
This increased the light yield by a factor of $\sim$50 with respect to water, allowing for the study of fundamental physics processes at lower energies and with greater resolution.
The scintillator phase will commence following the imminent completion of final commissioning tasks.
Finally, for the third operating phase, or ``tellurium phase'', the scintillator will be loaded with 3.9\,tonnes of natural tellurium.
This corresponds to 1.3\,tonnes of \isotope[130]{Te}, an isotope known to undergo two-neutrino double beta (\dbd{2}) decay.
In the tellurium phase, the experiment will measure the lifetime of the \dbd{2} decay,\footnote{This lifetime is currently measured to be $\left[8.2 \pm 0.2\,\mathrm{(stat.)} \pm 0.6\,\mathrm{(syst.)} \right]$\ee{20}\,y~\cite{cuore-0}.}
and perform a search for neutrinoless double beta (\dbd{0}) decay.

The deep underground location, high purity of materials used, and large volume make the SNO+ experiment ideally suited for studying several aspects of neutrino physics.
The following sections give a short overview of the physics goals pursued by the SNO+ collaboration and how these are achieved; a more quantitative discussion can be found in~\cite{snop2015}.

\subsection{Physics Goals}
\label{sec:physics}

\begin{figure}[htp]
    \centering
    \includegraphics[width=0.9\textwidth]{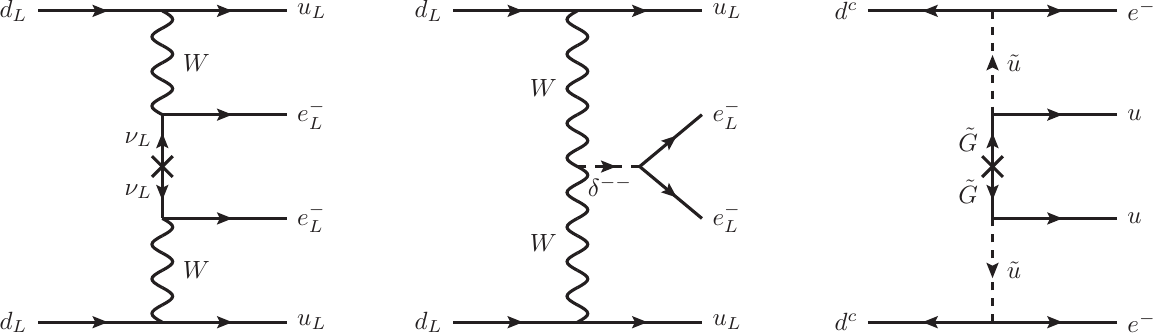}
    \caption{\dbd{0} processes ($\Delta L = 2$) via the introduction of Majorana neutrino masses (left), adding a Higgs triplet to the Standard Model (centre), or supersymmetric theories (right).
        Figure adapted from~\cite{lindner_talk}.
    }
    \label{fig:0vbb-feynman}
\end{figure}

The main goal of the SNO+ experiment is to search for \dbd{0} decay, a lepton number violating process that may happen if neutrinos are Majorana-type particles, \ie their own antiparticles~\cite{furry}.
This is predicted by various grand unification theories that extend the Standard Model by adding heavy right-handed neutrinos insensitive to electroweak interactions; such models are commonly known as seesaw mechanisms~\cite{seesaw1,seesaw2,seesaw3,seesaw4,seesaw5,seesaw6}.
The observation of \dbd{0} decay would indicate lepton number violation ($\Delta L = 2$), a key ingredient in the theory of leptogenesis~\cite{leptogenesis}.
Various theoretical processes leading to \dbd{0} decay are shown in \Cref{fig:0vbb-feynman}.
The process is often depicted as two simultaneous $\beta^{-}$ decays in which two neutrons are converted into two protons and two electrons, with the (anti)neutrinos from the two weak vertices being the same particle.
As a consequence, only the electrons are observed in the final state, carrying the entire energy freed in the process (the nuclear recoil is negligible).

The \dbd{0} decay signature is a peak at the Q-value of the process in the summed energy spectrum of the two electrons, with the half-life of the decay as the measured quantity.
Under the hypothesis that the \dbd{0} depends on the existence of Majorana neutrinos, the effective neutrino mass is defined as
\begin{equation}
    m_{\beta\beta} \equiv \left<m_{ee}\right> = \left| \sum_{k} m_k U_{e k}^{2} \right|,
\end{equation}
where $m_k$ denotes the neutrino mass eigenstates and $U_{e k}$ the Pontecorvo-Maki-Nakagawa-Sakata matrix elements.
Experimentally, $m_{\beta\beta}$ is derived from the \dbd{0} decay rate:
\begin{equation}
    \Gamma_{0\nu\beta\beta} = G_{0\nu} \left|\mathcal{M}_{0\nu}\right|^2 \left|m_{\beta\beta}\right|^2,
\end{equation}
which depends on the two-body phase space factor $G_{0\nu}$ and the nuclear matrix element $\mathcal{M}_{0\nu}$.
The latter is crucial for designing a \dbd{0} experiment since the target isotope mass required to reach a certain sensitivity for $m_{\beta\beta}$ is proportional to the fourth power of $\mathcal{M}_{0\nu}$~\cite{nme1}.
Different nuclear models for the calculation of $\mathcal{M}_{0\nu}$ exist, predicting a wide range of values~\cite{nme2}.
This leads to large uncertainties when converting the decay rate $\Gamma_{0\nu\beta\beta}$ to the corresponding neutrino mass $m_{\beta\beta}$.
Depending on the isotope and nuclear matrix calculation, a half-life on the order of $10^{26}$\,y corresponds to a neutrino mass range of about 30--200\,meV.
Sensitivity to such rare events requires detectors with a large mass of the target isotope and very low background rates.

\begin{figure}[htp]
    \centering
    \includegraphics[width=0.8\textwidth]{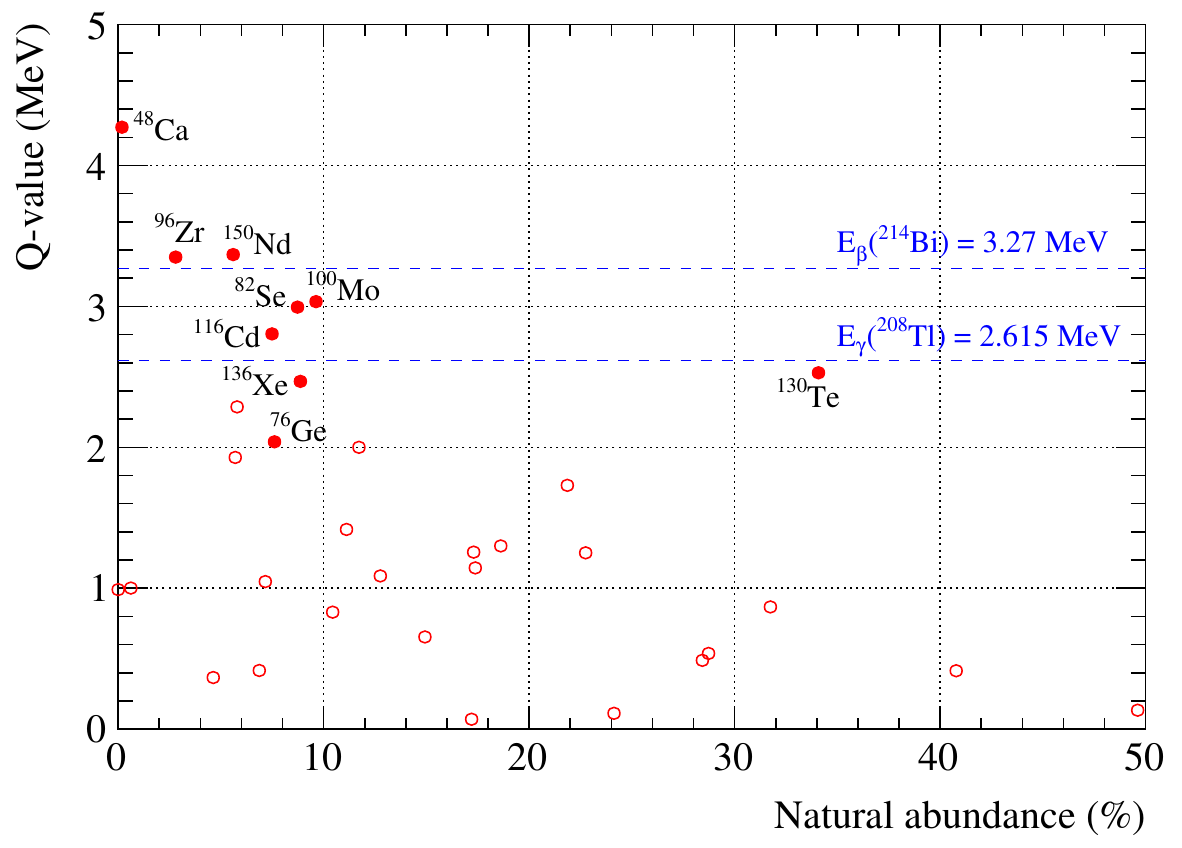}
    \caption{Isotopes capable of undergoing \dbd{2} decay, showing Q-value vs.\,natural abundance.
        Isotopes favourable for \dbd{0} decay searches are shown with solid circles.
        The most energetic particles emitted from the \isotope[238]{U} and \isotope[232]{Th} decay chains --- the two most dominant backgrounds --- are also shown.
        Values taken from~\cite{tretyak2002}; Figure adapted from~\cite{giuliani2012,simkovic2010}.
    }
    \label{fig:0vbb-isotopes}
\end{figure}

An evaluation of relevant properties suggests that \isotope[130]{Te} offers significant advantages over other possible double beta decay isotopes when used in liquid scintillator detectors.
Tellurium has no inherent absorption lines in the visible wavelength range,
allowing for the possibility of better optical transmission and the prospect of a relatively high loading~\cite{biller_2017}.
Another advantage of \isotope[130]{Te} is that the measured \dbd{2} decay half-life of 8.2\ee{20}\,y is among the longest of all the double beta decaying isotopes~\cite{pdg2018}.
Assuming light Majorana neutrino exchange and calculating $G_{0\nu}$ and $M_{0\nu}$ for different values of $m_{\beta\beta}$, the \dbd{0} decay rate can be predicted and compared to the \dbd{2} decay rate.
The ratio between the two is a good figure of merit for comparing different isotopes and indicates that \isotope[130]{Te} and \isotope[136]{Xe} are the optimal choices.
A smaller decay ratio between \dbd{2} and \dbd{0} decays helps minimize the impact of the intrinsic \dbd{2} decay background, which is particularly important when dealing with the relatively poor energy resolution of a typical liquid scintillator detector.
Furthermore, owing to the isotope's large natural abundance (34\%), enrichment is unnecessary, making it both economical and feasible to deploy tonne-scale quantities of \isotope[130]{Te}.
However, the Q-value of \isotope[130]{Te} (2.527\,MeV) is below both the highest natural radioactivity $\gamma$ line (2.615\,MeV)  and the $\beta$ decay Q-value of \isotope[214]{Bi} (3.270\,MeV), the highest $\beta$ and $\gamma$ energies~\cite{giuliani2012} that result from \isotope[222]{Rn} daughters.
This necessitates extremely careful control of radio-purity in the detector.
Existing liquid scintillator experiments have achieved sufficiently low levels of such impurities~\cite{borexino_CNO}.
\Cref{fig:0vbb-isotopes} compares the various isotopes known to undergo \dbd{2} decay as a function of Q-value and natural abundance.

After the SNO+ detector is filled with scintillator, it will be loaded with tellurium using a novel Te-diol loading method developed by the SNO+ collaboration in order to conduct a highly sensitive search for the \dbd{0} decay of \isotope[130]{Te}.
For an initial 0.5\% Te loading by mass, corresponding to 3.9\,tonnes of \isotope[nat]{Te}, this would be expected to yield a half-life sensitivity for \dbd{0} of 2.1\ee{26}\,y after five years of data-taking.
A significant advantage of the technique is its scalability; sensitivity to half-lives greater than $10^{27}$\,y could be readily and economically achieved by increasing the Te loading to several percent.
The SNO+ collaboration aims to pursue this course following the initial demonstration of 0.5\% loading.

Due to its large volume and high anticipated radio-purity, the SNO+ experiment can explore several other physics topics.
Measuring of geoneutrinos (antineutrinos emitted by natural radioactivity in the Earth) will help in understanding heat production in the Earth.
The radiochemical composition of the Earth can be assayed by looking at its geoneutrino emission, thereby providing constraints on the radiogenic portion of the Earth's heat flow and the radiochemical composition of the Earth's mantle and crust~\cite{geoneutrinos}.
The observation of reactor antineutrino oscillations will provide additional constraints on oscillation parameters.
The handful of nuclear power reactors in Ontario are each at similar distances from the SNO+ detector, and also farther (240--345\,km) than many of the reactors in Japan are from the KamLAND detector (140--210\,km)~\cite{kamland}.
This allows for a clean spectral shape, and provides good sensitivity to the second oscillation peak.

A measurement of the \isotope[8]{B} solar neutrino flux was made using data from the water phase~\cite{snop_solar}.
The low backgrounds in the SNO+ detector, the depth of SNOLAB~\cite{snolab_science}, and the lower energy threshold in the scintillator phase will also provide the opportunity to measure low energy solar neutrinos.
A precise measurement of the flux can probe the Mikheyev-Smirnov-Wolfenstein effect~\cite{wolfenstein_1978,msw_1986} as well as alternate models such as non-standard interactions~\cite{nsi_2013,nsi_2015}.
In addition, the spectral shape of the \isotope[8]{B} neutrinos can be used to search for new physics, and investigate the region between matter-dominated and vacuum-dominated oscillations.
Another open question in the solar neutrino field is related to the solar metallicity.
The Standard Solar Model was in excellent agreement with helioseismology until analyses suggested a metallicity of $\sim$30\% lower than the previous model~\cite{solar_metal1,solar_metal2,solar_metal3,solar_metal4,solar_metal5}.
This raised questions about the homogeneous distribution of elements heavier than helium in the Sun.
A measurement of the CNO neutrino flux could be used to complement those performed by Borexino~\cite{borexino_CNO} to solve the solar metallicity problem~\cite{solar_metal6}.
Furthermore, large liquid scintillator detectors serve as excellent supernova neutrino monitors.
The SNO+ collaboration plans to be a part of the SuperNova Early Warning System (SNEWS), an international network of experiments with the goal of providing an early warning of a galactic supernova~\cite{snews}.
Neutrinos and antineutrinos coming from supernova explosions would help answer unresolved questions in neutrino astronomy.
Finally, the SNO+ collaboration searches for exotic physics such as axion-like particles and invisible nucleon decay~\cite{snop_nd}.

\subsection{The SNO+ Detector}
\label{sec:overview}

\begin{figure}[htp]
    \centering
    \includegraphics[height=9cm]{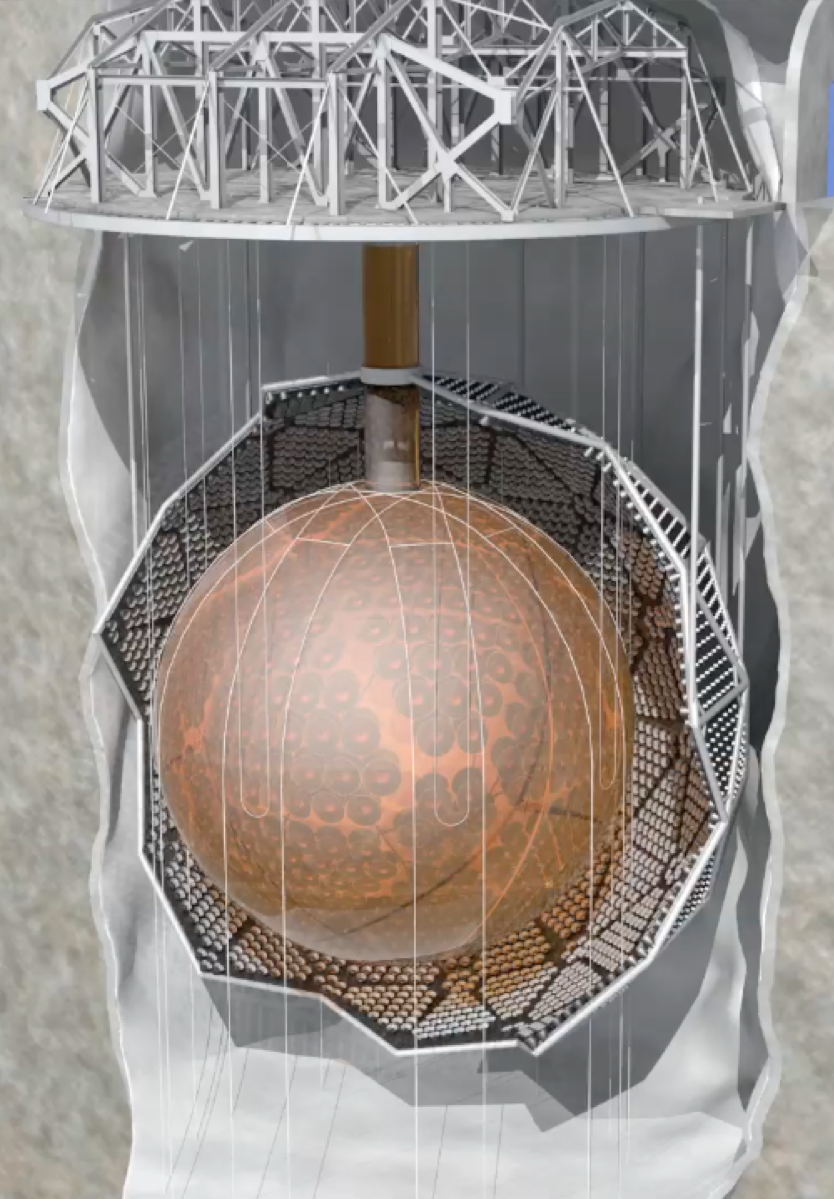}\quad
    \includegraphics[height=9cm]{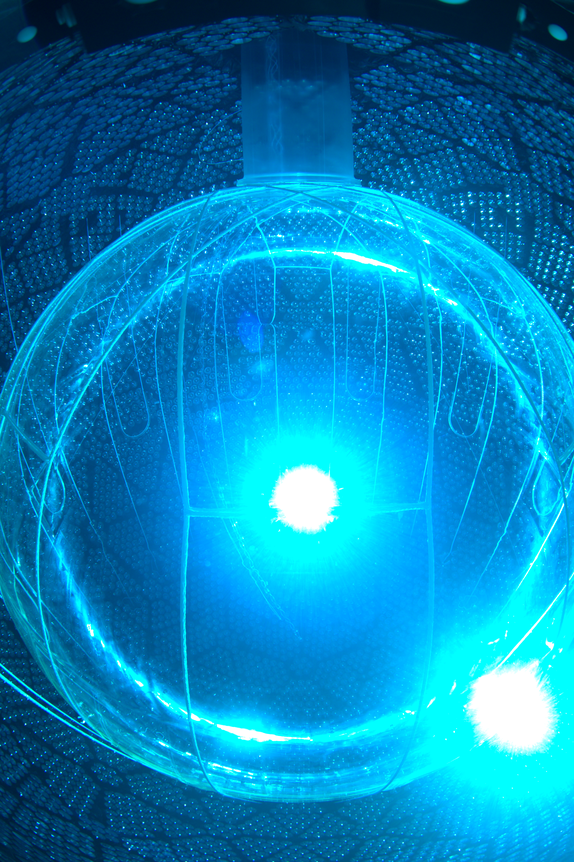}
    \caption{The SNO+ detector, shown as an artistic impression of the cavity containing the detector (left), as well as a photograph taken from inside the detector during the water phase (right).
    }
    \label{fig:cavity}
\end{figure}

The SNO+ detector (see \Cref{fig:cavity}) makes use of the SNO infrastructure~\cite{snonim} located at SNOLAB~\cite{snolab_facility} in Sudbury, Canada.
A flat rock overburden of 2070\,m provides a shield against cosmic muons.
The resulting muon rate is $\left(0.286 \pm 0.009\right)$\,$\mu$/m$^2$/d, corresponding to 6010\,m.w.e.~\cite{snoprd}.
The inner volume of the SNO+ detector consists of a 12-m diameter spherical acrylic vessel (AV).
The AV has a shell thickness of 5.5\,cm and contains the target medium.
9362 inward-facing photomultiplier tubes (PMTs) with low-activity glass and light concentrators are mounted on a concentric geodesic PMT support structure (PSUP), with their front faces $\sim$8.35\,m from the centre of the detector.
91 PMTs without light concentrators were also mounted facing outwards to detect light from muons and other sources in the region exterior to the PSUP.
The volume external to the AV is filled with 7000\,tonnes of ultra-pure water (UPW).
This external volume provides several meters of shielding for the AV from the PSUP and cavity walls, which are both sources of external radiation.

The three operating phases of the SNO+ experiment are distinguished by the target medium within the AV.
The water phase used UPW as the target medium.
The supply of UPW comes from the underground plant that was originally built for SNO and is now operated by SNOLAB.
The purpose of filling the AV with water was to re-commission the detector and mitigate radon ingress, since UPW contains significantly less radon than the surrounding mine air.
The water phase was used to collect physics data by operating SNO+ as a water Cherenkov detector from May 2017 to July 2019 while filled with 905\,tonnes of UPW.

The scintillator phase uses a new organic liquid scintillator developed by the SNO+ collaboration as the target medium.
Because of its lower density, the scintillator was introduced at the top of the AV while UPW was removed from the bottom, thus preventing exposure of the AV to radon-rich mine air during the transition from the water to scintillator phases.

The tellurium phase of the experiment will use tellurium-loaded liquid scintillator as the target material.
The tellurium will be added to the liquid scintillator using a novel metal-loading technique developed by the collaboration~\cite{snop_te}.

To transition from SNO to SNO+, there were four major items that required significant funding and/or development:
\begin{itemize}
    \item In SNO, the AV contained D$_2$O and was surrounded by UPW.
    As D$_2$O is denser than water, the AV needed to be held up by a system of ropes anchored to the top of the cavity.
    By contrast, the SNO+ liquid scintillator has a lower density than water, so the AV will be buoyant following the deployment of liquid scintillator.
    Thus, a hold-down rope-net for the AV was designed, cleanly manufactured, and installed (see \Cref{sec:ropenet}).
    In addition, the original hold-up ropes were replaced with ropes made from a less radioactive material.
    \item 780\,tonnes of liquid scintillator has been deployed within the AV.
    The necessary reagents required to produce and process this amount of the liquid scintillator were procured.
    \item In order to achieve the ultra-low backgrounds required for the physics goals of the experiment, a scintillator purification system was built (in place of the heavy water purification system that was used in SNO).
    This is described in \Cref{sec:scint_plant}.
    \item Two chemical processing plants were designed and constructed in order to purify the tellurium and subsequently deploy it into the liquid scintillator.
    This is described in \Cref{sec:te_plant}.
\end{itemize}
Further upgrades included a redesigned cover gas system in order to lower radon backgrounds, an upgraded electronics and data acquisition (DAQ) system required to handle the higher data rates from the increased light yield and higher event rates at lower energies, and the development and construction of new calibration deployment systems appropriate for the physics program and improved detector.

In the following sections, the infrastructure changes and upgrades are discussed.
\Cref{sec:targets} describes the development of the SNO+ liquid scintillator, as well as the purification and filling systems for the targets used in the three SNO+ operating phases.
\Cref{sec:psup} highlights mechanical changes to the detector infrastructure, while \Cref {sec:electronics} discusses electronics changes.
\Cref{sec:daq} introduces the data acquisition, monitoring, storage, and processing systems.
Finally, \Cref{sec:calibration} describes the various calibration techniques to be used during the SNO+ experiment.

    \section{Development of Active Target Materials}
\label{sec:targets}

To access the physics described in \Cref{sec:physics} at sufficiently low energy thresholds, the SNO+ detector requires a target material with a high light yield.
Liquid scintillator detectors have a significantly higher light yield than water Cherenkov detectors.
The application of large liquid scintillators for the study of neutrino physics at the MeV energy scale was pioneered by the Borexino~\cite{borexino} and KamLAND~\cite{kamland} experiments.

In addition to their high light yield, organic liquid scintillators confer several additional desirable properties.
First, it has been demonstrated that very low background levels of uranium, thorium, and potassium can be achieved in organic liquids~\cite{borexino_CNO}, since these ionic impurities do not readily dissolve into non-polar organic media.
Second, liquid scintillators have a fast timing response, on the order of nanoseconds.
The fast rise time provides effective position reconstruction of scintillation events by fitting the time-of-arrival of the scintillation photons detected by PMTs.
Position reconstruction algorithms are used in physics analyses to reject backgrounds in the detector from external origins; this is known as fiducialization.
A fast rise time also allows background discrimination and rejection by looking at time-correlated decays within the scintillator due to backgrounds such as radon.
Third, liquid scintillators have good linearity of energy response, and the light output and emission time profile depend on the ionization density of the deposited energy.
This enables liquid scintillators to have some capability to identify the type of particle that produced the energy deposit.
Previous liquid scintillator detectors have demonstrated that the different emission time profiles produced by different ionizing particles can effectively discriminate alpha particles from beta particles and, indirectly, also neutrons from gamma rays~\cite{kreslo_2011,wan_2015}.
Finally, techniques have been developed that enable the loading of various elements into liquid scintillators~\cite{buck_2016}.
Developing the metal-loading capability for the SNO+ liquid scintillator was a necessary precursor to the overall design of the experiment with the objective of searching for \dbd{0} decay.

Liquid scintillators are typically a binary solvent-solute ``cocktail'' with a scintillating solvent and a fluor as the solute.
The fluor is usually chosen to have a high quantum yield, to have emission at wavelengths longer than those that will be strongly absorbed by the solvent, and to have a fast emission decay time.
The addition of a fluor results in a liquid scintillator with an improved light yield while maintaining a fast timing response.

Despite their advantages, commonly used liquid scintillators prior to the SNO+ experiment were based on solvents that are not compatible with the acrylic in the AV.
This prompted the SNO+ collaboration to search for a new liquid scintillator, culminating in the identification of linear alkylbenzene (LAB) as an excellent solvent for use as the target in liquid scintillator neutrino detectors.
LAB is compatible with acrylic and has excellent transparency at relevant wavelengths, while conferring other advantages such as superior safety characteristics and a light yield that is competitive with other widely used scintillating solvents.
2,5-diphenyloxazole (PPO), the dominant fluor in previous liquid scintillators, was chosen as the primary fluor for the SNO+ liquid scintillator.

The SNO+ collaboration developed a liquid scintillator consisting of LAB doped with PPO to a concentration of 2\,g/L.
LAB-based liquid scintillators have been successfully deployed in a number of current detectors such as DayaBay~\cite{DYB}, RENO~\cite{reno}, and COSINE-100~\cite{cosine100} and will be used in future experiments including JUNO~\cite{juno} and SABRE~\cite{sabre}.
Some of the important characteristics of the SNO+ liquid scintillator are summarized in the following subsection.
A detailed discussion of the development and characteristics of the SNO+ liquid scintillator is found in ~\cite{snop_scint}.

\subsection{The SNO+ Liquid Scintillator}
\label{sec:scintillator}

Following a testing program that investigated a number of potential scintillating solvents, LAB was selected as the premier candidate for the SNO+ experiment based on its acrylic compatibility, good optical properties, ease of handling, relatively low cost, and logistical availability.

Acrylic compatibility is a critical requirement of the SNO+ liquid scintillator.
In order to confirm this compatibility, a series of tests that exposed stressed acrylic to LAB were initiated ~\cite{wright_2009}.
Destructive tests were periodically conducted on samples of the LAB-exposed acrylic pieces to measure the bulk modulus and yield strength.
The acrylic used in these tests were produced from a spare panel of the same material used to originally build the AV.
Long-term exposure tests are ongoing; to date, no significant changes have been observed in the mechanical properties of acrylic after 11 years of exposure to LAB~\cite{bartlett_2018}.

The SNO+ liquid scintillator has a light yield that is competitive with other widespread liquid scintillators.
The absolute light yield was measured to be $11900\pm600$ photons/MeV prior to purification.
Any liquid scintillator deployed within the detector has undergone an extensive purification process described in \Cref{sec:scint_plant}; this purification has been measured to increase the light yield by over 15\%.
 
It was desirable for the mean light attenuation length of the liquid scintillator to exceed 12\,m, the diameter of the AV.
Through a measurement of its Rayleigh ratio, LAB was calculated from the Rayleigh scattering cross section to have a mean attenuation length of $(72 \pm 14)$\,m at 546\,nm.
Pulse shapes have been measured for both alpha and beta excitation of the SNO+ liquid scintillator cocktail, and the ability to distinguish alphas and betas has been demonstrated.
Alpha identification has been incorporated into SNO+ background rejection schemes that look at delayed coincidences in the U and Th decay chains~\cite{snop2015}.

At the target PPO concentration of 2\,g/L, the emission decay time will be $\sim$5\,ns.
This is fast enough to prevent the decay time from adversely affecting position reconstruction, provided there is a sufficiently high light yield.
At higher PPO concentrations, the scintillation timing can be slightly faster, but this is offset by an increase in self-absorption.

The optical properties of the liquid scintillator were measured in order to fully characterize the SNO+ liquid scintillator.
These properties were used to provide inputs to a Monte Carlo simulation of the detector.
Some of these optical properties include:

\begin{itemize}
    \item absolute light yield as a function of PPO concentration,
    \item light absorption and scattering mean path lengths as a function of wavelength,
    \item non-radiative transfer efficiency as a function of PPO concentration,
    \item scintillation emission time constants for $\alpha$ and $\beta$ excitation,
    \item emission spectrum (which is the well-known PPO emission spectrum~\cite{buck_2015}), and
    \item Birks' constant measured with $\alpha$ particles and protons~\cite{vonKrosigk_2015,vonKrosigk_2013}.
\end{itemize}

A quantitative discussion of these properties and their measurements can be found in~\cite{snop_scint}.

\subsection{Active Target Deployment Systems}
\label{sec:process_systems}

Achieving ultra-low radioactive backgrounds in the scintillator cocktail is not only the main experimental challenge, but the primary requirement for maximizing the science output from the SNO+ experiment.
Consequently, minimizing exposure to sources of these backgrounds was the primary driving consideration behind the development and integration of all detector installations and upgrades.
In particular, great care was taken in the design and construction of the process systems used to purify all components of the liquid scintillator to the desired background targets.
The liquid scintillator was purified in a purpose-built purification plant (see \Cref{sec:scint_plant}).
Borexino and KamLAND both achieved residual internal backgrounds at the $\mathcal{O}(10^{-17})$\,g/g level for U and Th~\cite{borexino_background,kamland_reactor}, with recent results from Borexino showing backgrounds as low as $\mathcal{O}(10^{-19})-\mathcal{O}(10^{-20})$\,g/g for U and Th~\cite{borexino_bkg2}.
To benefit from previous experiences with scintillator purification, the SNO+ collaboration followed the Borexino approach to cleanliness and vacuum leak tightness in the design and construction of the SNO+ scintillator purification plant.

The ultimate objective of the SNO+ experiment is to conduct a search for \dbd{0} in tellurium.
To achieve this, the SNO+ collaboration developed a novel technique for loading tellurium into the liquid scintillator.
The tellurium phase will commence following the loading of the liquid scintillator to a target concentration of 0.5\% natural tellurium by mass, corresponding to 3.9\,tonnes of natural tellurium.

Two chemical processing plants were built for SNO+ tellurium operations (see \Cref{sec:te_plant}).
The tellurium was procured and brought underground as telluric acid (TeA), and will be processed in the purpose-built TeA purification plant.
The purified TeA will then undergo a condensation reaction with 1,2-butanediol (BD) in the BD synthesis plant, with the resulting water product vaporized and removed.
The resulting tellurium-butanediol (TeBD) complex will then be dissolved in LAB.
The tellurium loading technique was developed by the SNO+ collaboration~\cite{snop_te}; the technique and optical properties of the final cocktail are presented in~\cite{biller_2017}.

The following subsections describe the three main SNO+ process systems:
\begin{itemize}
\item the water systems (\Cref{sec:water_systems}),
\item the scintillator purification systems (\Cref{sec:scint_plant}), and
\item the tellurium process systems, consisting of the TeA and BD process plants (\Cref{sec:te_plant}).
\end{itemize}
A list of materials found to be compatible with the SNO+ liquid scintillator will also be provided (\Cref{sec:materials}).

\subsubsection{Water Systems}
\label{sec:water_systems}

The SNO+ experiment inherited the SNOLAB UPW process system originally designed for the SNO experiment~\cite{snonim}.
UPW is primarily used for three main purposes:
\begin{enumerate}
    \item shielding the AV from backgrounds in the PMT array and cavity walls;
    \item filling the AV for the water phase; and
    \item purifying the liquid scintillator through solvent-solvent extraction and steam stripping.
\end{enumerate}

The input water used in the UPW process system is sourced using municipal water from Lively, Ontario, which is provided to SNOLAB by Vale.
Water entering the laboratory is de-aerated and filtered to remove particulates.
It is then softened, charcoal filtered, and passed through a reverse osmosis unit.
This pre-treated water is then processed using the UPW system, which includes mixed-bed ion exchange to demineralize the water, ultra-violet filters to kill bacteria and ionize organic impurities, and a process de-gasser to remove dissolved gasses such as radon.
The water is then passed through a re-gasser to re-introduce a sufficient partial pressure of N$_2$, in order to prevent the water from inducing breakdowns in electrical connectors (see \Cref{sec:PMTrepair}).
Finally, the UPW undergoes a final polish through an additional reverse osmosis unit.
The UPW system also includes heat exchangers to cool the recirculated water, allowing the water temperature to be maintained in the 10--12\degC{} range.
This combats the effect of the natural rock temperature of 40\degC{} at this depth, lowers the PMT dark rate, and reduces the long-term stretching rate of the rope-net (see \Cref{sec:ropenet}).

In addition to purifying the water that is brought into SNOLAB, the UPW system is also used to re-purify water that is circulated from the cavity shielding volume outside the AV.
During the water phase, the UPW system was also used to re-purify water circulated from the AV.
The UPW system can treat water at a flow rate of about 130\,L/min.

The target of the UPW system is to produce water with intrinsic impurities at or below $\mathcal{O}(10^{-15})$\,g/g U or Th equivalent.
This is required since UPW not only serves to minimize direct external gamma backgrounds, but is also used in various stages of scintillator purification.
Though expressed as U g/g equivalent, levels of \isotope[222]{Rn} that can be out of secular equilibrium determine the amount of radioactivity seen further down the chain, such as \isotope[214]{Bi} decays.
During the first part of the water phase, Rn levels were higher in the AV, at a level of  $\mathcal{O}(10^{-14})$\,g/g for U.
Following the commissioning of the new cover gas system (see \Cref{sec:covergas}), the Rn levels in the AV water were reduced to $\mathcal{O}(10^{-15})$\,g/g equivalent for U.

\subsubsection{Scintillator Process Systems}
\label{sec:scint_plant}
\begin{figure}[htp]
    \centering \includegraphics[width=\textwidth]{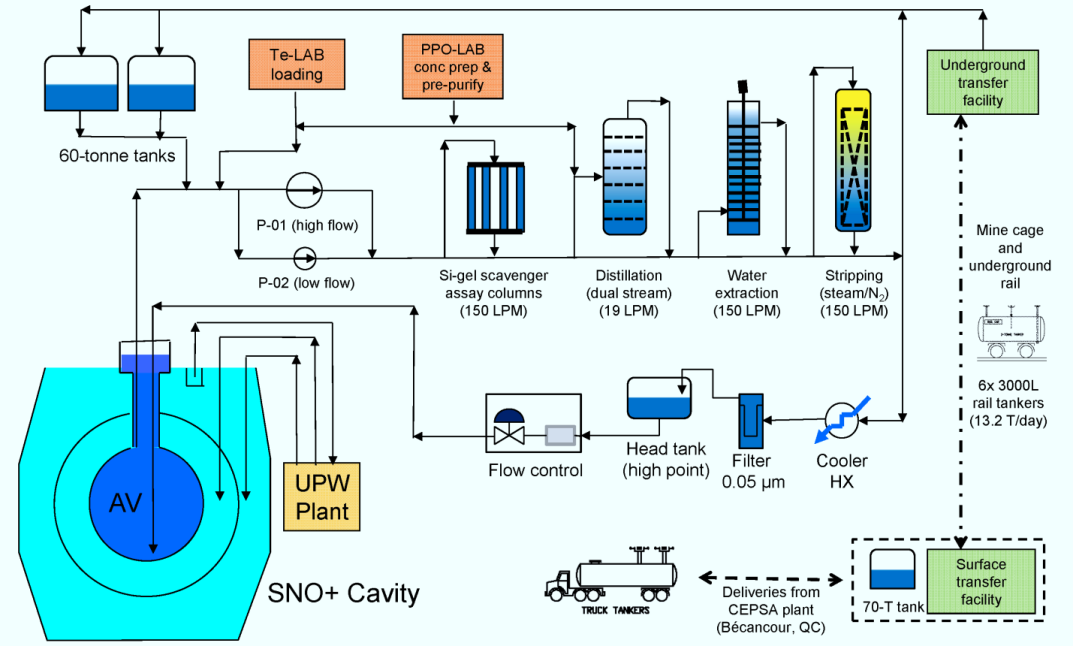}
    \caption[Scintillator liquid handling]{Simplified process flow diagram for the SNO+ liquid scintillator handling.
        Major plant areas are the surface transfer facility, underground transfer facility, 60-tonne tanks storage area, scintillator purification plant, and the AV flow control cabinet.
        The maximum operational flow rates for each process are listed in brackets.
        Reproduced from~\cite{ford_scintplant}, with the permission of AIP Publishing.
}
    \label{fig:scint_plant}
\end{figure}

The SNO+ liquid scintillator purification plant is a specialized implementation of standard petrochemical purification processes.
It was designed and built with strict cleanliness and vacuum-tight requirements to obtain ultra-high purity in the processed liquids.
The total air leak tightness of the system was required to be below 1\ee{-6}\,mbar$\cdot$L/s from all sources.
All surfaces in contact with the clean process fluid are electropolished 316L stainless steel,\footnote{MIL-STD-1246 level 50 cleanliness standard.}
which is chemically resistant to organic solvents such as linear alkylbenzene.
A simplified process flow diagram for the handling of liquid target materials is shown in \Cref{fig:scint_plant}.

The plant includes an extensive fluid handling system.
In addition to providing the piping, it controls interconnections between the purification plant (described below) and external plant systems comprising of:
\begin{itemize}
\item the surface transfer facility to which the LAB is delivered, \footnote{Delivered from CEPSA Qu\'{i}mica, PETRELAB 500-Q Linear Alkylbenzene (LAB). \href{https://chemicals.cepsa.com/en/chemical-products/petrelab-500-q}{https://chemicals.cepsa.com}}
\item the underground storage facility
\item the AV flow control cabinet,
\item the Te-LAB loading process, and
\item the LAB and PPO preparation process.
\end{itemize}
The process engineering, including design of the columns, vessels, heat exchangers, and piping, was done by Koch Modular Process Systems, LLC (KMPS).\footnote{KMPS also provided the design of the purification system for Borexino~\cite{scint_borexino}. \href{https://kochmodular.com}{https://kochmodular.com}}
The primary design challenges included the deep underground location of the SNOLAB facility and the safety considerations required for a system that heats a combustible fluid in an enclosed space.
The process design was also impacted by limitations on electrical power, cooling capacity, nitrogen supply, and available physical space for the infrastructure.

The purification plant is comprised of processes for multi-stage distillation, solvent-solvent water extraction, gas stripping, and metal scavenging.
The purification plant also facilitates the deployment of liquid scintillator and tellurium into the AV.

Distillation is a highly effective separation process, especially when implemented with a multi-stage fractionating column.
Distillation effectively removes lower volatility impurities including heavy metals (in particular K, Pb, Bi, Po, Ra, Th) in addition to fine particulates~\cite{scint_aip2011,scint_neutrino2010}.
Distillation also helps improve the optical transparency of the scintillator by removing oxygenated products such as carboxyl groups and 1,4-benzequinone.
As exposure to oxygen is known to reduce optical clarity of the liquid scintillator, this is critical for obtaining a sufficient light yield and detector energy resolution~\cite{snop_scint_2011}.
Distillation is used for initial purification of the scintillator as part of the LAB delivery and detector filling process.
PPO also undergoes a purification through distillation.
PPO powder is loaded into purified LAB in a high concentration solution ($\sim$120\,g/L).
This solution is pre-purified with nitrogen vacuum purging and three cycles of solvent-solvent water extraction before it is distilled.
This high concentration solution is then mixed into purified LAB during AV filling operations to reach the desired concentration of 2\,g/L.

Solvent-solvent extraction is an equilibrium stage process where two immiscible solvents are brought into close phase contact before coalescing and re-separating.
When water is one of the solvents, the process is known as water extraction.
The process is based on the differences in solubility between water and LAB for various chemical species.
To implement water extraction, water and LAB are flowed counter-current within a column that is separated by baffles into multiple levels, making this a multi-stage process.
Laboratory bench testing has shown water extraction to be highly effective at removing ionic heavy metal species such as K, Pb, Ra, Th, and U, as well as suspended ultra-fine particulates~\cite{scint_aip2011}.
Water extraction is intended to be used to polish and re-purify the liquid scintillator following the completion of the scintillator fill to further suppress Rn daughters in the AV (see \Cref{sec:psup}).

Gas stripping is another standard equilibrium mass-transfer process that is used in the scintillator purification system for the removal of volatile gases and liquids (in particular Rn, Kr, Ar and water).
At equilibrium, impurities will partition between the liquid and gas phases at the liquid-gas interface according to their relative solubilities, which is equivalent to their relative partial pressures as given by their Henry coefficients.
The stripping gas acts to flush away the impurity gases to result in a non equilibrium partial pressure, such that impurities are further driven from the liquid to the gas phase.
The process uses counter-current flow in a packed column to maximize the liquid-gas surface area.
The scintillator is fed from the top of the column and falls through the gas phase due to gravity, while the stripping gas is injected under pressure at the bottom of the column and vents to the top.
The best efficiency is obtained at high temperatures (which decreases the gas solubility in liquid) and under vacuum (which decreases the sum of all partial pressures), while the overall efficency is dominated by the number of equilibrium stages, which is increased with column height and stripping gas flow rate.
The efficiency is different for each impurity, but most challenging for Rn which has a high solubility in aromatic solvents such as LAB; a removal efficiency in the 95--98\% range is expected for Rn.
Nitrogen was used as the stripping gas during the initial addition of scintillator into the AV, which also resulted in the thorough de-watering of the scintillator.
When the scintillator is recirculated for further purification at a relatively higher flow rate, super-heated UPW steam is used as the stripping gas to meet the larger gas flow requirement.
Since the UPW used to produce the steam was nitrogen/vacuum stripped in the water systems, it has a reduced Rn content and therefore a higher Rn removal efficiency when used as the stripping gas.
The steam injector is located above the bottom stage of the column, which is stripped with a smaller flow of pure nitrogen gas designed to provide humidity control at a $\sim$40 ppm target water concentration.

The final system in the scintillator purification process is the metal scavenging system.
Several process columns have been designed to use functional metal scavengers in a silica-gel matrix to polish the LAB following the scintillator fill.
The design was based on extensive testing with QuadraSil-AP,\footnote{QuadraSil-AP is from Reaxa LTD, now owned by Johnson Matthey PLC.} which has an aminopropyl functional group on a 50\,\textmu{}m mesh silica gel.
This was found to be highly effective for Pb and Ra removal~\cite{scint_aip2011}.
The motivation for developing the scavenger purification was to provide an alternate and complementary process to solvent-solvent water extraction that is also expected to be effective for Bi removal.
A feature of QuadraSil-AP is that it can be regenerated with HCl acid; the metals can be recovered and subsequently analyzed with coincidence beta-alpha counting.
This provides a method for \emph{ex situ} assays to explicitly determine the contamination level from \isotope[224]{Ra} and \isotope[226]{Ra}, as was done in SNO~\cite{scint_nima2009}.

Filling the SNO+ detector with liquid scintillator was accomplished by removing water from the bottom of the AV while simultaneously adding purified scintillator to the top, which remained separated from the water due to differences in liquid density.
At the water-scintillator interface, ionic impurities were not efficiently partitioned from water to scintillator.
This was due to the large relative solubility of ionic compounds in polar molecules such as water when compared to non-polar molecules such as LAB.
This behaviour was first demonstrated in the radio-purification of liquid scintillator in~\cite{scint_purification}.

Following the completion of the SNO+ scintillator fill, the material is being continuously recirculated within the process systems.
This process removes residual moisture content from the water phase.
The systems allow for the recirculation of the entire detector volume in $\sim$100\,hours.

\subsubsection{Tellurium Process Systems}
\label{sec:te_plant}
One of the main advantages of using liquid scintillator as the active medium is the possibility to dissolve heavy metals with long-term stability and good optical properties.
For the tellurium phase, an innovative technique will be used to load tellurium into LAB.
As with all SNO+ instrumentation, the major challenge driving the design of the tellurium processes was the need to minimize the presence of isotopes with decays around the Q-value of \isotope[130]{Te} (2.527\,MeV), the SNO+ region of interest (ROI).
The final purity requirement of the tellurium-loaded scintillator cocktail for the \dbd{0} decay is at the $<\mathcal{O}(10^{-15})$\,g/g level for U and Th.
This implies that the target for trace contamination in the tellurium is $<$1\ee{-13}\,g/g for U and $<$5\ee{-14}\,g/g for Th.

The tellurium was procured as TeA in a crystallized form.\footnote{Delivered from ABSCO Limited, Telluric Acid (TeA) CAS 7803-68-1. \href{https://www.absco-limited.com/det/10784/Telluric-Acid/}{https://www.absco-limited.com}}
As cosmic ray spallation on tellurium can produce long-lived isotopes with decays around the ROI such as \isotope[60]{Co}, \isotope[110m]{Ag}, \isotope[126]{Sn}, \isotope[88]{Y}, \isotope[124]{Sb}, and \isotope[22]{Na},
the TeA has been stored in the SNOLAB underground facility since 2015 to mitigate this effect.
Inductively Coupled Plasma Mass Spectrometry (ICP-MS) impurity analysis of the procured TeA measured both U and Th at the $\mathcal{O}(10^{-11})$\,g/g level, therefore requiring further purification of the tellurium with a reduction factor of $\sim$200 and $\sim$600 for these elements, respectively.
Purification requirements for cosmogenic isotopes can be higher, reaching $10^6$ in some cases~\cite{Lozza_2015}, though storage underground for many years has resulted in many cosmogenic isotopes to have decayed to negligible levels prior to any purification.

The tellurium is purified and loaded using the tellurium process systems, consisting of two purpose-built underground chemical plants:\,the TeA purification plant cleans the TeA, while the BD synthesis plant converts it into a chemical form that is soluble in LAB.
As TeA is a weak acid, both the tellurium synthesis plant and the telluric acid purification plant used polypropylene (PP) and fluoropolymer (PFA/PTFE) in the construction of all wetted surfaces such as vessels and piping to avoid the leaching of metals into the tellurium.
In order to further minimize the risk of contamination, all major process equipment was leached with ultra-pure acid and rinsed with UPW prior to operation of the plants.
During the tellurium loading of the AV, the two plants will operate in parallel to produce 75\,kg batches of tellurium.
The entire loading process will require 52 batches, and can occur within 36 weeks.
In spike-tests and pilot-plant runs of the tellurium purification and diol synthesis procedures, it has been demonstrated that the target background concentrations of $<$1\ee{-13}\,g/g for U, $<5$\ee{-14}\,g/g for Th, and $<$7.5\ee{-12}\,Bq/kg for $^{60}$Co can be achieved.

Purification of the tellurium will occur in the TeA plant.
The solubility of TeA in water is a function of pH and temperature.
Therefore, tellurium will be purified by dissolving TeA in hot UPW before recovering the TeA through both acid and thermal re-crystallization.
The TeA-UPW solution will first be passed through a fine particulate filter to remove insoluble contaminants.
The TeA will then be recrystallized by reducing the temperature and lowering the pH through the addition of nitric acid, leaving soluble contaminants in the supernatant solution to be filtered away.
The remaining pure TeA crystals will be rinsed with clean nitric acid, redissolved with hot UPW, and pumped to the next stage of processing.
The pH based re-crystallization will be repeated twice.
The product will then be polished by recrystallizing and rinsing using only UPW and temperature cycling in order to minimize residual nitric acid in the product.
This method is fully explained in~\cite{hans_2015}.

The dissolved, purified product will be transferred to the BD synthesis plant.
In this second plant, the TeA will be heated under vacuum with purified 1,2-butanediol, producing a LAB soluble species referred to as tellurium butanediol (TeBD).
The product will then be mixed with liquid scintillator extracted from the AV at a ratio of 1:1 and transferred to the scintillator plant for further dilution prior to addition into the detector.

The final SNO+ cocktail to be used for the beginning of the SNO+ tellurium phase is LAB with 2\,g/L PPO and loaded with TeBD to a concentration of 0.5\% natural tellurium by mass.
LAB scintillator containing TeBD at the target 0.5\% Te concentration maintains both good optical properties and a high light emission level.
At increased Te concentrations, the transparency of the cocktail is still excellent, but the intrinsic light yield starts to quench.
This effect can be mitigated by the addition of N,N-Dimethyldodecylamine (DDA), which also helps to stabilize the TeBD-LAB cocktail (increasing the light yield by 15\% or more, depending on the Te concentration).
A Te concentration of several percent natural tellurium by mass is feasible.
A secondary wavelength shift, 1,4-bis(2-methylstyryl)benzene (bis-MSB) will be added to the SNO+ scintillator cocktail at a concentration below 15\,mg/L.
This will shift the scintillation light to longer wavelengths, reduce absorption-reemission by PPO, and further increase the detected light yield.

\subsection{Material Compatibility}
\label{sec:materials}

Materials that potentially come into contact with the active medium were confirmed to be compatible with the LAB scintillator, both with and without the TeBD complex.
Extensive compatibility and stability studies were performed in multiple tests, with some over 100\,months in duration~\cite{wright_2009,bartlett_2018,snop_scint}.
The materials tested included elements used in the detector, the deployable calibration sources, and the chemical process systems.
A list of some recently tested items can be found in \Cref{tab:materials}.
Note that only some of these materials are used in the plants, based on the test results.

\begin{table}[htp]
    \centering
    \caption{Materials tested for compatibility and stability when in prolonged contact with the scintillator cocktail, both before and after the addition of the TeBD complex.}
    \label{tab:materials}
    \begin{tabular}{| l | l | l |}
        \hline
        \textbf{Metals} & \textbf{Polymers and Plastics} & \textbf{Miscellaneous} \\
        \hline
        Anodized aluminum & Acrylic & Activated aluminum oxide \\
        Galvanized steel & Cable ties (nylon 66, white) & Fibre reinforced ECTFE\tablefootnote{John Brooks, Innomag TB/U-Mag high purity pump parts. \href{https://www.johnbrooks.ca/resource/innomag-mag-drive-pumps/}{https://www.johnbrooks.ca}} \\
        Nickel (SNO NCD name plate\tablefootnote{Some name plates used to identify NCD anchors during the AV construction were not removable due to their location.}) & Delrin\tablefootnote{DuPont. \href{https://www.dupont.com/brands/delrin.html}{https://www.dupont.com}} (black/white) & Silicon carbide from pump \\
        Silver-plated copper & FEP encapsulated O-ring\tablefootnote{McMaster-Carr. \href{https://www.mcmaster.com/o-rings}{https://www.mcmaster.com/o-rings}} & Quadrasil-AP\tablefootnote{Alfa Aesar, QuadraSil Aminopropyl 46303. \href{https://www.alfa.com/en/catalog/046303/}{https://www.alfa.com/en/catalog/046303/}} \\
        Silver-plated VCR\tablefootnote{Swagelok, SS-12-VCR-2. \href{https://www.swagelok.com/en/catalog/Product/Detail?part=SS-12-VCR-2}{https://www.swagelok.com/en/catalog/Product/Detail?part=SS-12-VCR-2}} & FFKM O-ring\tablefootnote{COG, perlast-g80a. \href{https://www.cog.de/uploads/tx_datenblattgenerator/pdf/en/perlast-g80a.pdf}{https://www.cog.de/uploads/tx\_datenblattgenerator/pdf/en/perlast-g80a.pdf}} (black) & \\
        Stainless steel 316/316L & FFKM O-ring\tablefootnote{TRP Polymer Solutions Ltd. \href{https://trp.co.uk/ffkm-hub/ffkm-o-rings-and-seals/}{https://trp.co.uk/ffkm-hub/ffkm-o-rings-and-seals/}} (white) & \\
        & HDPE & \\
        & Nylon compression fitting\tablefootnote{McMaster-Carr. \href{https://www.mcmaster.com/compression-fittings}{https://www.mcmaster.com/compression-fittings}} & \\
        & PFA tubing\tablefootnote{Crist Group, Ametek ultra high purity PFA. \href{http://cristgroup.com/distribution/fluoropolymer-tube-pipe/pfa-tubing-pipe/}{http://cristgroup.com}} & \\
        & POM (Delrin, rough/smooth) & \\
        & PTFE (tubing, clear/white) & \\
        & PTFE tape\tablefootnote{McMaster-Carr. \href{https://www.mcmaster.com/ptfe-tape}{https://www.mcmaster.com/ptfe-tape}} & \\
        & Tensylon\tablefootnote{BAE Systems (DuPont since 2012). \href{https://www.dupont.com/products-and-services/personal-protective-equipment/vehicle-armor/products/dupont-tensylon.html}{https://www.dupont.com}} ropes & \\
        & Tygothane tubing\tablefootnote{McMaster-Carr, Tygothane C-­210-­A. \href{https://www.mcmaster.com/tygothane-tubing}{https://www.mcmaster.com/tygothane-tubing}} & \\
        \hline
    \end{tabular}
\end{table}

    \section{Detector Hardware Additions and Upgrades}
\label{sec:psup}

The AV and PSUP are both suspended in a barrel-shaped cavity approximately 30\,m high and 21\,m wide.
A deck is mounted to the cavity walls, allowing for protection from seismic motion.
The AV is suspended from the deck by ten rope loops that pass through external grooves inside acrylic panels, which are mounted on the equator outside the AV.
The PSUP is a 92-sided geodesic polygon suspended by cables from the same deck.
The spherical AV has a 7-m high neck that reaches the level of the deck.
This hardware has been generally unchanged since the SNO experiment~\cite{snonim}.
During original construction of the AV, the bonds were checked with polarized light and visually inspected for voids and signs of crazing.
While transitioning to SNO+, every bond was re-inspected for additional signs of crazing or changes in any existing voids.
Although some light crazing was expected and observed, the changes in the AV were determined by independent engineers to be of no concern for the running of SNO+.
A survey was performed on the AV, which concluded that the dimensions and sphericity did not change during the operation of SNO.

After the completion of the SNO experiment in 2006, the heavy water was drained from the AV and the UPW was drained from the surrounding cavity.
During this process, the AV was contaminated, and the Urylon~\cite{sno_urylon} liner at the bottom of the cavity was found to be damaged.
Consequently, the AV was cleaned and the cavity floor was repaired.
The inside of the AV was cleaned using two custom pieces of equipment.
The top half of the inner AV was cleaned using a rotatable cantilevered platform, supported by a tower that was mounted to the deck and passed through the neck of the AV.
The platform was lowered into the AV in pieces and assembled \emph{in situ}.
The bottom half of the AV was cleaned using a rotating ladder mounted on a small acrylic block.
The cleaning was done using Alconox\footnote{Sigma-Aldrich Alconox\textregistered{} detergent. \href{https://www.sigmaaldrich.com/CA/en/sds/aldrich/z273228}{www.sigmaaldrich.com}} with multiple passes over all surfaces.
The cleaning efforts were successful; assays performed on the UPW from the AV during initial filling operations show that U and Th concentrations from Ra and Bi measurements are in agreement with those obtained from the water directly from the UPW plant.
Following the initial water fill, the cleaning efforts were completed by recirculating and repurifying the UPW deployed in the AV, thus further reducing possible contaminants.

Following the cleaning of the AV, substantial upgrades and improvements were made to the detector hardware, which are discussed further in the following sections.
Rn daughters had deposited onto the AV during the cleaning period, and much work was done to evaluate the subsequent backgrounds (\cref{sec:leaching}).
Because liquid scintillator is less dense than water, a new hold-down rope-net was designed and installed to counteract the buoyant forces on the AV when filled with scintillator.
The installation of this rope-net required access to the bottom of the cavity, the neck of the AV, and the top of the PSUP itself (\cref{sec:ropenet}).
Both the installation of the hold-down rope-net and initial filling operations required access to the cavity, which allowed for the removal, repair, and re-installation of PMTs that failed during SNO operations (\cref{sec:PMTrepair}).
In addition, a new optical calibration system was mounted to the PSUP, described in \Cref{sec:opsources}.
After all work was completed and the detector was filled with water, a newly designed cover gas system was installed and commissioned to ensure that radon levels in the active target area of the acrylic vessel are suppressed to below the specified levels (\cref{sec:covergas}).

\subsection{Leaching of Radon Daughters}
\label{sec:leaching}

During its initial construction and refurbishment, the surface of the AV was exposed to mine air, which has a \isotope[222]{Rn} concentration of 123\,Bq/m$^3$.
\isotope[222]{Rn} ($T_{1/2} = 3.8$\,d) decays into \isotope[218]{Po} ($T_{1/2} = 3.1$\,min), which subsequently decays into \isotope[214]{Bi} ($T_{1/2} = 19.9$\,min) and eventually into \isotope[210]{Pb} ($T_{1/2} = 22.2$\,y).
Radon daughters exist in ionic form and tend to deposit on to the acrylic, resulting in a sizable fraction being implanted $\sim$100\,nm below the surface itself.
As a consequence, long half-life \isotope[210]{Pb} atoms are still embedded in the vessel surface.

A constraint on the embedded \isotope[210]{Pb} activity was obtained in 2013 by measuring the \isotope[210]{Po} activity in different parts of the vessel inner surface, yielding an average contamination of $(2.4\pm0.8)$\,Bq/m$^2$, which corresponds to an activity of about 1\,kBq for the entire vessel inner surface.
This number can be further verified by counting $(\alpha,n)$ events in water and scintillator phases.
While \isotope[210]{Pb} is not a background for the SNO+ experiment, its daughters \isotope[210]{Bi} and \isotope[210]{Po} represent a non-negligible source of background for various physics analyses.
In addition to the surface activity, embedded Rn daughters can leach through contact from the surface of the AV into the detector media.
The leaching rate, defined as the relative activity removed per unit time, depends on the temperature of the medium in contact with the AV (rate increases with temperature) and its chemical composition (water is expected to have a larger leaching rate than LAB)~\cite{khaghani_2016}.

Leaching rates for the SNO+ experiment have been studied in benchtop experiments.
The same type of acrylic used for the AV was exposed to concentrated Rn gas in order to implant Rn daughters onto the acrylic surface.
After washing, the acrylic was placed in contact with the liquid of interest and the activity in the liquid was counted after 1--2 months of exposure using silicon and high-purity germanium detectors.
The measurement was repeated for temperatures between 25\degC{} and 12\degC{}, the latter being the expected operating temperature during data taking.
A leaching rate of $(1.05^{+0.18}_{-0.19})$\ee{-3}/day is expected for UPW at 18\degC{}, which would correspond to an activity in water of 0.40\,Bq/m$^3$ after one year exposure of the AV surface to UPW.
Leaching rates in scintillator are measured to be about an order of magnitude smaller, depending on the temperature.
\emph{In situ} and \emph{ex situ} assays during all phases will be used to cross check expectations.
The major mitigation strategy for the leaching backgrounds is the \emph{in situ} purification of the deployed target material.
During the water phase, the UPW plant was used to recirculate the water inside the AV for purification.
The water outside the AV separately undergoes regular recirculation.

\subsection{Hold-Down Rope-Net}
\label{sec:ropenet}

\begin{figure}[htp]
    \centering
    \includegraphics[width=0.8\textwidth,trim={0 0 0 0},clip]{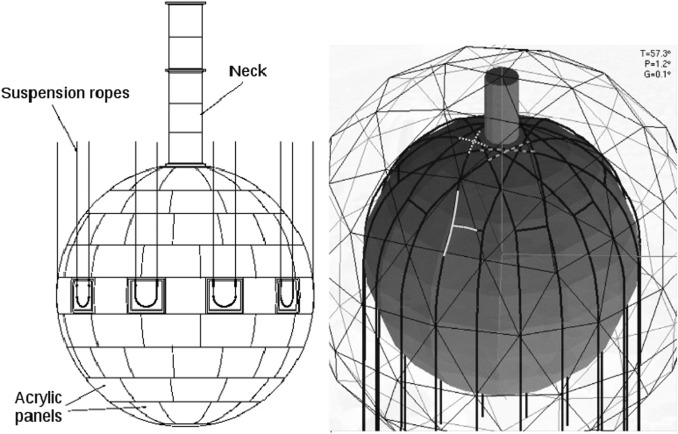}
    \caption{Left: Suspension ropes of the SNO acrylic vessel that are located in special U-shaped grooves in the acrylic at the equator of the vessel.
        Right: Isometric view of the SNO+ hold-down rope-net, highlighting the different locations of the knots.
        Suspension ropes are not shown in the right figure.
        Reprinted from~\cite{snop_ropes}, with permission from Elsevier.
    }
    \label{fig:rope-net}
\end{figure}

The SNO+ liquid scintillator has a lower density $(0.86$\,g/cm$^3$ at 12\degC{}) than the surrounding UPW, requiring a new system to compensate for the 1.25\ee{6}\,N of buoyant force.
The rope-net system (see \Cref{fig:rope-net}) was designed based on Finite Element Analysis (FEA) to calculate the amount of stress on the acrylic induced by the ropes, hydrostatic pressures, and gravity.
The details of the FEA and rope net analysis are detailed in~\cite{snop_ropes}.
The mechanical properties of the acrylic were standard values, derated for long term service and temperature~\cite{stachiw_acrylics}.
The FEA was validated, with particular attention to the rope-acrylic interface by laboratory measurements of strain while the acrylic was subject to rope loads.

The hold-down rope-net consists of five sub-nets, each composed of two rope pairs passing on either side of the acrylic vessel neck.
Every rope intersection takes the form of a single Brummel knot.
Each of the 20 rope ends is anchored to the bottom of the cavity and equipped with a load cell used to monitor the force on each rope.
Drilling to install the rock bolts that secure these anchor points underneath the PSUP while maintaining detector cleanliness required a number of precautions.
An umbrella-like structure was installed below the PSUP to protect the clean detector hardware from any potential contamination caused by the work taking place on the cavity floor.
In addition, small tents were erected around each of the drill sites to prevent the rock dust from contaminating the rest of the cavity.
The anchoring rock bolts were drilled at least 1.8\,m into the cavern floor.
The hardware is covered with a new Urylon floor liner, making the entire cavern water tight.

The Tensylon\texttrademark{} hold-down ropes are made out of high-purity, high-performance ultra-high-molecular weight polyethylene fibres that have intrinsically low levels of radioactivity, high strength, low creep, and a low coefficient of friction.\footnote{Yale Cordage. \href{http://www.yalecordage.com/custom-and-specialty-ropes}{http://www.yalecordage.com/custom-and-specialty-ropes}}
Each 3.96-cm diameter rope is fabricated out of $\sim$4800 Tensylon fibres.
The original hold-up ropes have also been replaced with thinner 1.85-cm diameter Tensylon ropes to further reduce radioactive contamination, especially from \isotope[40]{K} present in the previous Vectran ropes.

In FEA calculations, a simplified model of the rope intersection points was used.
To validate this approximation, it was necessary to experimentally determine a correction factor to the calculated stresses.
Consequently, a test stand was constructed and used to measure the force on the acrylic due to a rope intersection point, and the ratio between measured and FEA calculated strains was found to be 1.3~\cite{snop_ropes}.
The overall behaviour of the rope-net was tested as the detector and cavity were being filled with UPW prior to the water phase.
At one point, the cavity was filled to a significantly higher level than the AV, producing a buoyant load on the rope-net equivalent to what is expected when the AV is filled with scintillator and the cavity is filled with water.
The rope-net responded as expected to the applied load.

The stretch of the ropes depends on the temperature of the surrounding water.
Higher water temperatures increases static creep~\cite{rope_stretch}.
To minimize creep in the rope-net, and hence the upward motion of the AV during the course of the experiment, the water in the cavity is recirculated through the UPW plant.
This removes heat from the surrounding rock and maintains the temperature of the cavity water at 12\degC{}.

\subsection{PMT Repairs}
\label{sec:PMTrepair}

The SNO+ experiment uses the Hamamatsu R1408 PMTs that were purchased for the SNO experiment~\cite{snonim}.
Throughout the lifetime of SNO, $\sim$800 PMTs (8.5\%) ceased to operate, with a higher failure rate during the first few years.
Most failures were attributed to an electrical short in the high voltage (HV) divider base, as determined by diagnostics performed on PMTs that were removed during the draining phase of the SNO experiment.
Less than 10\% of the failures were related to the PMTs themselves.
During the SNO+ construction phase, over 600 failed PMTs were removed and repaired.
Components in the bases were replaced and the PMTs were re-installed into the PSUP.

The PMTs are installed with their individual ABS plastic housings in groups of 9--16 on PMT panels.
To install the hold-down rope-net, a number of PMT panels needed to be modified; 2--4 PMTs needed to be removed for each hold-down rope to pass through to the cavity floor.
Each PMT is equipped with a 27-cm diameter concentrator that consists of small aluminum coated petals, which increases the effective photocathode coverage of the detector to about 54\%.
The reflectivity of some of these concentrators degraded and became diffuse after being immersed in UPW over time~\cite{snoprc}.

The PMT repair process involved their removal from the PSUP, transportation through the mine to the repair facility at Laurentian University, and transportation back; the average turn-around time was one week.
Fixed PMTs were re-installed at their initial positions on the PSUP whenever possible.
This allowed the original high-voltage mapping to be preserved, which is used to ensure that all tubes have a gain as close as possible to $10^{7}$.
The PMTs at the bottom of the PSUP were accessed and repaired during hold-down rope-net installation and early filling operations.
PMTs at higher locations had to be accessed using boats during filling operations while the water level rose.
As such, it was unavoidable that some PMTs could not be returned to their original location.
In these cases, other PMTs replaced those that could not be repaired in time.

To access and replace the broken HV splitter base for each PMT, the water-tight housing and silicon filling were removed.
After the base had been replaced from stock leftover from SNO, the housing was refilled and resealed.
Fewer than 10\% of the repaired PMTs failed again in the first month after being re-installed, which was consistent with expectations and allowed the SNO+ experiment to start taking data with more PMTs than the number at the end of SNO operations.

Four High Quantum Efficiency Hamamatsu R5912 PMTs were also installed to better understand the performance of potential future SNO+ phases.
They were installed in locations where original SNO PMTs were removed due to failure.

The SNO+ experiment started data-taking with 9453 functional PMTs installed in the PSUP, 91 of which were mounted facing outwards to detect light from muons and other sources in the region exterior to the PSUP.
One to two PMTs are lost each week due to electronic components breaking.
This happens uniformly across the detector, and is not expected to be an issue during the lifetime of the experiment.
The failed PMTs are taken into account by detector simulations (see \Cref{sec:simulation}) to track their expected impact on the position and energy resolution.

\subsection{Cover Gas Systems}
\label{sec:covergas}

The SNO+ detector has two inert cover gas systems.
The cavity cover gas is the volume above the shielding UPW in the cavity surrounding the AV.
This volume is a flow-through system and is constantly flushed with high purity nitrogen gas at the rate of 5\,L/min, as described in~\cite{snonim}.
The AV cover gas is the volume above the liquid scintillator inside the neck of the AV.
The latter is a newly developed and completely sealed system.
Both systems protect the detector from radon contamination.
The nitrogen used for the cover gas is rated to have a maximum oxygen concentration of 2\,ppm and water concentration of 3\,ppm.

In principle, the short-lived \isotope[222]{Rn} daughter \isotope[214]{Bi} could be a background to the \dbd{0} decay measurements.
The diffusion time through the static LAB column in the neck of the AV is much longer than the half-life of \isotope[222]{Rn} ($T_{1/2}=3.8$\,d) or of the intermediate short-lived nuclei \isotope[218]{Po} and \isotope[214]{Pb}.
Therefore, \isotope[214]{Bi} will decay before reaching the spherical part of the AV, though some convective transport is possible.
However, long-lived \isotope[222]{Rn} daughters lower in the chain are the main background when measuring low-energy solar neutrinos.
Since \isotope[222]{Rn} daughters cause significant backgrounds to the pep/CNO solar neutrino analyses, the target level for the whole system was set to 650 decays/day in the upper Universal Interface, described in \Cref{sec:UI}.
The scintillator cover gas was thus designed to reduce the ingress from the radon-rich air present in the lab (123\,Bq/m$^3$) into the detector volume, and all materials in the scintillator cover gas system were carefully chosen for both their low radon emanation and permeability in order to reach these low levels.

As a sealed system, the scintillator cover gas includes protections against over- or under-pressurizing the detector volume.
A differential pressure between the inside and outside of the detector could result from the changes in mine air pressure during operations, which could cause damage to the AV.
The permissible pressure difference of $\pm0.28$\,psi between inside and outside the AV is defined by the constraint on the mechanical integrity of the AV~\cite{snop_ropes}.

\begin{figure}[htp]
    \centering
    \includegraphics[height=80mm,trim={10mm 10mm 10mm 10mm},clip]{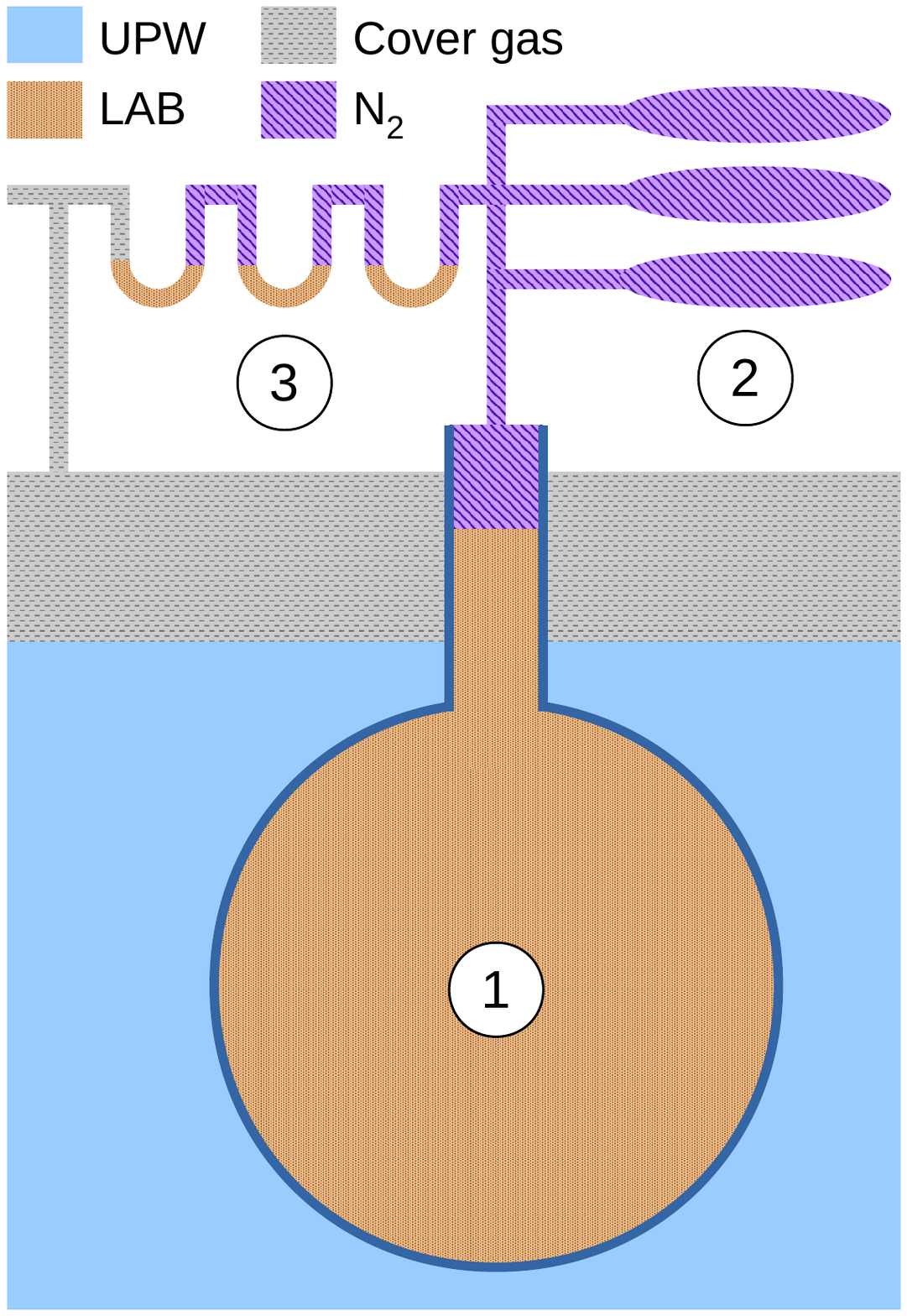}
    \qquad
    \includegraphics[height=80mm]{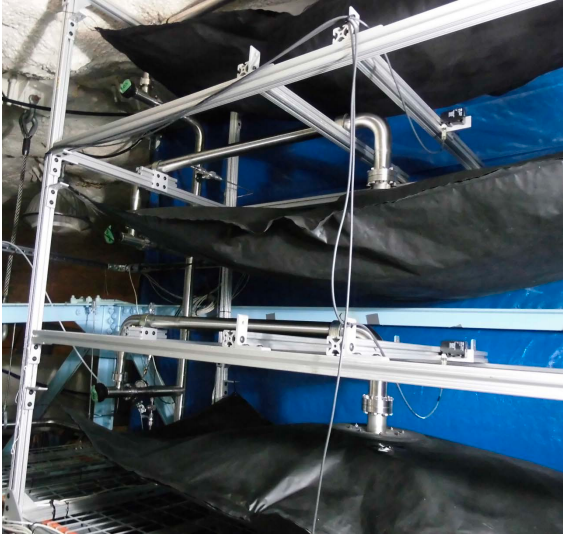}
    \caption{Left: Schematic of the AV cover gas system (not to scale), showing the AV (1) filled with LAB, the AV cover gas bags (2) filled with N$_2$, and the U-traps (3) which contain both LAB and N$_{2}$ and are connected to the cavity cover gas.
        Right: Photo of AV cover gas bags (black).
    }
    \label{fig:covergas}
\end{figure}

To protect the AV while preventing mine air from entering, three specially designed bags are coupled to the UI and form part of the cover gas volume (see \Cref{fig:covergas}).
Combined, they provide up to 360\,L of expansion volume that compensates for the small variations in the pressure typically observed at SNOLAB.
Additionally, a two-way pressure safety device (PSD), which consists of three U-traps connected in series, has also been installed.
The lower parts of these traps are filled with LAB; the fluid level defines a threshold on the maximum differential pressure before gas is allowed to bubble through the system.
A large change in pressure that surpasses the capacity of the cover gas bags will cause gas to pass through the PSD in whichever direction is necessary to equilibrate the pressure.
The three U-traps decrease the possibility of contamination from dissolved gases transported through the device by diffusion and convection.
A diagram and photo of the system can be seen in \Cref{fig:covergas}.

A dedicated electrostatic radon monitor with a sensitivity $\mathcal{O}$(1)\,mBq/m$^3$ was built and installed to continuously monitor the radon concentration inside the scintillator cover gas system.
The system has been in operation since October 2018.
The monitor shows that the cover gas system gives a reduction factor of below 1.9\ee{-4} in radon content relative to mine air.
Efforts are underway to further reduce Rn ingress into the AV cover gas headspace.

    \section{Electronics Upgrades}
\label{sec:electronics}

The SNO+ trigger and readout electronics record PMT signals with nanosecond-level timing resolution and charge thresholds well below one photoelectron per channel.
Although the data rate is several kHz during normal operations, the read-out electronics can handle short bursts of much higher rates, such as would be expected in the case of a galactic supernova.
Compared to SNO, the primary difference in the electronics requirements for the SNO+ experiment is the higher average channel occupancy as a result of the higher light yield of the liquid scintillator target.
The trigger is an analog system, where DC-coupled current sums are proportional to the rate of photon detection.
The expected higher rates in SNO+  necessitate both a higher-bandwidth readout and modifications to reduce power in the system.
Experience from SNO led to analog trigger cards upgrades which provide a more steady baseline.
A new multipurpose trigger utility board was developed and installed.

\begin{figure}[htp]
    \centering
    \includegraphics[width=0.8\textwidth]{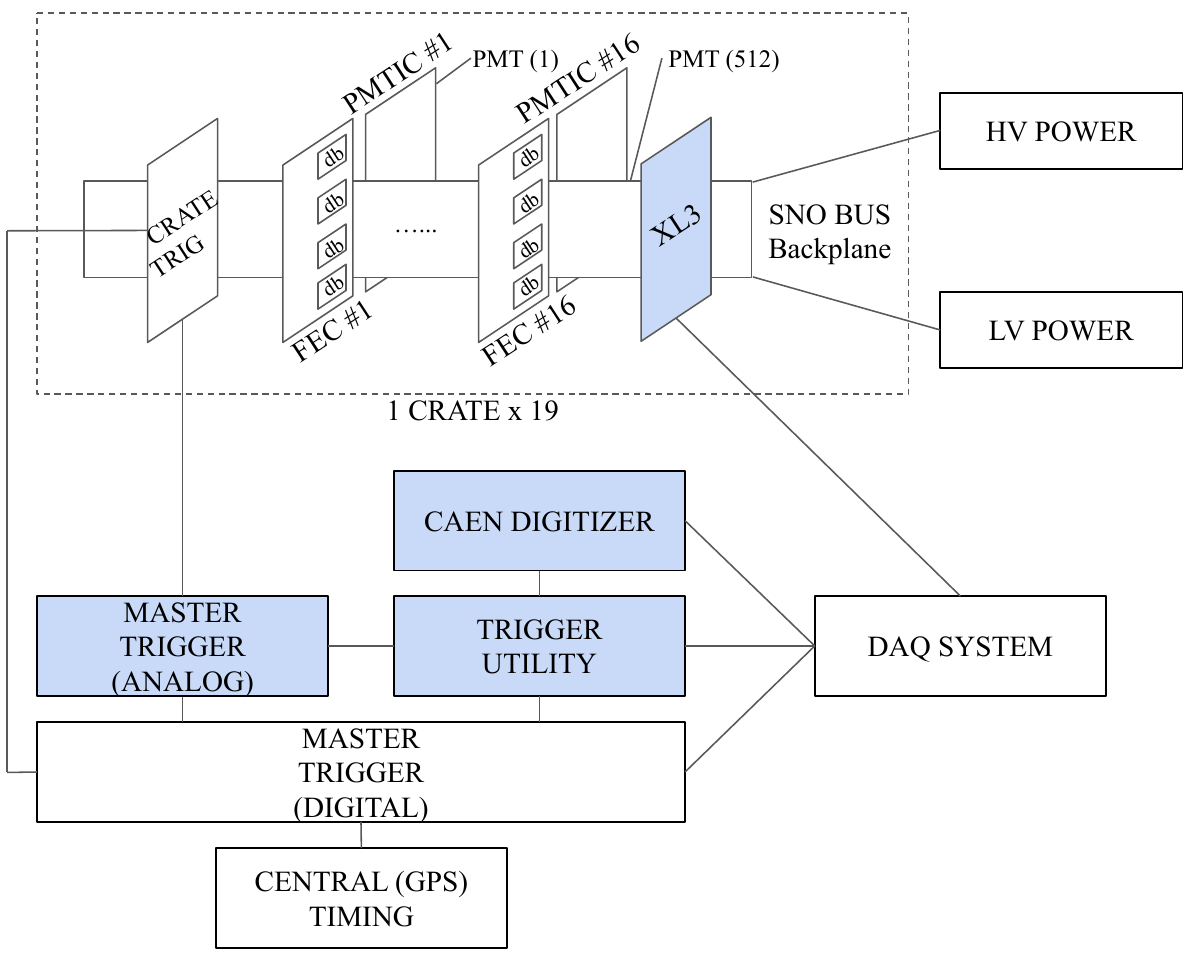}
    \caption{A schematic of the SNO+ electronics.
        Adapted from~\cite{bonventre_2014}, with the updates with respect to SNO highlighted.
    }
    \label{fig:electronics}
 \end{figure}

The SNO+ experiment reuses the SNO readout and trigger electronics~\cite{snonim} wherever possible.
The front-end readout cards (FECs) responsible for isolating and digitizing PMT signals, measuring channel hit timing, and generating trigger signals are unchanged.
Also unchanged is the master trigger system, which includes a set of majority triggers (sum of discriminated inputs) with multiple coincidence windows, calibration interfaces, and a pulser.
The GPS time synchronization system is also preserved from SNO.
A schematic of the electronics, with the new components highlighted, is shown in \Cref{fig:electronics}.

The readout system consists of 9728 electronics channels, 32 of which are reserved for calibration inputs.
Channels are distributed across 19 front-end crates.
Each crate carries 16 FECs that hold 32 channels, a trigger board and a crate control board.
Communication within the crate is over a custom interface that is VME-like but includes additional grounding.
Prior to commissioning the SNO+ detector, the SNO electronics were cleaned, refurbished, and repaired as necessary.
Additionally, small modifications were made to address known issues and proactively replace failure-prone components.
Maintenance on these electronics is ongoing as the experiment runs.

The SNO+ electronics record the time and three integrated charges for each PMT hit that has crossed a certain individual channel threshold, 0.25\,photoelectrons.
This threshold is known from data obtained using deployed calibration sources (see \Cref{sec:radsources}).
Due to upgrades to the stability of the electronics and a resulting reduction in noise, events can be discerned when fewer PMTs generate a signal within the trigger window than in SNO.
In the water phase, the detector could be set to trigger with only seven out of the $\sim$9000 PMTs, which roughly corresponds to 1\,MeV.
Weekly electronics calibrations of each PMT channel for charge and timing constants are also performed.
These constants are integrated into the data processing.
The recorded time has an offset due to the cable delay and internal electronics paths.
Furthermore, the time resolution of the PMTs are degraded by an effect arising from the combination of the fixed discriminator level and variations in the pulse sizes (``time-walk'').
Regular PMT calibrations remove the time offsets and account for time-walk.
PMT calibration data are taken by injecting nanosecond-width light pulses from fixed locations in the detector (see \Cref{sec:opsources}).

\subsection{Refurbishing the SNO Electronics}

During the course of SNO, it was noticed that build-up of a corrosive agent on the printed circuit boards could lead to low-impedance electrical connections between traces or components as well as the complete erosion of some traces.
In order to study this corrosion, tests were performed using silver and copper plates placed at different locations in SNOLAB.
These tests indicated that the corrosion product was a copper sulphide compound, 
which was the result of the electronics being exposed to low levels of H$_2$S in the mine air~\cite{snolab_sulfide, snolab_h2s}.
The time to failure due to this corrosion varied widely across the printed circuit boards, and was typically on the order of years.
Corrosion was also observed on other underground components, such as disk drives and cable connectors.
Because this corrosion could lead to board failures --- and in extreme cases would actually permanently damage the circuit boards --- a procedure was developed to clean all the electronic boards, which included removal of excessive heat compound from the sides of the heat sinks, ultrasonic cleaning, and oven baking at 80\degC{} for two hours.
In addition, all of the cards in one crate were conformally coated to study its use in the prevention of corrosive build up.
Carbon filters have also been added to air handling systems to suppress H$_2$S and help reduce this issue.

In addition to cleaning all FECs and eight-channel charge integration daughter cards, several modifications were made to each board to prevent known failures that occurred during SNO data taking.
This included small changes to regulator voltage monitoring circuits, replacement of frequently faulty passive components, and upgrades to the high voltage cards in order to improve individual PMT supply current monitoring and mitigate breakdowns, as described in \Cref{sec:PMTrepair}.
Each board underwent a rigorous testing procedure prior to re-installation.

\subsection{Readout System Upgrade}

In order to handle the increased data rates expected with the deployment of the liquid scintillator target, the SNO front-end readout electronics have been replaced with a new system for the SNO+ experiment.
In the SNO architecture, the central DAQ readout was bandwidth-limited to $\sim$2\,Mbps~\cite{snonim}.
In SNO+, a new crate controller card (XL3) polls the memory of FECs within a crate, assembles the data into Ethernet packets, and pushes the data to a central server over a 100\,Mbps link using TCP/IP.

The interface to the crate backplane and high voltage control circuitry are identical to SNO, with the readout logic implemented in an evaluation board\footnote{Xilinx ML403. \href{https://www.xilinx.com/support/documentation/boards_and_kits/ug080.pdf}{https://www.xilinx.com/support/documentation/boards\_and\_kits/ug080.pdf}} that integrates an FPGA\footnote{Virtex-4 XC4VFX12-FF668-10. \href{https://www.xilinx.com/support/documentation/data_sheets/ds112.pdf}{https://www.xilinx.com/support/documentation/data\_sheets/ds112.pdf}} and PowerPC processor.
The FGPA runs a state machine that pulls any available digitized data from FECs and stores it into a shared memory buffer.
The PowerPC runs a C program that handles Ethernet communication, sending assembled data packets to the central DAQ.
The hybrid combination of high-level C code and a low-level VHDL state machine provides a cost-effective solution for reading out the custom SNO electronics via TCP/IP using commodity network hardware.
The total bandwidth for parallel readout of the full detector with this new system is approximately 400\,Mbps, about a factor of 100 greater than the expected nominal data rate.

\subsection{Analog Trigger Upgrade}
\label{sec:trigger}

In addition to the front-end readout, the trigger system required upgrades in order to accommodate a higher rate of hit channels.
The primary SNO+ physics triggers are comprised of a set of analog majority triggers.
Each hit channel contributes a current pulse, and the detector-wide sum is compared to an adjustable threshold.
Each PMT threshold crossing generates two square current pulses, $\sim$100\,ns and $\sim$40\,ns wide, and a shaped copy of the PMT pulse.
These pulses are summed separately detector-wide for inward- and outward-looking PMTs.
Multiple thresholds can then be applied to each of these sums.
Together with pulsers and external inputs, these channels define a rich set of triggers from which readout conditions are built.

The final summing of these analog pulses and the threshold comparison is performed on a set of central analog master trigger cards (MTC/A).
The increase in the rate of hit PMTs expected with scintillator would increase the power through the SNO MTC/A to unacceptable levels.
The MTC/A+ is an upgraded analog master trigger card used in SNO+.
It works on the same principle as the MTC/A, but replaces a sequence of current mirrors with high-speed, ultralow-noise current-feedback operational amplifiers.\footnote{Texas Instruments THS3001. \href{http://www.ti.com/product/THS3001}{http://www.ti.com/product/THS3001}}
The MTC/A+ can perform measurements of an attenuated version of the large amplitude signal (600\,\textmu{}A per channel) from the upstream legacy SNO trigger system.
The signals are summed with three different gains, allowing single-channel resolution up to $\sim$100\,hits and approximately three-channel resolution over a 3000-channel dynamic range.
Each of these three gain paths are compared to an independent user-defined threshold before output for digitization and storage with event data.
Additional upgrades included installing programmable reed relays to remove individual crates from the trigger sum, improving uptime and diagnostic capabilities, and implementing active compensation for long-term DC drifts in the input baselines.

The trigger logic is now implemented in a complex programmable logic device\footnote{Xilinx XC2C512 CPLD. \href{https://www.xilinx.com/support/documentation/data_sheets/ds096.pdf}{https://www.xilinx.com/support/documentation/data\_sheets/ds096.pdf}} rather than discrete logic, improving flexibility.
The firmware uses feedback from the digital master trigger system (MTC/D) to automatically re-trigger for events extending beyond the nominal 400\,ns event window, avoiding dead time by generating back-to-back triggers.
It also includes the ability to force a fixed number of automatic re-triggers for any trigger channel, opening a threshold-less gate of fixed duration.
Programmable firmware allows for future enhancements as well, such as using signals from two of the discriminator thresholds to form an energy window trigger and open a low-energy trigger gate for coincidence searches.

The analog waveforms that are summed on the MTC/A+ cards are read out by a digitizer,\footnote{CAEN v1720 12-bit digitizer. \href{https://www.caen.it/products/v1720}{https://www.caen.it/products/v1720}} which is configured to read out a 416\,ns window around each event with a 4\,ns sampling time.
It is synchronized with the rest of the DAQ electronics through its LVDS front panel inputs and read out using its VME interface.
The digitized waveforms are used for removing instrumental effects and assessing data quality.

To provide an interface between the optical calibration systems discussed in \Cref{sec:opsources} and the trigger system, a custom auxiliary trigger board called the Trigger Utility Board Mark  \lowercase\expandafter{\romannumeral2} (\tubii) was developed.
\tubii expands on the role of the Trigger Utility Board (TUB) from SNO.
In addition to providing I/O for the calibration systems, \tubii has general purpose trigger inputs that are fed into a daughter card called the MicroZed.\footnote{Xilinx MicroZed 7020 SoM. \href{http://zedboard.org/product/microzed}{http://zedboard.org/product/microzed}}
The MicroZed, much like the programmable device on the MTC/A+, allows for a flexible logical trigger system.
\tubii has access to a large number of possible trigger inputs, allowing for a wide variety of potential trigger conditions.
\tubii also transforms the analog signal from the MTC/A+ cards to fit within the digitizer's peak-to-peak range of 2\,V.
It does so by optionally clipping or attenuating the signal, allowing one to maintain resolution or the full dynamic range on any digitized signal.
Other utilities provided by \tubii include a programmable discriminator threshold and tunable delays for trigger timing to reduce readout dead-time.

    \section{Data Acquisition and Processing}
\label{sec:daq}
The SNO+ detector is operated at all times by a member of the SNO+ collaboration, relying on a suite of custom software tools for detector monitoring, data acquisition (DAQ), and data processing.
Offline analysis systems are responsible for simulation of the detector response using Monte Carlo (MC) techniques and analysis of the data.
The following sections briefly introduce the tools for control and monitoring of experiment subsystems (slow control), DAQ software, real-time data quality analysis (nearline), detector simulations, and data processing.

\subsection{Slow Control}
Detector conditions that vary on time scales of seconds to minutes are monitored with the SNO+ slow control system.
This system consists of four I/O servers,\footnote{Acromag IOS-7200. \href{https://www.acromag.com/resource/ios-7200-user-manual}{https://www.acromag.com/resource/ios-7200-user-manual}}
the DeltaV process monitoring and control system,\footnote{DeltaV\texttrademark{} software. \href{https://www.emerson.com/en-us/catalog/deltav-monitor-and-control-software}{https://www.emerson.com/en-us/catalog/deltav-monitor-and-control-software}}
and a web interface for operator monitoring.

The I/O servers contain analog input cards for measuring various voltage readings,\footnote{Acromag IOS-320. \href{https://www.acromag.com/resource/ios320-user-manual}{https://www.acromag.com/resource/ios320-user-manual}} as well as a card for digital output.\footnote{Acromag IOS-408. \href{https://www.acromag.com/resource/ios-408-user-manual}{https://www.acromag.com/resource/ios-408-user-manual}}
The system measures and monitors the status of detector components on the deck.
All power supply voltages in the PMT electronics racks and timing electronics rack are recorded by the I/O servers.
The servers also monitor and control 14 magnetic compensation coils embedded in the cavity walls.

DeltaV is a distributed control system used to monitor and control many aspects of detector operation.
Monitoring data relevant for detector operations and data quality validation include the AV hold-up and hold-down rope tensions, acrylic vessel position, cavity water depth and temperature, deck humidity, AV liquid levels, cover gas pressures, the status of the uninterruptible power supply, and operational parameters related to cavity filling/recirculation.

All slow control data from each detector component are replicated to a single database for alarm posting and operator viewing in a web interface.
This provides detector operators a single online location that can be used to view the data and status of all slow control components.
Additionally, alarms are issued on the overview page and in the alarm GUI whenever a detector component is operating outside predefined conditions.
Any alarm issued is also broadcast via email to detector experts and stored in the slow control database for data quality checks.

\subsection{Data Acquisition and Monitoring Software}
A new DAQ software system has been developed for the SNO+ experiment to accommodate changes to the front-end readout scheme.
A set of DAQ server processes handle low-level data flow from front-end crates, the trigger system, and calibration sources.
The XL3s in each crate asynchronously send TCP/IP packets with PMT data to a central ``data server'' program.
Independently, the trigger information is read out from the MTC/D by a single board computer over a VME backplane, and is also sent to the data server.
The data server then passes this data to an event builder system that aggregates these fragments into event records based on their global trigger identification number and writes complete events to disk.

The builder turns the continuous data stream from detector readout electronics into discrete physics events, each associated with a trigger.
This process is identical to that followed by the SNO experiment and is detailed in~\cite{snonim}.
The resulting ``Level 1'' (L1) raw data files contain all the detector information such as PMT charge, hit times, and trigger settings, and are stored in the ZEBRA data bank (ZDAB) format~\cite{zebra}.
An SQL database records hardware settings on a run-by-run basis for use in simulations and data analysis, as well as run quality verifications.

The hardware configuration and run control is implemented in the ORCA GUI~\cite{ORCA}.
Low-level hardware-oriented interfaces access all configurable parameters of the detector electronics, \eg individual PMT channel thresholds, while a high-level standard operator interface automatically configures the detector according to standard run definitions.
The ORCA GUI also includes configuration interfaces for much of the \emph{in situ} optical fibre calibration hardware described in \Cref{sec:opsources}.

DAQ monitoring is performed in a set of online and nearline systems.
Detector operators use a variety of custom GUI and web-based tools to monitor low-level metrics such as the trigger rate, DAQ and event builder status, and use a live event display with a latency on the order of seconds to track detector performance and diagnose any issues.
An integrated alarm system aggregates alarms from across all subsystems and rapidly alerts the operator.

Continuous data quality monitoring is performed in a nearline system, where recently completed runs are processed with low-level analysis modules.
These provide continuous detector performance monitoring such as signal gain, channel occupancy and trigger pulse shape, among other metrics.
Higher-level nearline data quality checks such as continuous trigger level and channel gain monitoring allow operators to immediately identify completed runs unlikely to pass analysis criteria and perform the appropriate corrective actions.

Additional monitoring tools include a remotely accessible digitizing oscilloscope showing the summed analog trigger signals, a network of webcams in the underground area, and a speaker connected to the global trigger signal which produces a rate-dependent sound similar to a Geiger counter.
About 80\% of detector time is spent in taking physics data, with the remaining time spent performing calibrations or regular maintenance.

\subsection{Data Processing}
\label{sec:processing}
In addition to the L1 files produced by the builder, the data pipeline can produce ``Level 2'' (L2) files containing events that satisfy additional criteria.
This feature was designed to manage the large data rates expected in the scintillator phase of the experiment.
The trigger threshold is continuously adjusted to stay within the data rate capacity.
In the initial water phase, both L1 and L2 files were produced and archived in triplicate.
In the future, it is anticipated that L2 files will be archived along with only a fraction of the L1 files.

When a data run is completed and its files produced, a run metadata object is created and stored in a database.
The metadata object contains the number of files produced, their names, and timestamps of the beginning and end of the run.
In addition to the run metadata, the database also accumulates information on detector geometry, conditions, and calibration constants including: individual PMT states, PMT noise levels, active triggers, and trigger thresholds.
This database information can be accessed at any point of the data pipeline, and can be used to configure subsequent analyses and simulations.

Data files are transferred from SNOLAB to replica sites using the CERN File Transfer Service (FTS),\footnote{\href{https://fts.web.cern.ch/}{https://fts.web.cern.ch/}}
which ensures that the file is copied to three separate storage locations before removing the copy on the SNOLAB computers.
The replica sites are ComputeCanada resources at Simon Fraser University (SFU), Fermi National Accelerator Laboratory (FNAL), and the Rutherford Appleton Laboratory (RAL), the last of which has been designated as the permanent archive for raw data files.

\subsection{Detector Simulation}
\label{sec:simulation}
The primary offline analysis infrastructure is based on RAT,\footnote{RAT (is an Analysis Tool) User’s Guide. \href{https://rat.readthedocs.io/en/latest/index.html}{https://rat.readthedocs.io/en/latest/index.html}}
which was originally developed for the Braidwood reactor experiment~\cite{braidwood} and is used by the MiniCLEAN~\cite{miniclean} and DEAP~\cite{deap} dark matter experiments, among others.
In RAT, detailed detector simulation is handled by GEANT4~\cite{geant4,geant4-2} and is based on GLG4sim~\cite{glg4sim}, a generic liquid scintillator package originally developed by the KamLAND~\cite{kamland} experiment.
Initial particles in an event are simulated using a series of generators: a vertex generator, which simulates the physics interaction, including the particles and their kinematics; a position generator, which places the particles in an appropriate location in the detector; and a time generator, which attaches a suitable timestamp to the event.
Custom high-level generators that integrate all three steps have been implemented for specific physics processes and backgrounds, such as certain radioactive decay chains, solar and reactor neutrinos, and calibration processes.

Once the initial event has been generated, it is passed to GEANT4, which simulates the propagation of particles and their daughters through the detector materials.
The processes modelled include the production of optical photons via Cherenkov and scintillation processes, scattering, absorption, and reemission from wavelength shifters, and PMT charge and timing response.
A detailed description of the SNO+ detector has been implemented in RAT, down to each pipe, rope, and the internal structure of the PMTs, with particle tracking terminating at the PMT dynode stack.

Detector and trigger conditions for specific data runs are stored in a database.
This data can then be employed by RAT to simulate a given data-taking run.
In this way, a full MC dataset of many runs, incorporating varying conditions as recorded by the detector itself, can be produced for analysis alongside the corresponding data.
As a last step, the front-end electronics, trigger system, and event builder are simulated, producing simulated data files which can be passed through the same processing pipeline as the raw data files.

\subsection{Grid Processing}
The processing of raw data and production of MC data is performed primarily on ComputeCanada resources at SFU.\footnote{\href{https://www.computecanada.ca/}{https://www.computecanada.ca/}}
Ganga is used for the submission and management of jobs and storage via a custom GangaSNOplus plugin.\footnote{\href{https://ganga.readthedocs.io/en/stable/}{https://ganga.readthedocs.io/en/stable/}}
A CouchDB database is used to track data files and jobs.
Job requests are stored as documents in this database, which are then accessed by a Grid worker node.
Typically based on job priority, the node will select a subset of jobs to run if prerequisite conditions are met.
Output data are catalogued in the same database, and archived in Grid storage and/or further databases.
The processed data and produced MC datasets are stored primarily at SFU.

The input data are then processed in order to remove known electronics effects and physics events that are not of interest.
A processing job achieves this in two steps using the RAT infrastructure.
The first step applies low-level data cleaning criteria which are designed to flag instrumental artifacts, and produces database records that are used in subsequent processing steps.
One such use of database records is to associate events that share a common topology such as muons, which generate sufficient light in the detector to result in multiple triggers.
The second step builds on the first step by applying further data cleaning criteria based on those database records, \eg to flag said muon events as well as secondary events caused by them.
This step also converts the ZDAB files into a structured ROOT-based~\cite{root} format known as the ``RAT data structure'' (RATDS).

From this point, there are additional processing steps that depend on the type of data, such as physics or specific calibration runs.
These later steps apply algorithms to reconstruct physical quantities such as position, time, and energy, and calculate figures of merit for event classification.
At the same time, PMT calibrations (see \Cref{sec:opsources}) and a data-blinding procedure can be applied.
The SNO+ reconstruction algorithms determine the event position and energy, as well as other event characteristics.
The tail of late photon signals in scintillator allows for discrimination between events caused by $\alpha$s, $\beta$s or $\gamma$s.
The time difference between successive events can be used to tag specific decays of short-lived isotopes, and thus identify parts of internal radioactive decay chains.
Finally, the time difference of the signals between groups of PMT hits in the same event can be used to reject pile-up from different events in the same trigger window.

The end results of the processing job are both a RATDS file and a smaller summary ROOT ntuple file.
The former contains the raw detector data and all quantities calculated during the different processing steps.
The latter contains only the subset of information necessary for most physics analyses, and is the most commonly used because of its much smaller size.
Specific output files for calibration runs are also produced.

    \section{Calibration Systems}
\label{sec:calibration}

%
%

The SNO+ detector is calibrated using optical and radioactive sources temporarily deployed inside the AV and in the region between the AV and the PSUP (see \Cref{fig:cavity}), as well as optical sources permanently mounted on the PSUP.
The optical sources are used to measure \emph{in situ} the optical absorption and scattering of the detector media in order to calibrate the PMTs' timing and charge response, as well as their total and relative efficiencies.
The radioactive sources are used to calibrate and determine the systematic uncertainties associated with reconstructed quantities such as energy, position, and direction.

In order to match the very stringent radiopurity requirements described in \Cref{sec:targets} and material compatibility requirements with both the scintillator and UPW, the SNO+ calibration source deployment hardware system was fully redesigned with respect to the system used in SNO while maintaining the same operating principle.
As was the case with all hardware used in the SNO+ experiment, great care was taken in the design of the sources in order to minimize the possibility of contamination.

In order to determine the positions of the calibration sources deployed inside the detector (independently of the PMT array data) and monitor the position of the AV and the hold-down rope system, a system of six cameras in clean underwater enclosures was installed onto the PMT array.

In this section, the calibration program and hardware are described.
An overview of requirements of the calibration program is given, followed by a description of the systems that allow the deployment of the internal calibration sources.
Subsequently, a description of the camera system is presented.
Finally, the optical and radioactive calibration sources are described.

%
%

\subsection{Calibration Goals}
\label{sec:cal_goals}

The SNO+ collaboration has developed an extensive \emph{in situ} calibration program aimed at measuring parameters describing the production, propagation and detection of scintillation and Cherenkov light in the detector, as well as evaluating the performance and systematic uncertainties of the reconstruction algorithms, including particle identification.
The main detector parameters to be determined \emph{in situ} are the scintillation light yield, absorption and scattering coefficients, as well as PMT gains, time offsets and efficiencies.
These calibrations will complement extensive \emph{ex situ} measurements characterizing the PMT array~\cite{snonim,snonim_pmt} and the scintillator properties~\cite{snop_scint,vonKrosigk_2015, OSullivan_2012}.
Validation of the detector model includes the regular monitoring of the stability of those parameters, as well as the verification of the detector geometry description --- in particular the position of the AV and of the hold-down rope-net.

The calibration plan takes advantage of the sequence of SNO+ data taking phases.
The water phase was used to extensively characterize the PMT array and AV, profiting from the higher absorption and scattering lengths of the UPW with respect to the scintillator.
In fact, several deployments with optical and radioactive sources were carried out inside and outside the AV, in addition to an extensive use of the fixed optical sources.
The use of deployed optical sources inside and outside the AV allowed for the calibration of the PMT and reflector array up to high incidence angles of $\sim$60\textdegree{}, as well as a detailed parametrization of attenuation in the AV and external water regions.
Radioactive source events were simulated using the \emph{in situ} measured light propagation and detection parameters and compared to water-phase calibration data, confirming the validity of the optical calibration parameters.
All of these measurements will be used in the scintillator phase, where the scintillator properties will also be characterized.

%
%

\subsection{Radon Mitigation for Calibration Hardware}
\label{sec:radon}

The calibration deployment system can affect the radon and radon daughter content in the AV in two ways: through contamination of the cover gas and through direct contact with the scintillator.
As stated in \Cref{sec:covergas}, the allowed limit for the number of \isotope[222]{Rn} decays in the cover gas above the scintillator volume is 650 counts per day.

To maintain the cover gas purity by preventing exposure to the measured 123\,Bq/m$^{3}$ activity in mine air, each of the calibration system elements were stringently tested for gas tightness and radon emission.
The maximum leak rate for the entire system is less than $\mathcal{O}(10^{-6})$\,mbar$\cdot$L/s.
Leak tests determined the leak rate to be an order of magnitude lower for all the permanently connected elements, and of the same order of magnitude for elements that will be connected to the cover gas only during calibrations.
The materials used in the calibration systems have been measured directly for both compatibility with the scintillator (see \Cref{sec:materials}) and emanation of radon.
Where possible, electropolished stainless steel was used to provide the cleanest possible surfaces.
Measurements of the radon emanation for the materials used in the Universal Interface (UI, see \Cref{sec:UI}) have been conducted; results are shown in \Cref{tab:nasimEman}.
Of these components, only the calibration side ropes and the UI are in permanent contact with the active target material.

Exposure of the scintillator cocktail and cover gas to the calibration deployment systems is reduced by minimizing the time it stays in contact with the respective components.
The Umbilical Retrieval Mechanism (URM, see \Cref{sec:urm}) and the gloves used to mount the sources (see \Cref{sec:UI}) are isolated when not in use.
While the gloves are the largest source of Rn, they were chosen for their compatibility with LAB~\cite{barnard_2014}.
Diffusion of air from the sealed space in which the side ropes are stored is minimized by the length of the tubes through which the ropes enter the UI.

\begin{table}[htp]
    \centering
    \caption{Measured radon emanation from components of the source manipulator and cover gas systems.}
    \label{tab:nasimEman}
    \begin{tabular}{| c | c |}
        \hline
        Component & Emanation (atoms/day) \\
        \hline
        Bags (total) & $230\pm83$ \\
        UI & $\sim250$\\
        URM aluminum & $<219$ \\
        URM feedthrough    & $160\pm40$ \\
        URM box leaks & $<240$ \\
        Gloves & $\sim11520$ \\
        Umbilical & $9\pm9$ \\
        Calibration side ropes & $66\pm56$ \\
        Calibration central rope & $45\pm42$ \\
        \hline
    \end{tabular}
\end{table}

%
%

\subsection{Source Manipulator System}
\label{sec:sms}

Even though the detector is spherically symmetric to first order, the presence of the neck, hold-down rope-net, hold-up ropes with their associated grooved AV plates, and gaps in the PMT array (see \Cref{sec:psup}) all create asymmetries that affect the observation of light.
In addition to this, the difference in individual PMTs, the presence of inactive PMTs, and the ability of the AV to move relative to the PSUP create further asymmetries in the detector response.
For these reasons, it is essential to allow the deployment of sources in many different positions, especially beyond the central vertical axis.

The SNO+ Source Manipulator System (SMS) is largely based on the one used in SNO~\cite{snonim}.
It allowed flexibility in the variety of deployed sources, a large phase space of sampled positions,
and an almost complete degree of automation in source movement.
For these reasons, the SNO+ SMS uses the same design principle, but with several new adaptations and redesigns aimed at meeting the SNO+ scintillator compatibility and radiopurity requirements.
The source insertion system must be gas-tight with respect to the laboratory air, and all construction materials must either have low radon emanation, or be used in a way to limit their exposure time with the scintillator or cover gas.

\begin{figure}[htp]
    \centering
    \includegraphics[width=0.6\textwidth]{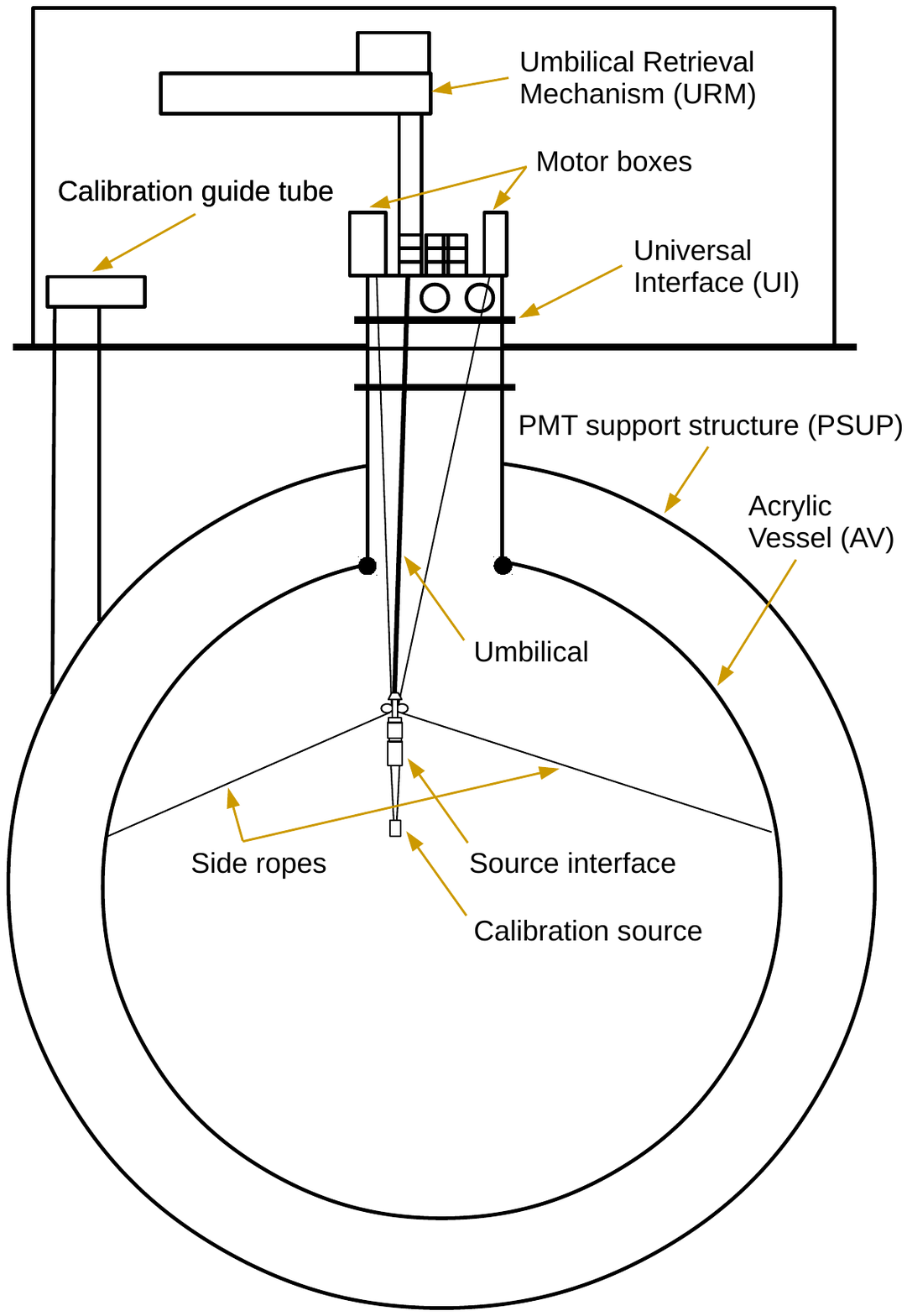}
    \caption[Source Manipulator System]{An overview of the Source Manipulator System (not to scale) showing the main components.
        Only one set of side ropes is shown, as the other set of ropes is in the plane orthogonal to the cross section shown here.
        Note that the central rope securing the source is not shown separately; it follows the same path as the umbilical.
    }
    \label{fig:sms_overview}
\end{figure}

The SMS is shown in \Cref{fig:sms_overview}.
Signals are sent to deployable sources through an umbilical (\Cref{sec:umbilical}), which is manipulated along the central vertical axis of the detector using the URM (\Cref{sec:urm}), and off-axis within two vertical perpendicular planes using the side rope manipulator system (\Cref{sec:UI}).

Sources can also be deployed in the region between the AV and PSUP using one of six calibration guide tubes.
This region has less stringent requirements on bulk radioactivity compared to the AV.
Therefore, as the SNO SMS is also compatible with UPW, it is used for the deployment of the sources in this volume.

\subsubsection{Umbilical Retrieval Mechanism}
\label{sec:urm}

\begin{figure}[htp]
    \centering
    \includegraphics[width=0.9\textwidth]{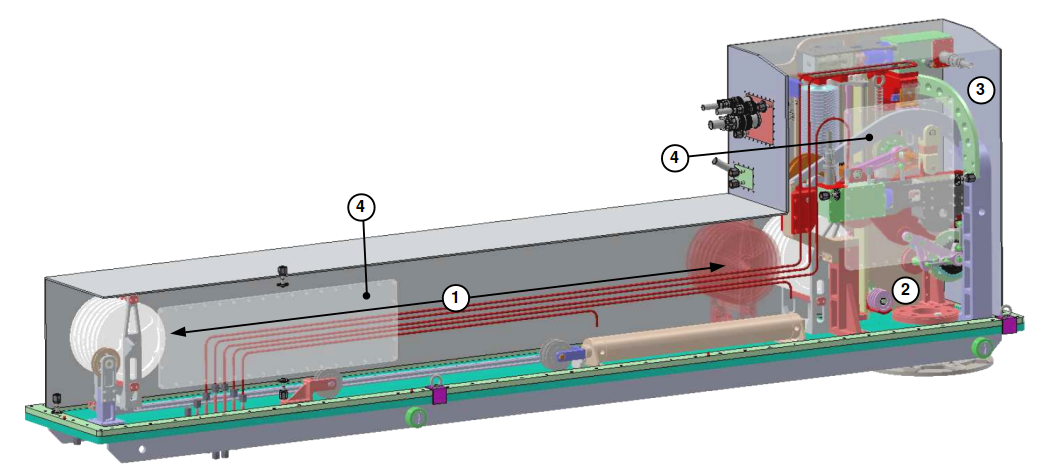}
    \caption[The URM]{Schematic of the Umbilical Retrieval Mechanism (URM), showing (1) the umbilical storage system, (2) its drive system, (3) the rope mechanism, and (4) the acrylic viewing ports.}
    \label{fig:urm}
\end{figure}

The URM drives the motion of the deployed calibration sources and uses both an umbilical and a central rope.
The umbilical allows the power cables, signal cables, and/or gas capillary lines required by the sources to connect to equipment on deck within a closed system.
The rope supports the source weight to avoid a mechanical load on the umbilical.
The URM is comprised of an umbilical storage system, an umbilical drive system, and a rope drive mechanism, all of which are contained within a large box and connected to the detector by a single lower opening (see \Cref{fig:urm}).
Source movement is driven by two motors with ferrofluidic feedthrough connections.
Controls include motion encoders and load cells, which are all read out externally.
Since some of the 30-m long umbilicals contain optical fibre bundles, it is necessary to avoid twisting the umbilical and to keep its bending radius above 7.6\,cm.

The umbilical storage mechanism consists of two sets of six co-axial 10\,cm radius pulleys.
The umbilical is wound around those two sets, one of which is movable, so the umbilical is fully retracted into the URM when the distance between the pulley sets is maximal.
A pneumatic piston pulls on the movable pulleys to keep the umbilical constantly under tension.
Safety guards hold the umbilical in place close to the pulley in case of loss of tension.

The umbilical drive mechanism had to be updated with respect to SNO because the SNO umbilical is not compatible with LAB, and the tygothane material chosen for the SNO+ umbilicals has much higher slippage in the pulleys than the SNO umbilicals.
Wetting the umbilical with LAB also increases the slippage.
The larger pulleys that drive the umbilical have a V-shape and small grooves on the surface in order to improve grip.
Instead of a single drive pulley, the umbilical is forced around two drive pulleys of different diameters that have their movement synchronized by a set of gears and chains.

The central rope mechanism is unchanged with respect to SNO.
The 30-m long rope is stored in a single custom-threaded drum and deployed using a winder mechanism.

All these mechanisms are enclosed in a 2.75-m long aluminum box and sealed with a double O-ring closure.
Two acrylic viewing ports are included that allow inspection of the internal mechanisms without breaking gas-tightness.
Helium leak-checking indicated a leak rate of 8\ee{-6}\,mbar$\cdot$L/s for the whole box.

Once retrieved from the detector, the umbilical and the central rope return to the URM soaked in LAB, so any part that they come into direct contact with was built from materials tested for LAB compatibility (see \Cref{sec:materials}): POM, PTFE, 304 and 316 stainless steel.
Compatibility tests included prolonged soaking and optical absorbance measurements of the residual LAB.
The base of the URM was designed with draining grooves to prevent any LAB inside the URM from flowing back into the detector.
This precaution serves to remove any LAB that might have come in contact with other materials such as aluminum, which was used to build other parts of the URM in order to minimize weight.

The requirements for the new URM were set to limit the radio-contamination in the detector while calibrating and ensuring chemical compatibility with the scintillator.
Emanation of \isotope[222]{Rn} from the aluminum was estimated by $\gamma$ counting \isotope[226]{Ra} from a production sample and applying a scaling factor that provides the emanation from the bulk activity.
This factor was obtained in two ways: a calculation based on the diffusion length from~\cite{mamedov} and from a more direct measurement using ZnS(Ag) coated decay cells in~\cite{artradon}.
Both the central rope and umbilical drive motors have a high radon emanation, and are therefore isolated from the rest of the box atmosphere by two gas-tight boxes equipped with ferrofluidic vacuum-grade rotary couplings.
The radon budget for the URM is listed in \Cref{tab:nasimEman}.

\subsubsection{Interface with Acrylic Vessel}
\label{sec:UI}

\begin{figure}[htp]
    \centering
    \includegraphics[width=0.75\textwidth,trim={12mm 12mm 12mm 12mm},clip]{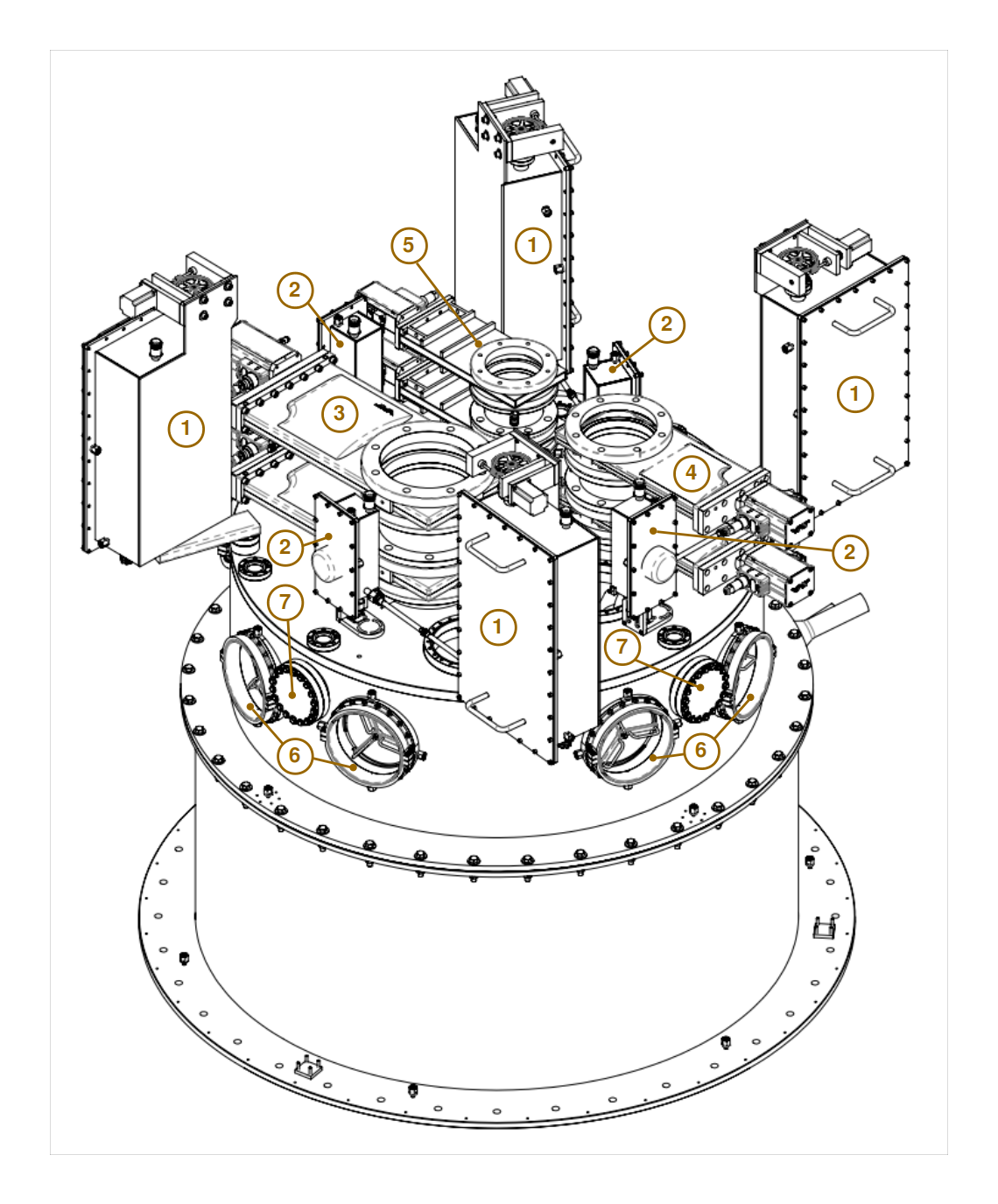}
    \caption[The Universal Interface]{Schematic of the cylinder-shaped Universal Interface on top of the AV, showing
        (1) the four side rope mechanism boxes,
        (2) the extension boxes that connect the rope mechanism boxes to the UI cover gas space,
        (3--5) the three flanges for source insertion (accessed by gate valves),
        (6) the (visible) glove ports and
        (7) the UI viewing ports with covers.
    }
    \label{fig:ui}
\end{figure}

The interface between the AV and the SMS is achieved through a large stainless steel cylinder called the Universal Interface (UI, shown in \Cref{fig:ui}), which is sealed to the top of the neck of the AV.
The URMs can be attached to any of three ports located on the top of the UI.
The attachment of the sources to the side ropes is done through glove ports in the UI.
The butyl gloves were chosen for their relative compatibility with LAB~\cite{barnard_2014}.
Special measures are taken to reduce the potential radon exposure from the gloves by nitrogen purging the glove volume before use and reducing the total use time.
The glove ports are closed when not in use to isolate them from the UI.
This isolation capability also allows the gloves to be exchanged if necessary.

One end of each side rope is mounted to an acrylic anchor block located on the inner AV above the equator.
The other end of each rope exits the top of the UI and enters a box that contains a drive mechanism to adjust both the length of rope deployed and the tension on that rope.
Since a substantial portion of the ropes remain in the AV permanently, they have more stringent radioactivity requirements than any external or non-permanent parts.
The Vectran rope used in SNO had a \isotope[40]{K} activity above the SNO+ target level.
This rope was therefore replaced with Tensylon, the same material chosen for the new hold-down rope system described in \Cref{sec:ropenet}.

In addition to the calibration hardware, the UI also serves as the interface for all the services that pass into the AV: piping for filling and draining the AV, piping for assays of the detector media, the cover gas system (see \Cref{sec:covergas}), and connections for monitoring devices needed inside the AV.
A change from SNO is that all seals are either stainless steel VCR fittings, ConFlat\textsuperscript{\textregistered{}} flanges, or double O-ring seals with pump/purge capability between the seals.
The devices used by the manipulator system to adjust the rope length are mounted directly on the UI.
The components and materials used in these devices are subject to the same material compatibility requirements as the cover gas system.
Double O-rings seal the covers of the rope boxes, and VCR fittings are used to conduct the rope from the UI to the spooling system in the rope box.
Because the motors that drive the spools are known to emanate radon, they are excluded from the interior of the rope box.
The motors are coupled to the rope spools through ferrofluidic feedthrough devices to prevent exchange of gas between the cover gas an exterior environment, as is done with the motors for the URM rope and umbilical.

\subsubsection{Source Umbilical}
\label{sec:umbilical}

\begin{figure}[htp]
    \centering
    \includegraphics[width=0.4\textwidth]{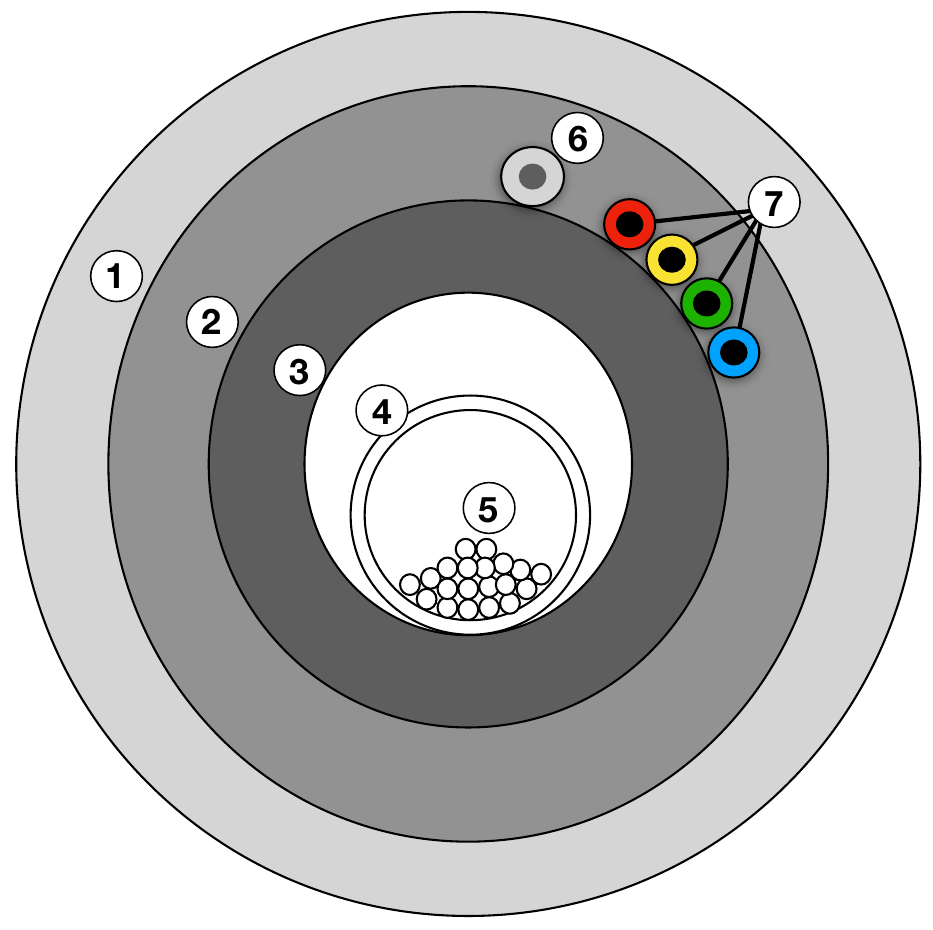}
    \caption[The Umbilical Cable]{Cross-section of a SNO+ umbilical cable.
        The outer layer (1) is a Tygothane tube that contains a LDPE tube (3).
        The inner tube is helically wrapped with four hook-up wires (7) and one coaxial cable (6).
        The volume in between (2) is filled with a silicone gel.
        In certain configurations, optical fibres (5) contained in their own protective tube (4) can be fed through the inside of the inner tube.
    }
    \label{fig:umbilical}
\end{figure}

The umbilical cables play an integral role as part of the calibration system.
Each cable is $\sim$30\,m long and encloses gas capillaries, optical fibres, and signal wires such that a calibration device may be connected to the deck.
A cross-sectional drawing of an umbilical cable is shown in \Cref{fig:umbilical}.

The inner tubing is used to protect optical fibres for the laserball (see \Cref{sec:laserball}) or to provide gas feed and return lines for the Cherenkov (see \Cref{sec:cherenkov}) and \isotope[16]{N} (see \Cref{sec:n16}) sources.
Surrounding that are four 24-AWG hook-up wires and one 30-AWG coaxial cable, helically wrapped around the tube to allow for bending of the umbilical.
During the source deployment, the 0.5-inch diameter outer layer will be in direct contact with the scintillator, and therefore must be chemically compatible and have low radon emanation.
Tygothane tubing\footnote{
    Tygothane\textsuperscript{\textregistered{}} Precision Polyurethane Tubing. \href{https://cdn2.hubspot.net/hub/108912/file-16501745-pdf}{https://cdn2.hubspot.net/hub/108912/file-16501745-pdf}
}
is compatible with the SNO+ liquid scintillator (see \Cref{sec:materials}), with radon emanation tests yielding 2.2\,mBq/kg.
The volume between the inner and the outer tubes around the wires is filled with a silicone gel\footnote{Wacker SilGel\textsuperscript{\textregistered{}} 612~A/B. \href{http://willumsen.se/silicones/wp-content/uploads/2014/11/SilGel_612.pdf}{http://willumsen.se/silicones/wp-content/uploads/2014/11/SilGel\_612.pdf}} with an added inhibitor that extended the cure time of the gel.
This allowed for the complete filling of the 30-m umbilical before the gel set.

Prior to mounting on the URM, the umbilicals are thoroughly cleaned using a UPW rinse and subsequent LAB soak.
This process is repeated multiple times to remove contaminants and monitor their rate of desorption from the umbilical surface.

\subsubsection{Umbilical Flasher Object}
\label{sec:ufo}

In order to determine the position of a deployed source using the camera system described in \Cref{sec:cameras} while the PMTs remain at HV, the Umbilical Flasher Object (UFO) was developed.
It consists of eight LEDs contained within a cylindrical acrylic shell
that will be permanently attached to the top of the source connector described in \Cref{sec:source_connector} and shown in \Cref{fig:source_assembly}.
Thus, it will have a fixed position and be deployable with any source.
An enclosed driver board pulses the LEDs sequentially using an external trigger (pulse rate 0--35\,kHz), causing them to emit short pulses (FWHM 5--20\,ns) of red light ($655\pm5$\,nm).
As the sequence always starts at the same LED, the relative orientation of the source can be determined.
The LED intensity can be tuned over a wide range, since the PMT quantum efficiency for this wavelength is very low ($<$1\%).
Conversely, the camera system (see \Cref{sec:cameras}) was selected for infrared sensitivity.
Thus, the light pulses can be intense enough to be seen both by the cameras and the PMT array without risk to the latter.
This will provide a unique signal in two subsystems, allowing the reconstruction software to pinpoint the location and orientation of the UFO and therefore the attached source.

\subsubsection{Source Interface}
\label{sec:source_connector}

\begin{figure}[htp]
    \centering
    \includegraphics[width=0.9\textwidth]{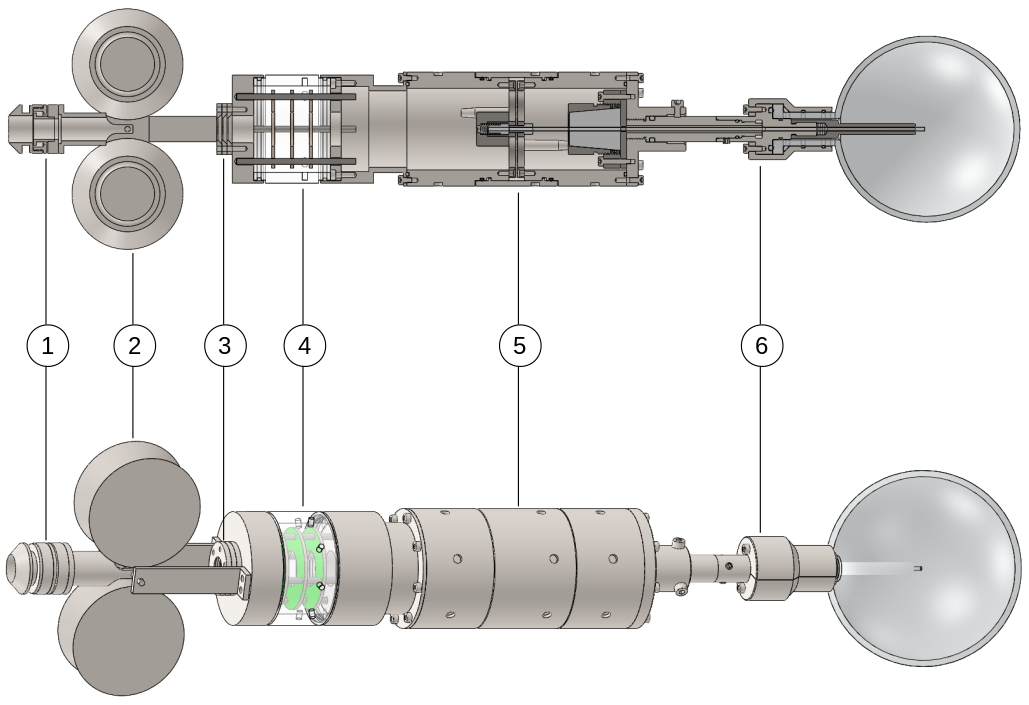}
    \caption[Full source assembly]{Preliminary schematic of the full source assembly for deployment during the scintillator phase, shown as cross section (above) and in 3D (below):
        (1) pivot assembly, (2) pulley system, (3) O-ring stack for umbilical seal, (4) Umbilical Flasher Object (UFO), (5) source connector, (6) attached source, depicting the laserball as an example.
        The image is rotated by 90\textdegree{} counter-clockwise compared to how it will be deployed in the detector.
    }
    \label{fig:source_assembly}
\end{figure}

The source interface must meet all requirements related to preserving scintillator cleanliness.
A connector system shown in \Cref{fig:source_assembly} was developed that allows
\begin{itemize}
    \item only electropolished stainless steel to be in contact with the active liquid,
    \item use of an elastomer seal that is compatible with LAB,
    \item all service lines (gas, optical fibre and electrical lines) to be inside and able to connect and disconnect simultaneously, and
    \item an easy connection and disconnection action that can be performed in a glove box environment.
\end{itemize}

\begin{figure}[htp]
    \centering
    \includegraphics[width=\textwidth]{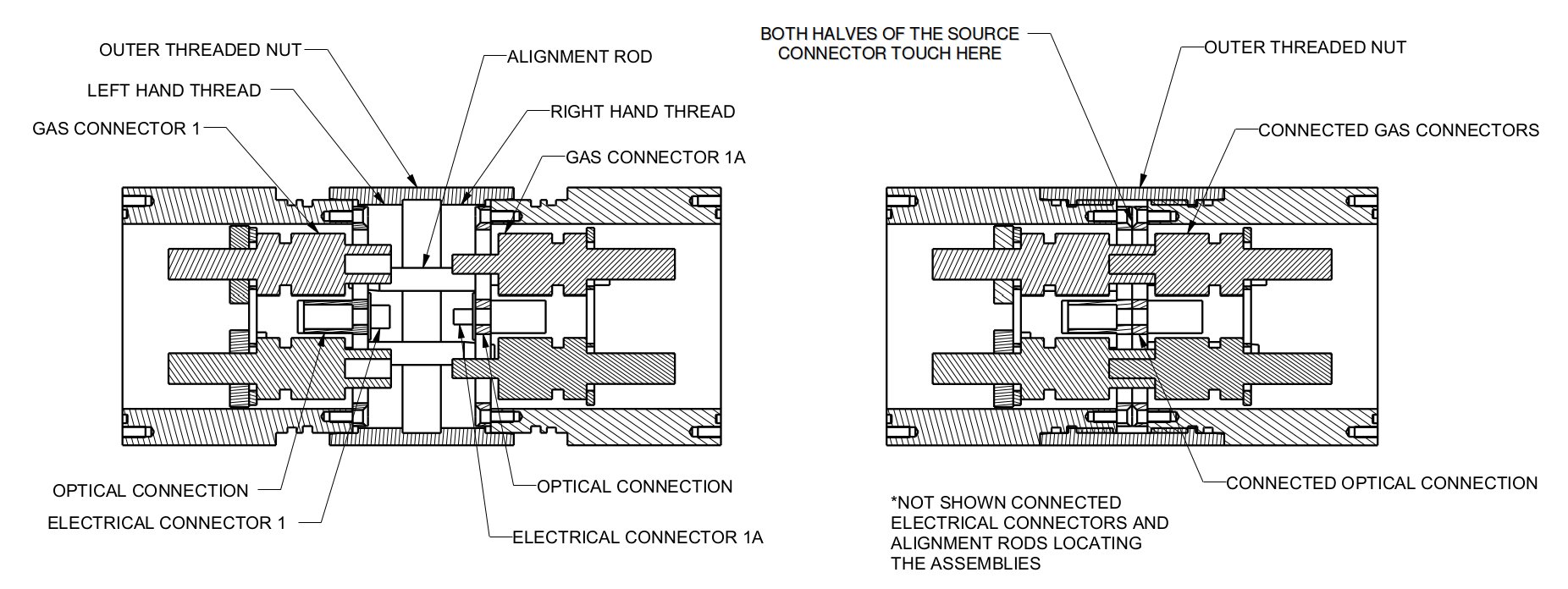}
    \caption[The source connector]{Cross-sectional view of the source connector before (left) and after connection (right).}
    \label{fig:source_connector}
\end{figure}

The source interface allows for the attachment or detachment of a source to the deployment hardware with a single operation.
When attached, all electrical, gas, and optical interfaces are connected.
The design uses internal threads to separate these interfaces from the active liquid while the source is deployed.
A schematic of the source connector can be seen in \Cref{fig:source_connector}.

%
%
\subsection{Camera System}
\label{sec:cameras}

A system of six underwater cameras has been installed in the SNO+ detector.
The cameras, which are mounted on the PMT support structure, deliver data in a reference frame that is independent of the AV and the deployed sources mechanisms.
The system is used to measure the position of calibration sources with respect to the PMT array and to monitor the movement of the AV and the rope systems over the course of the experiment.
It provides additional information and a cross-check for the equator monitors and neck sense-ropes, detailed in~\cite{snonim}.
The cameras can also verify the position of the calibration sources using the signal from the UFO (see \Cref{sec:ufo}).

\begin{figure}[htp]
    \includegraphics[width=0.53\textwidth]{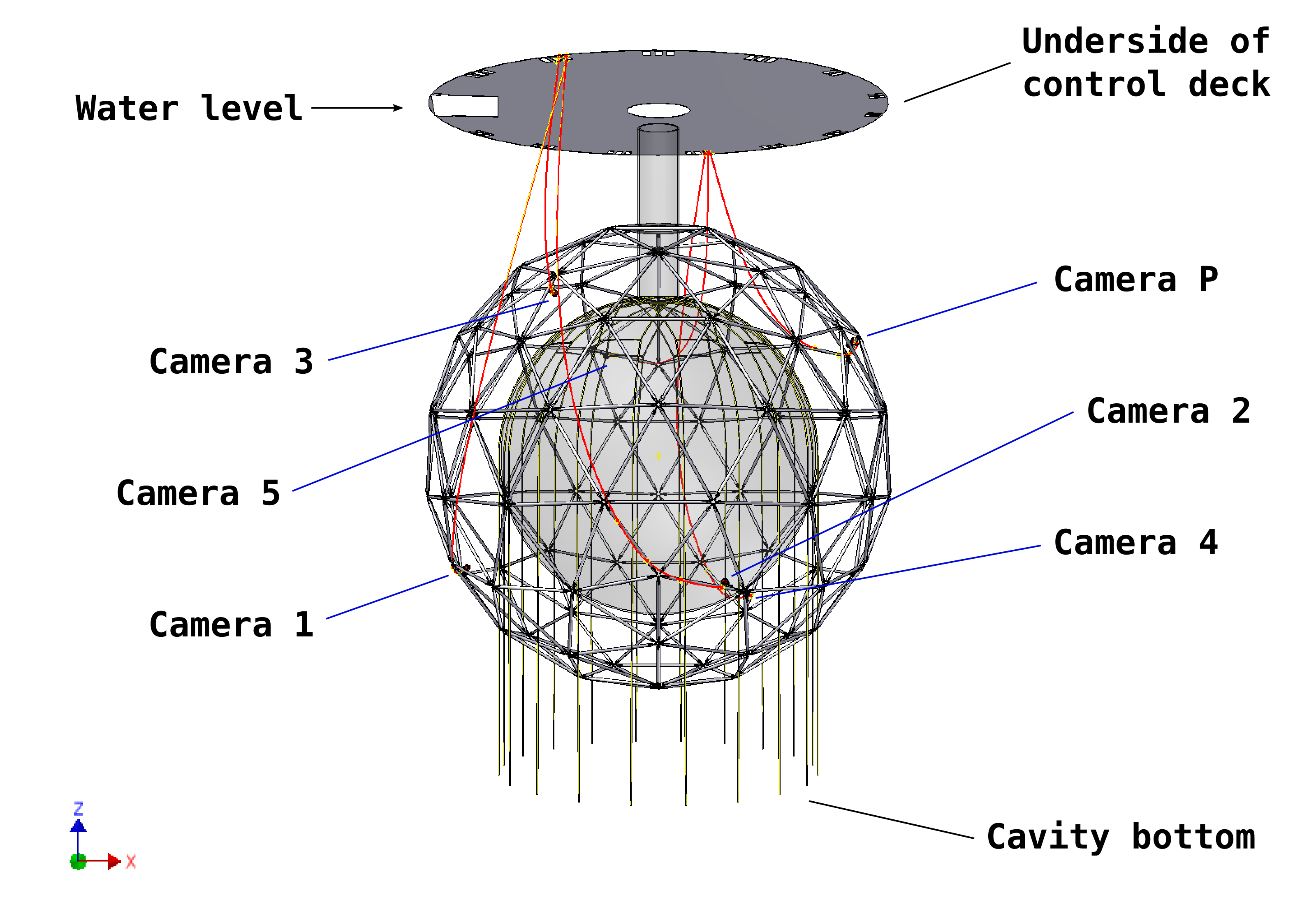}
    \includegraphics[width=0.46\textwidth]{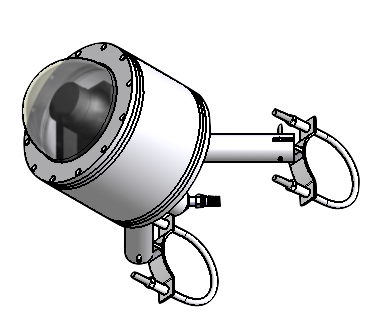}
    \caption[Camera locations and enclosure]{The figure on the left shows the locations of the cameras after installation.
        On the right, the camera enclosure is displayed.
        It is made of stainless steel and has an acrylic viewing port.
        The flanges are sealed with double O-rings to make it watertight.
        Power, signal, and an N$_2$ gas flush system (not shown) pass through the plug located on the back of the enclosure.
    }
    \label{fig:locations}
\end{figure}

Each node of the camera system has a Nikon D5000 camera equipped with a fisheye lens.\footnote{Nikon AF-S DX 10.5\,mm F2.8G fisheye lens. \href{https://en.nikon.ca/nikon-products/product/camera-lenses/af-dx-fisheye-nikkor-10.5mm-f\%252f2.8g-ed.html}{https://en.nikon.ca/nikon-products/product/camera-lenses/}}
Each camera is housed in a watertight stainless steel enclosure with an acrylic viewport (\Cref{fig:locations}, right).
Power and signal cables pass through a watertight hose and connect to the deck, allowing for remote control of the cameras.
A nitrogen gas flush system removes moisture and condensation from the camera housings.

The camera placement is shown in~\Cref{fig:locations} on the left.
The three lower hemisphere cameras were installed in October 2011, while the upper hemisphere ones were mounted in November 2016.
The connections to cameras \#2 and \#4 have been lost over time, possibly as a result of water entering the enclosure.

Several underwater tests were done to determine the functionality of the watertight enclosures, remote camera control and calibration.
These studies suggested that the system with four working cameras is capable of reaching a positional accuracy of 1.5\,cm for an object located in the centre of the AV~\cite{petriw_2012}.

The cameras in the system are calibrated \emph{in situ} using a model with 11 free parameters describing the lens distortion, the position, and orientation of each camera.
The calibration makes use of the known location of PMTs in the picture to extract these parameters.

%
%

\subsection{Optical Calibration Sources}
\label{sec:opsources}

\begin{figure}[htp]
    \centering
    \includegraphics[width=0.5\textwidth,trim={1mm 1mm 1mm 1mm},clip]{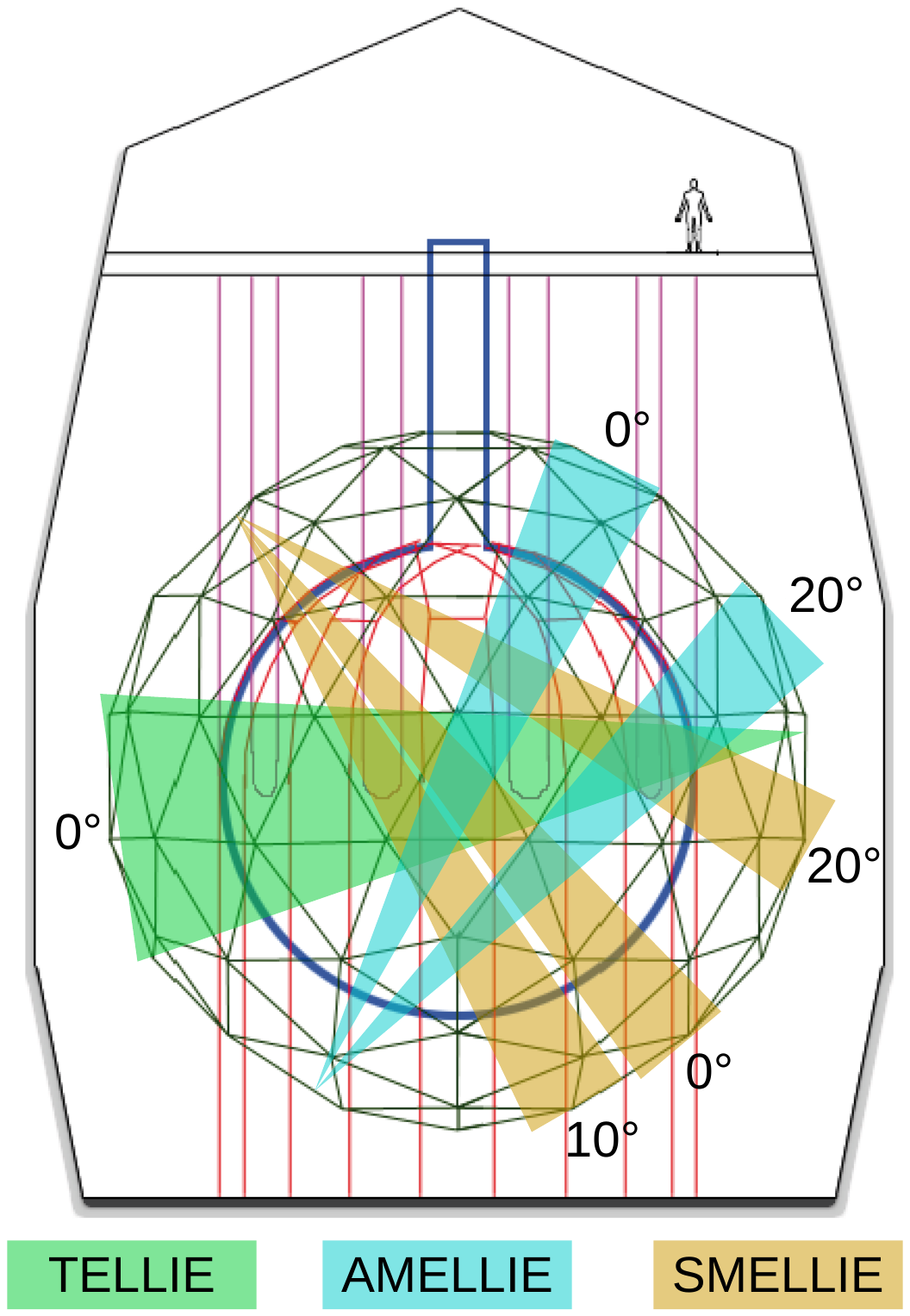}
    \caption{Sketch of the SNO+ detector, with the AV shown in blue.
        The PSUP (dark green) holds both the PMTs and the fibre ends of the ELLIE calibration system.
        Three of the many possible light injection points are also shown:
        Light from the SMELLIE (three orange beams) and AMELLIE subsystems (two cyan beams) can be injected at different angles, while light from TELLIE (one green beam) is always aimed at the centre of the AV.
        Figure adapted from~\cite{ellie_nuphys_2016}.
    }
    \label{fig:ellie-schematic}
\end{figure}

In SNO, the laserball (see \Cref{sec:laserball}) was the primary optical calibration source and was regularly deployed inside the AV.
Because of the more stringent radiopurity requirements of SNO+, it is undesirable to immerse external sources into the AV with the same regularity.
The Embedded LED/Laser Light Injection Entity (ELLIE) was developed to reduce the frequency of laserball deployment by using fixed light injection positions around the PSUP to inject light across the active volume of the detector.
A schematic is shown in \Cref{fig:ellie-schematic}.

This section starts with a description of the ELLIE calibration system.
ELLIE consists of three subsystems: the Timing system (TELLIE) for PMT hit-time and gain calibration, the Attenuation Module (AMELLIE) for monitoring relative changes in attenuation lengths of materials in the detector, and the Scattering Module (SMELLIE) for measuring the optical scattering cross sections of these materials as functions of scattering angle and photon wavelength.
This is followed by a description of the laserball, a uniform optical calibration source that can be deployed inside the detector and was well understood in SNO.
This source can be coupled to a separate laser driver, which can mimic a supernova event in order to test the response of the SNO+ detector under such conditions.
Finally, a Cherenkov source is described.
As the scintillator light yield and the detector efficiency are strongly correlated in the measurement of the detected photoelectron yield, this source makes it possible to test the understanding of the absolute detector efficiency, independently of scintillator effects.

\subsubsection{TELLIE}
\label{sec:tellie}
The SNO+ event position reconstruction uses photon time-of-flight and is therefore directly linked to the time calibration of the entire detector.
The timing module named TELLIE~\cite{snop_calib_jinst} provides an optical timing calibration source designed to accurately measure the gain and charge response and the relative time offset of the PMTs.

The main aim of TELLIE is to synchronize the time response of each PMT relative to all other PMTs in the array for the case of global, instantaneous single photoelectron generation.
This synchronization is heavily influenced by electronics delays along the signal propagation path, including paths through both the detector readout electronics and the trigger systems, and must be established at a level of 1\,ns or better.
A precision of $\mathcal{O}(1)$\,ns is necessary for the synchronization systematics to be sub-dominant with respect to the unavoidable timing uncertainties: the transit time spread on the PMTs' signal generation (1.7\,ns standard deviation~\cite{snonim}), and the decay time of the scintillator signal (4.6\,ns for $\sim$0.5\,MeV electrons in deoxygenated, pre-distilled LAB at 2\,g/L of PPO wavelength shifter~\cite{snop_scint,snop_scint_2011,lombardi_2013}).

In addition to the timing calibrations, the TELLIE source will be used to calibrate the gain and charge response of the PMTs, an important factor for energy reconstruction.
Time and charge information are measured for each PMT pulse if it crosses the leading edge discriminator with a fixed threshold, as described in \Cref{sec:electronics}.
For each channel, the discriminator is set at approximately 25\% of the average peak voltage for single photoelectron pulses.
If the gain changes, the photoelectron detection efficiency of that channel also changes.
Although the gain of the PMT array is expected to be stable over time, offline analysis on TELLIE data can determine and calibrate for any unexpected changes.

TELLIE is composed of 92 injection points spread around the PSUP, pointing towards the centre of the detector.
LEDs were selected as the light sources\footnote{Brite-LED Optoelectronics, BL-LUCY5N15C series. \href{http://www.brite-led.com/PDF/BL-LUCY5N15C\%20series\%20datasheet.pdf}{http://www.brite-led.com/}}, as using a single laser source coupled to an optical fibre switch would have introduced the risk of a single-point failure in the system.
The use of LEDs provides an economical solution in the case of single channel failures, as the LED or its driver board can easily be replaced.
A 500\,nm wavelength was chosen to match the lowest opacity region of the Te-loaded scintillator absorption spectrum, while remaining in a region where the PMT quantum efficiency is still relatively high ($>$80\% of maximum efficiency)~\cite{snop_calib_jinst}.

The optical fibres used are 1\,mm diameter, duplex, PMMA step-index fibres.
With a numerical aperture of 0.5, they provide a wide beam (22\textdegree{} half-opening angle in water at 450\,nm).
This ensures good overlap of beam spots, maximizes coverage, and allows redundancy.
Ten spare fibres have additionally been installed as backup.
The fibres are 45\,m in length, each connected to the patch panel on the deck.
From there, 2-m long fibres connect the detector fibres to the corresponding driver box, each containing 8 LEDs.
The TELLIE system consists of 12 boxes (96 channels); 95 LEDs were installed in total.

At the end of the fibres, the optical pulse width produced by the system is around 4\,ns (FWHM).
Each PMT will see a large number of pulses ($N$).
Since the statistical precision on the mean time of the pulse is proportional to $\mathrm{FWHM}/\sqrt{N}$, the statistical uncertainty quickly becomes insignificant.
It was found that a timing calibration of the PMTs of order 0.1\,ns per PMT is achievable.

\begin{figure}[htp]
    \centering
    \includegraphics[width=\textwidth]{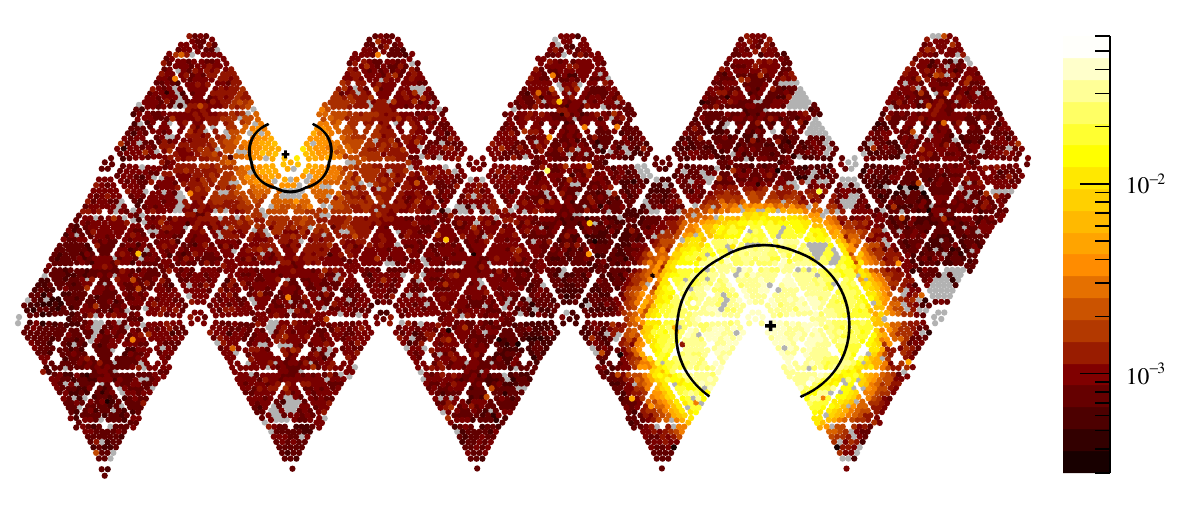}
    \caption{Typical TELLIE run in SNO+ water phase, shown as flat-map with each dot indicating a PMT position.
        The colours represent the relative number of times the PMTs were hit (log scale), with grey indicating offline PMTs.
        The direct light spot is on the bottom right side, while the light reflected off the AV can be seen on the upper left side.
        Both light spots are fitted to a 2D Gaussian, with the fit centre and width shown in black.
    }
    \label{fig:tellie-flatmap}
\end{figure}

In the water phase, TELLIE was run at a high rate (1\,kHz), with each channel being pulsed for 200\,s.
The summed PMT hits from a typical run are shown in \Cref{fig:tellie-flatmap}.
For the scintillator phase, efforts are being made to automate the pulsing at a lower rate (50\,Hz) to enable automatic calibration data to be taken during physics runs.

\subsubsection{AMELLIE}
\label{sec:amellie}
The attenuation module AMELLIE provides a calibration source designed to monitor the stability of the optical attenuation in the detector volume over time.
Attenuation, also often referred to as extinction, indicates the removal of a photon from its initial path, which can happen via scattering or absorption.
The attenuation coefficient (reciprocal of the attenuation length) is therefore the sum of the absorption and scattering coefficients, and is typically wavelength dependent.
Understanding the attenuation lengths of the optical paths through the detector is vital for event reconstruction throughout the detector volume.

Light injected along a well-defined beam through the detector can be used to study the optical properties of the detector media \emph{in situ}.
PMTs falling in the direct paths within the beam will be predominantly hit by prompt, unattenuated light.
Light hitting PMTs outside these direct beam paths can be mainly attributed to scattering and absorption followed by re-emission, as well as reflection off the boundary surfaces of the detector media.
Depending on the angle between the observed photon and the beam and the timing delay relative to prompt light, each of those optical effects will give varying contributions.
Changes in optical attenuation length can be deduced by monitoring the long-term stability of relative amounts of light hitting different regions of PMTs (relative to the beam) at different times (relative to prompt times).

AMELLIE utilizes step-index multimode fibres with 200-\textmu{}m diameter fused-silica cores attached to LED sources.
With a numerical aperture of 0.22, they produce a beam of 9.47\textdegree{} half-opening (at 450\,nm in water), considerably narrower than that of the TELLIE fibres, allowing for better discrimination between the different light components, yet being wide enough to average over many PMTs in the direct beam.
Eight 45-m long fibres are attached to four different light injection points on the PSUP.
In order to sample a range of different optical paths, two points are located near the equator, and one each near the top and bottom.
At each injection point, one fibre points at 0\textdegree{} and the other at 20\textdegree{} with respect to the centre of the detector, providing different ratios of path lengths in scintillator vs.\,water.

AMELLIE uses the same driver electronics as the TELLIE system, and occupies one box containing eight channels.
The detector fibres are connected to the diodes in the same manner via 2-m long patch fibres.
Currently, AMELLIE uses LEDs identical to those installed in TELLIE.
In the future, they can be replaced to allow sampling at different wavelengths.

\subsubsection{SMELLIE}
\label{sec:smellie}

The scattering module SMELLIE is designed to measure the scattering characteristics of the detector media \emph{in situ}.
It can rapidly measure the wavelength and angular dependence of the scattering cross-section of visible photons in the detector media and thus track rapid changes.
Optical scattering at relevant wavelengths is expected to be dominated by Rayleigh scattering, which has a well-known wavelength and scattering angle dependence.
Therefore, different wavelengths can be used to understand the nature of scattering in different optical media and discern scattering from absorption and re-emission.
As scattering indicates a change in angle of the photon, the initial angle must be well-defined using appropriate beam collimation.
The longer paths and travel times of scattered photons can be detected with fast and precise timing.
These wavelength, time and position measurements allow the system to distinguish scattered photons from reflected, refracted or re-emitted ones.
For these reasons, SMELLIE uses five short-pulsed laser systems coupled into low-dispersion fibres in order to inject light into the detector.
Short-pulsed lasers can have a high pulse-to-pulse variation of the number of photons emitted, in particular when operated at low laser gain.
Therefore, the SMELLIE system also monitors the energy of every laser pulse with a separate monitoring PMT that can determine variations in the number of photons emitted at the sub-percent level.

There are five light injection nodes for SMELLIE, each holding three optical fibres.
The injection nodes are widely spread over the detector.
The three fibres at each node are pointing at 0\textdegree{}, 10\textdegree{} and $-$20\textdegree{} with respect to the centre of the detector, resulting in three different sets of path lengths in the different detector media (see \Cref{fig:ellie-schematic}).
This allows the system to disentangle the otherwise degenerate effects that external water and the medium inside the AV can have on the distribution of scattered photons.

The SMELLIE fibres are Corning InfiniCor SXi, graded-index, 50-micron core, telecommunications fibres, originally designed for minimum modal dispersion around 800\,nm.
Budget constraints excluded the use of fibres optimized for short pulses in the visible range, but the large shot energies and narrow spectral bandwidth of the laser allowed for the use of much lower cost telecommunications fibres with significantly increased absorption and mildly increased modal and spectral dispersion in the visible wavelength range.

Each of the optical fibres can be connected to any of the four pulsed diode laser heads and one super-continuum (SK) laser via a 5\,$\times$\,15 fibre switch.
The wavelength distributions of the diode lasers peak at 375\,nm, 405\,nm, 440\,nm and 495\,nm with spectral FWHM from 1.5--4\,nm.
The pulses from the diode lasers are very fast with FWHM from 50--200\,ps and can have repetition rates from single shots to 20\,MHz.
Manually-variable attenuators are in-line with each laser to match the range of shot energies available via electronic drive pulse adjustment to the range of the number of photons desired in the detector.
The SK laser pulses are passed through a wavelength selector producing pulses with a 10-nm spectral bandwidth in the 400--700\,nm region.
The pulses from this laser are longer with a FWHM of $\sim$2\,ns.
The maximum repetition rate is 24\,kHz, but the SNO+ DAQ limits stable acquisition of SMELLIE data to rates below 2\,kHz.

To study the scattering angles in the detector media SMELLIE requires a good determination of the direction of light emitted into the fibre.
The angular spread must however not be too small to allow a sufficiently large number of PMTs to measure the central beam intensity and arrival time with high accuracy.
SMELLIE therefore uses partially collimated beams.
These are produced by shortened, cylindrical, graded index lenses that allow the collimators to be built with only flat surfaces.
This simplifies a gapless construction with zero water ingress.

AMELLIE and SMELLIE perform complementary measurements of the detector media.
While AMELLIE does not reveal fundamentally new information in regards to optical processes within the detector media, the system provides a cost-effective solution to ensure some redundancy and cross-check SMELLIE data over time.

\subsubsection{Laserball}
\label{sec:laserball}

Further \emph{in situ} measurements of the optical properties of the SNO+ detector are conducted using the laserball, a spherical light diffuser.
As in SNO, a pulsed nitrogen laser (337\,nm) drives a selectable dye laser system, which produces light at six wavelengths with well-characterized spectra.
Light is conducted to the laserball source via an umbilical containing an optical fibre bundle, as well as an electrical cable with which an independent LED light source may be run.
The purpose of the LED light source is to independently determine the orientation of the laserball.
The implementation of this system as deployed in SNO has already been described in detail~\cite{MOFFAT2005255}.

Prior to being passed into the dye cells, a subset of the light is picked off using a beam splitter and sent to a photodiode.
The signal from the photodiode is processed using a constant fraction discriminator in order to produce a pulse which is fed into the DAQ as an external, asynchronous trigger.
Triggering in this way provides a consistent reference time with which to measure timing-related detector constants, such as absolute time delays and PMT time-walk.

\begin{figure}[htp]
    \centering
    \includegraphics[width=\textwidth]{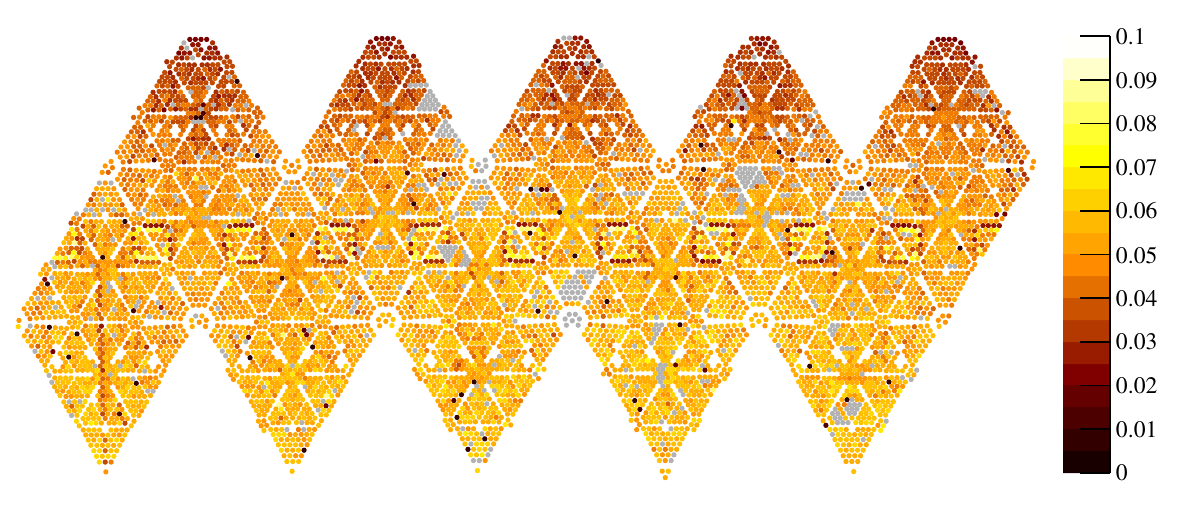}
    \caption{Typical laserball run in SNO+ water phase, shown as flat-map with each dot indicating a PMT position.
        The colours represent the relative number of times the PMTs were hit (linear scale), with grey indicating offline PMTs.
        The occupancy is lower near the top of the detector as a result of shadowing by the source deployment system.
        The rectangular shapes around the equator are caused by lensing effects of the acrylic plates supporting the hold-up ropes.
        The shadow from the AV pipes can be seen as vertical line in the leftmost section.
    }
    \label{fig:laserball-flatmap}
\end{figure}

The laserball is typically pulsed at 10--40\,Hz while deployed at various positions inside the AV.
Accumulated data from a typical 2.5-hour water phase run at the centre of the AV is shown in \Cref{fig:laserball-flatmap}.
The intensity is tuned such that a given PMT is hit less than 5\% of the time to ensure that the PMTs are operating in single photoelectron emission mode.
Data from central deployments are used to calibrate the time offsets between PMTs (see \Cref{sec:electronics}).

\begin{figure}[htp]
    \centering
    \includegraphics[height=10cm,trim={4cm 1cm 4cm 0},clip]{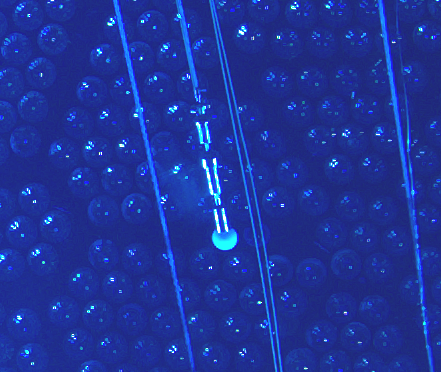}
    \hspace{2cm}
    \includegraphics[height=10cm]{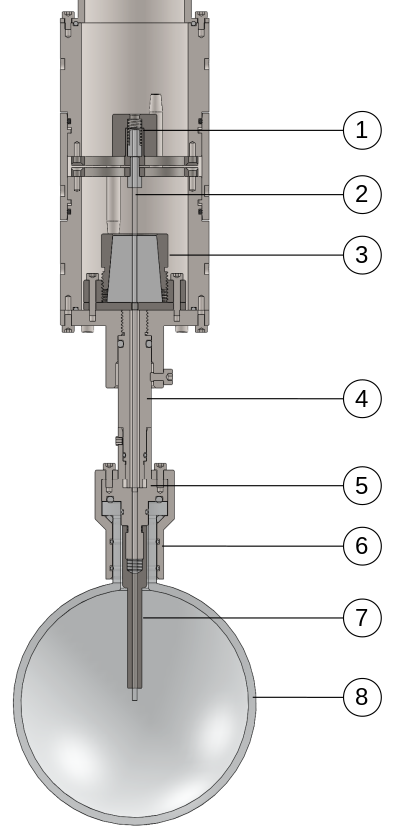}
    \caption[Laserball photo and schematic]{\textit{Left:} Photo of the SNO laserball as deployed during the SNO+ water phase.
        \textit{Right:} Schematic overview of the new laserball developed for the scintillator phase: (1) cable fibre lock, (2) quartz rod light guide, (3) PTFE clamp assembly, (4) neck tube, (5) neck bung, (6) neck clamp, (7) quartz support extension, (8) diffuser flask.
    }
    \label{fig:laserball}
\end{figure}

The increased purity requirements in the SNO+ scintillator phase required a new laserball to be developed (shown in \Cref{fig:laserball}).
Features of this new source include an improved neck design to reduce self-shadowing and a solid quartz rod with an adjustable extension stage that actively adjusts the position of the light injection point within the quartz flask.
This allows for an improved polar isotropy of the laserball.
The diffuser consists of a spherical quartz flask (11\,cm diameter) filled with hollow glass micro-spheres (20--70\,\textmu{}m diameter, 1\,\textmu{}m thickness) suspended homogeneously in 600\,mL of silicone gel.
The concentration of beads and the light injection point in the flask are optimized to produce a quasi-isotropic light distribution with low temporal dispersion.
Within the source connector (see \Cref{sec:source_connector}), light is injected from the optical fibre into the quartz rod (2\,mm diameter, $\sim$22\,cm length).
The connection to the optical fibre is achieved via a central fibre lock.
The \emph{in situ} testing of this new laserball will occur during the scintillator phase.

\subsubsection{Supernova Source}
\label{sec:SNC+}
\begin{figure}[htp]
    \centering
    \includegraphics[width=0.9\textwidth]{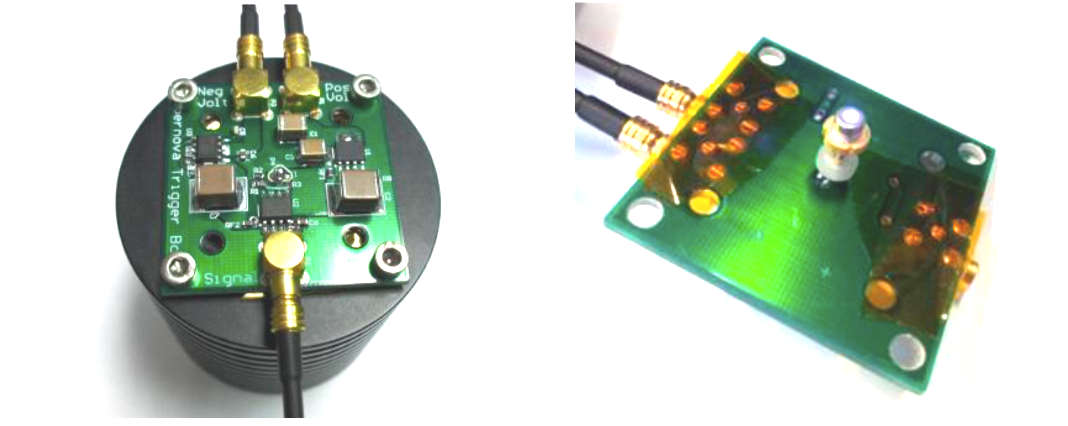}
    \caption{\textit{Left:} Laser diode PCB attached to the heat sink collimator.
        \textit{Right:} Underside of the PCB showing the laser diode in its socket.
    }
    \label{fig:SNC_images}
\end{figure}

A supernova produces a burst of high energy neutrinos with an intensity that can potentially be observed in a detector like SNO+.
The supernova calibration source simulates sequences of light pulses produced by these neutrino interactions in the detector.
This is achieved by directing pulsed light from a high-powered (120\,mW), blue-violet (405\,nm) laser diode into the laserball (see \Cref{sec:laserball}).\footnote{US-Lasers, D405-120 laser diode, part no.\,38-1035-ND. \href{https://www.digikey.ca/product-detail/en/us-lasers-inc/D405-120/38-1035-ND/3438595}{https://www.digikey.ca}}
The laser diode is inserted into a socket and mounted on a custom-designed printed circuit board (PCB) which contains the circuitry for driving the laser diode (see \Cref{fig:SNC_images}).

Using a set of realistic pulse lengths of a supernova neutrino signal~\cite{mirizzi_2016,huedepohl_2010}, the source can deposit bursts of photons equivalent to a total deposition energy of up to 100\,MeV within the scintillator volume.
This results in all PMTs in the detector being hit by multiple photons, which would be highly unusual during typical physics or calibration running, but is expected for supernova events.
Stress-testing the data acquisition system in such a way will characterize the detector's ability to process and save data from a nearby supernova event in which a large number of neutrino events occur within seconds in the detector.

To mimic a realistic supernova neutrino burst, each event is characterized by the pulse energy and pulse-to-pulse timing separation of the laser diode.
The expected supernova energy distribution is sampled to provide the pulse energy, whereas a Poisson distribution corresponding to the total rate of neutrino interactions at the event time is sampled to provide the pulse-to-pulse timing separation.
Both pulse energy and timing are encoded together as 32 bits, and all events for a supernova burst are written to a single ``burst'' file.
Burst files are calculated in advance for various supernova models, neutrino oscillation cases, and distances.

An FPGA interprets the contents of the burst files into logical signals which are sent to the laser diode driver board.
The timing information sets a timer which is counted down using a 12.5\,MHz clock.
The clock frequency permits realistic event pile-up in the 400\,ns trigger window.
The energy information is converted via dedicated energy look-up tables to a current which flows through the laser diode once the timer finishes counting down.
Light from the laser diode is then transmitted via optical fibre bundle to the laserball, which distributes the light isotropically from the centre.
A shutter exists between the laser diode and the connecting fibre optic bundle and is normally closed to protect the detector and operators from exposure to laser light.
This shutter is only open during calibration running once the system is verified to be connected in a safe state.

\subsubsection{Cherenkov Source}
\label{sec:cherenkov}

A source of Cherenkov light was designed to calibrate the SNO+ PMT optical response independently of the light yield, absorption, and remission properties of the scintillator.
Decoupling these effects is important in order to fully understand the detector response --- in particular, as a function of time and position within the target volume.
It is known from SNO that the global PMT collection efficiency decayed over time~\cite{snoprc}, due at least in part to degradation of the concentrator reflectivity.
This effect will have a very different time dependence than the changes in scintillator properties and thus must be independently monitored and modelled.

AMELLIE (\Cref{sec:amellie}) and SMELLIE (\Cref{sec:smellie}) monitor stability of the scintillator absorption and scattering.
These can be done using relative measurements, and thus do not require stability in the overall light levels.
However, monitoring of the overall light yield and PMT efficiency requires a stable source of light, with a known, constant yield.
The uniquely stable nature of Cherenkov light generation provides such a source, and the well-understood underlying physical process means the result can be modelled and predicted very accurately via simulations.
A Cherenkov light source can therefore be used to monitor PMT light collection behaviour over time.
Using this to define the PMT response, other calibration sources can be used to characterize and monitor the liquid scintillator light yield.

\begin{figure}[htp]
    \centering
    \includegraphics[width=0.8\textwidth, trim={0 35mm 0 0}, clip]{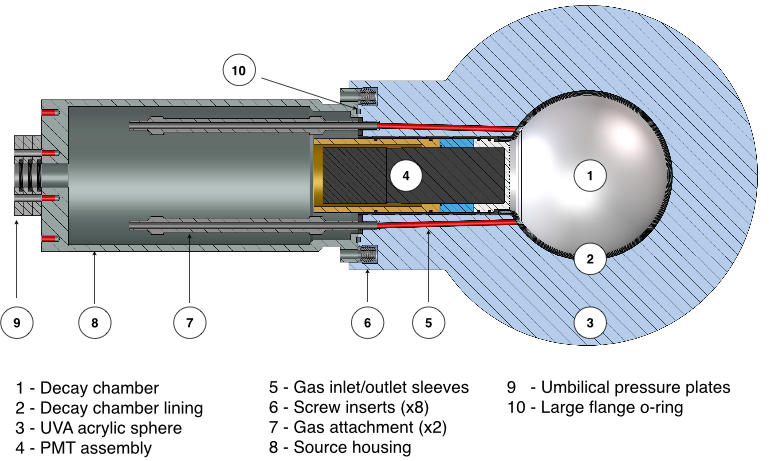}
    \caption{The SNO+ Cherenkov source design: (1) decay chamber, (2) decay chamber lining, (3) UVA acrylic sphere, (4) PMT assembly, (5) gas inlet/outlet sleeves, (6) screw inserts, (7) gas attachment, (8) source housing, (9) umbilical pressure plates, (10) large flange O-ring.
        Note that the top flange shown here corresponds to the water-phase umbilical connection.
    }
    \label{fig:cherenkovsource_design2}
\end{figure}

The Cherenkov source was modelled around the SNO \isotope[8]{Li} calibration source~\cite{Tagg:2002}, which was used to provide a source of electrons similar in spectrum to those from \isotope[8]{B} solar neutrinos.
\isotope[8]{Li} is produced by the \isotope[11]{B}(n,$\alpha$)\isotope[8]{Li} reaction in a deuterium-tritium (DT) neutron generator, and transported using a carrier gas (primarily helium) into a spherical decay chamber.
\isotope[8]{Li}\,decay produces a $\beta$ and two subsequent $\alpha$ particles from the decay of the very short-lived \isotope[8]{Be} daughter nucleus.
The SNO+ collaboration has adapted this source by replacing the decay chamber with a 6\,cm thick acrylic shell, in which 
$\beta$s produce Cherenkov light before being absorbed (see \Cref{fig:cherenkovsource_design2}).
The $\alpha$s cause the helium inside the decay chamber to scintillate.
This scintillation light is used to tag events, using a small built-in PMT above the decay chamber.
The ultra-violet absorbing (UVA) acrylic emits photons only above $\sim$400\,nm, thus making the light largely unaffected by absorption and reemission effects in the scintillator.
This is due to the low optical absorption of the liquid scintillator in this wavelength region.
The $\alpha$ scintillation light must be contained within the decay chamber to prevent contamination of the Cherenkov light.
A black opaque lining is therefore deposited on the inside of the decay chamber.
On top of this layer, a reflective titanium dioxide (TiO$_2$) lining and the wavelength shifter tetraphenyl butadiene (TPB) are deposited.
These improve the tagging efficiency by increasing the probability that a photon will hit the tag-PMT, and by shifting the $\sim$80\,nm helium scintillation light to $\sim$420\,nm, respectively.

%
%

\subsection{Radioactive Calibration Sources}
\label{sec:radsources}

In addition to the optical sources described above, radioactive sources can be deployed inside the detector for energy calibration.
As the SNO+ requirements on internal radioactivity are extremely strict, only sources with a robust double encapsulation are permitted.
In practice, this means neither $\alpha$ or $\beta$ sources could be practically constructed to meet the strict requirements on limiting the risk of contaminating the sensitive volume of the experiment.
The radioactive sources considered for the SNO+ experiment are shown in \Cref{tab:rad_sources}, covering the energy range from 0.1\,MeV to $\sim$10\,MeV.
This energy range covers the expected energy threshold for most planned physics topics.

\begin{table}[htp]
    \centering
    \caption{List of planned radioactive sources to be deployed in the SNO+ detector.
        The total $\gamma$ energy is given, and where appropriate, the individual energies of the $\gamma$s emitted.
        The bracketed $\gamma$ energies for the AmBe source are from deexcitations of nuclei following neutron capture.
    }
    \label{tab:rad_sources}
    \begin{tabular}{ |c|c| }
        \hline
        Source & Gamma energy (MeV) \\
        \hline
        $^{16}$N & 6.1 \\
        AmBe &  4.4 [2.2, 7.6, 9.0] \\
        $^{46}$Sc & 2.0 (0.89, 1.12) \\
        $^{48}$Sc & 3.3 (0.98, 1.04, 1.31) \\
        $^{137}$Cs & 0.66 \\
        $^{57}$Co & 0.14 \\
        \hline
    \end{tabular}
\end{table}

Prior to source deployment in the scintillator phase, and as a cross-check for other calibrations, it will be possible to use simulated event samples to fit for specific features in the low-energy spectrum in the observed data.
These will be sensitive to an offset and scale in energy response as well as the energy resolution.
In particular, a fixed energy peak from \isotope[210]{Po} and the $\beta$ decay continuum from \isotope[210]{Bi} are expected to be present from radon daughters in the scintillator (see \Cref{sec:psup}).
The collaboration plans to periodically assay the liquid scintillator, allowing a cross-check measurement with \emph{in situ} analysis techniques.

The rest of this section describes the deployable radioactive gamma and neutron sources developed by the SNO+ collaboration.

\subsubsection{\texorpdfstring{\isotope[16]{N}}{N-16} Calibration Source}
\label{sec:n16}
The SNO \isotope[16]{N} source was the main energy calibration source for the water phase.
There are no plans to deploy the \isotope[16]{N} source within the AV during the scintillator phase in order to reduce the risk of contamination, since the energy of the \isotope[16]{N} decay is above the main region of interest in scintillator phase physics goals.
However, the source will be deployed in the external water to obtain calibration data that can be directly compared to data taken during the water phase.
\isotope[16]{N} $\beta$ decays to \isotope[16]{O}, with a 6.1\,MeV (7.12\,MeV) $\gamma$ being produced in 66.2\% (5\%) of all decays.
The $\gamma$ primarily Compton scatters, producing electrons that may be observed within the detector medium.
The resulting spectrum is broad (3.5--6\,MeV) and peaks at 5\,MeV; it is well modelled in simulation~\cite{snop_nd}.
The $\beta$ is used to tag the decay event through interaction with a scintillator block and PMT inside the source.
\emph{Ex situ} measurements have shown that the detection efficiency of this tag signal is better than 95\%.
Further details can be found in~\cite{Dragowsky:2001ax}.

\subsubsection{Americium Beryllium Calibration Source}
\label{sec:ambe}

The americium-beryllium (AmBe) source used by the SNO+ collaboration was inherited from SNO~\cite{loach_2008}.
The source contains a mixture of powdered \isotope[241]{Am} and \isotope[9]{Be} within a welded stainless steel capsule surrounded by Delrin encapsulation.
The long-lived \isotope[241]{Am} $\alpha$ decays ($T_{1/2}$ = 432\,y) and emits a 59.5\,keV $\gamma$ with a branching ratio of 84.6\%~\cite{TabRad_v5}.
The $\alpha$s are absorbed by the \isotope[9]{Be} with a capture efficiency of $\mathcal{O}(10^{-4})$~\cite{Liu_2020}.
The two dominant processes are:
\begin{align*}
    & \isotope[9]{Be}\left(\alpha,n\right)\isotope[12]{C} & \mathrm{Q = 5.701\,MeV};\\
    & \isotope[9]{Be}\left(\alpha,n\right)\isotope[12]{C}^* & \mathrm{Q = 1.261\,MeV};\\
    & \hspace{20mm} \rotatebox[origin=c]{180}{$\Lsh$}~\isotope[12]{C} + \gamma & \mathrm{E_{\gamma} = 4.44\,MeV}.
\end{align*}

The second process has a branching ratio of approximately 60\%, resulting in a prompt 4.4\,MeV $\gamma$ to be emitted from the deexcitation of \isotope[12]{C}.
The neutron thermalizes and is captured mostly by protons in the scintillator and other detector media, producing a 2.2\,MeV $\gamma$.\footnote{Neutron captures on heavier nuclei such as \isotope[12]{C} are also possible, resulting in higher $\gamma$ energies.}
This coincidence signal mimics an antineutrino event and is crucial for neutron detection efficiency calibration.
The AmBe source also provides a high energy calibration point at 4.4\,MeV, complementing the 6.1\,MeV point from the \isotope[16]{N} source.

\begin{figure}[htp]
    \centering
    \includegraphics[height=6cm,trim={2cm 0.5cm 4cm 3cm},clip]{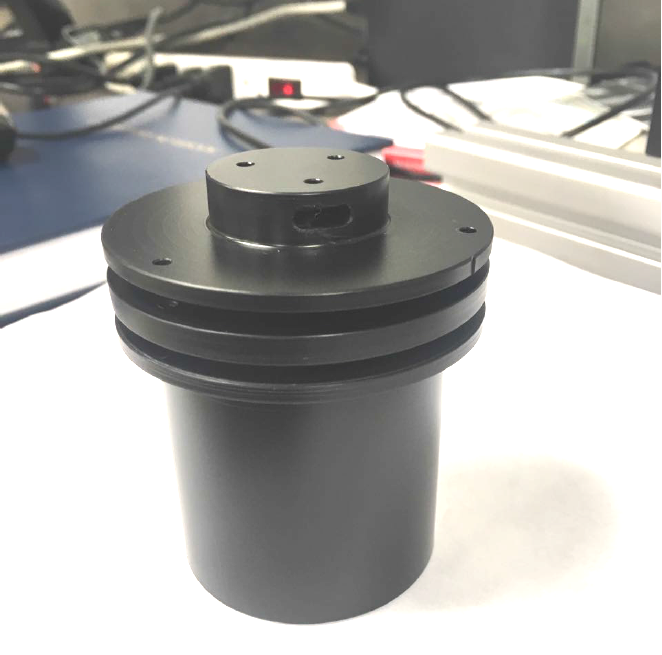}
    \caption[AmBe source]{The AmBe calibration source as used in the SNO+ water phase.}
    \label{fig:ambe_source}
\end{figure}

The AmBe source was shipped to SNO in 2005, arriving doubly encapsulated and leak tested.
The source strength was measured by SNO~\cite{loach_2008}, and calculated to be $(67.39\pm0.73)$\,Hz at the time of SNO+ water phase deployment in 2018.
Additional layers of encapsulation have been subsequently added by both the SNO and SNO+ collaborations.
The source is shown as deployed in the water phase in \Cref{fig:ambe_source}.
It was used to determine the unprecedentedly high neutron detection efficiency of $\sim$50\% achieved in the water phase~\cite{snop_ncapture}, made possible by the upgrades to the electronics triggers (see \Cref{sec:electronics}) and DAQ system (see \Cref{sec:daq}).

For the scintillator phase, additional source shielding layer must be added to block the 59.5\,keV X-rays that are emitted alongside the $\alpha$ decays.
Two encapsulation layers will be added on top of the source shielding layer to ensure cleanliness and compatibility with the scintillator.
The water phase kinematic energy threshold was $\sim$260\,keV, corresponding to the Cherenkov threshold of electrons in water, rendering the X-rays undetectable.
The kinematic energy threshold is lower in the scintillator phase, and these X-rays produce a very high rate of low energy events that obscures the desired higher energy neutron signals being calibrated.
The challenge of the new source shielding layer must be to maximize the absorption of these low energy $\gamma$s without sacrificing the neutrons and tags~\cite{semenec_2017}.
The new encapsulation layer (see \Cref{fig:ambe_source}) will ensure compatibility between the shielding layer and the scintillator.
Ultimately, 2--3.6\,mm of lead was chosen for the source shielding.
At this thickness, simulations show that only 0.004\% of the 59.5\,keV X-rays are transmitted, compared to 94.1\% of neutrons.

Neutron captures on heavier isotopes cause $\gamma$s in the 6--10\,MeV energy range to be produced~\cite{neutron_capture}.
To extend the range of the energy calibration, the potential of adding such isotopes close to the AmBe source was investigated.
\isotope[56]{Fe} (92\% natural abundance) and \isotope[58]{Ni} (68\% natural abundance) deexcite via 7.6\,MeV and 9.0\,MeV $\gamma$ emission following neutron absorption, respectively.
Iron is already abundant in the stainless steel encapsulation.\footnote{Type 316L stainless steel contains 62--69\% iron and 10--14\% nickel.}
With a stainless steel shielding thickness of 10\,mm, $\sim$3.2\% of neutrons capture on \isotope[56]{Fe}.
This can be used as a third, higher energy calibration point.
The entire source is encapsulated in an acrylic layer with thickness of 5\,mm, for cleanliness considerations.

\subsubsection{\texorpdfstring{\isotope[46]{Sc}}{Sc-46} Calibration Source}

The energy distribution of backgrounds in the \dbd{0} decay region of interest (ROI) is extremely asymmetric because of the \dbd{2} decay background.
If this background is mismodelled such that the high energy tail in the ROI is increased by 1\%, this could lead to a 5\% systematic error on the \dbd{0} decay lifetime measurement~\cite{sibley_2016}.

The \isotope[46]{Sc} tagged calibration source is designed to work in the scintillator and tellurium phases to test the energy scale and Gaussian response of the detector.
The isotope \isotope[46]{Sc} decays ($T_{1/2}=83.29$\,d) via the emission of a $\beta$ particle (111.8\,keV) and two $\gamma$s (889.3\,keV, 1120\,keV) 99.98\% of the time.
The $\beta$s reach the scintillator $<$0.002\% of the time, while $(98.5\pm0.5)$\% of the $\gamma$s deposit their full energy in the scintillator.

\begin{figure}[htp]
    \centering
    \includegraphics[height=6cm,trim={5mm 0mm 135mm 0mm},clip]{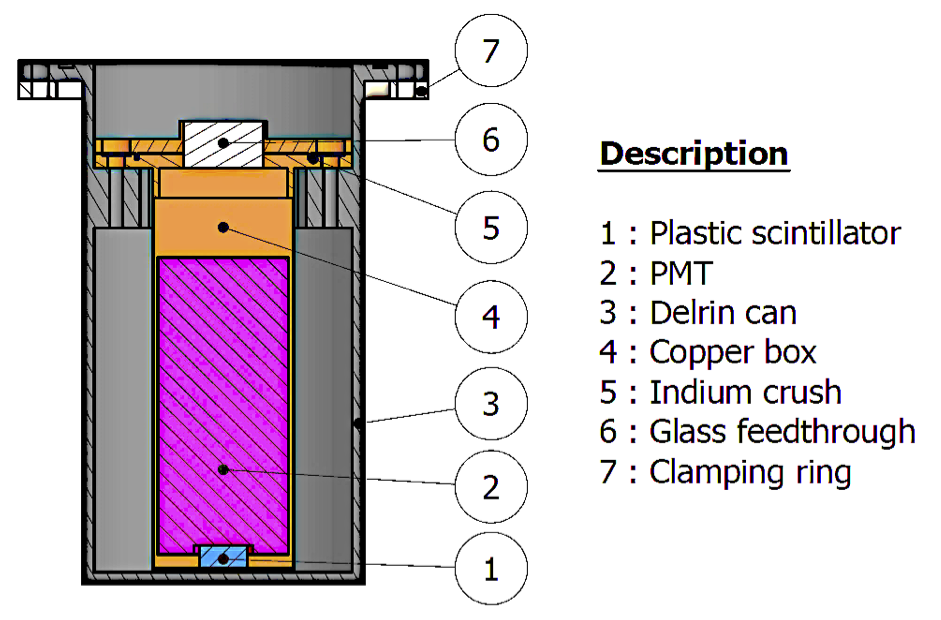}
    \qquad
    \includegraphics[height=6cm]{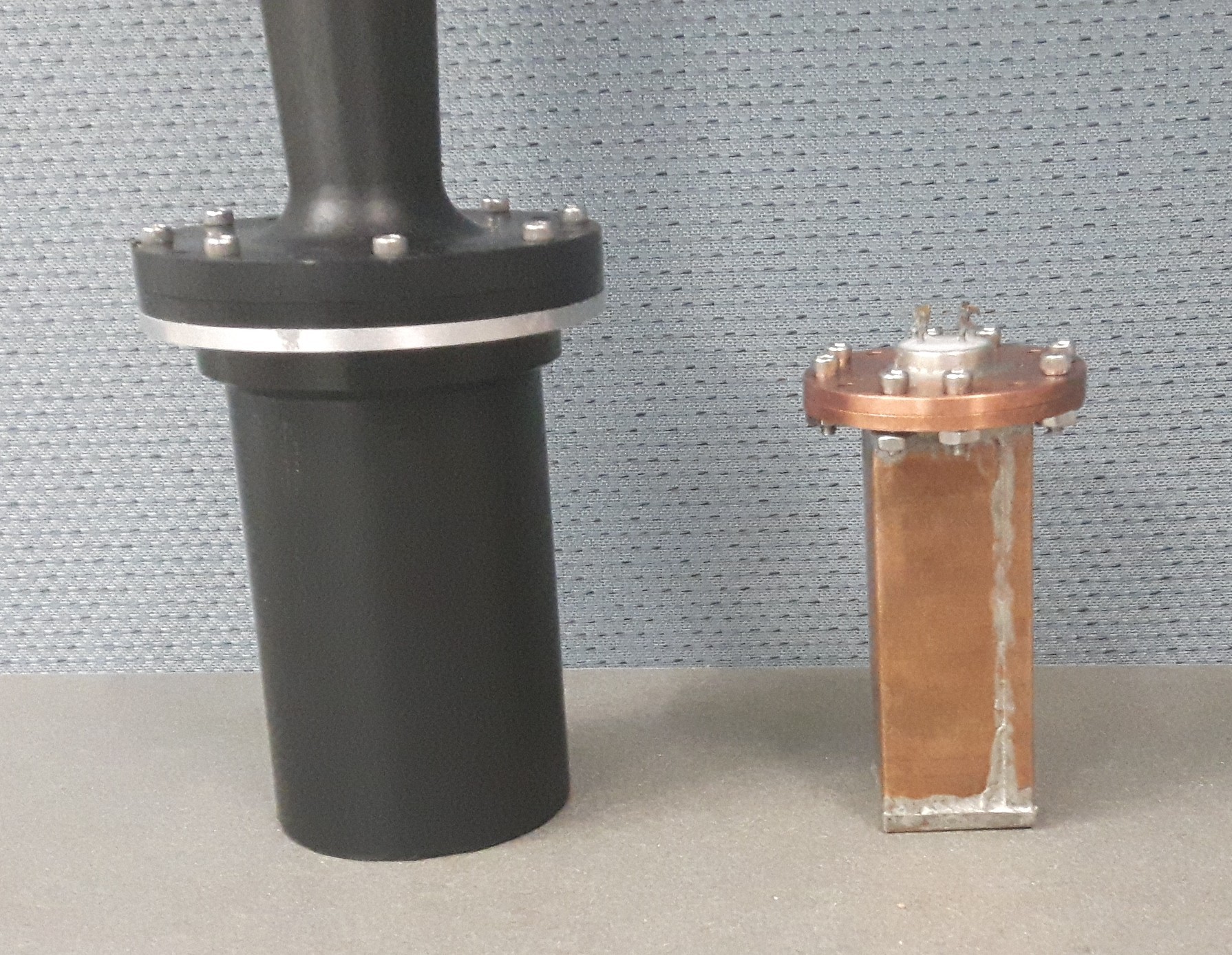}
    \caption[Sc-46 tagged source]{The \isotope[46]{Sc} tagged source.
        \textit{Left:} Schematic of the source: (1) plastic scintillator, (2) PMT, (3) Delrin can, (4) copper box, (5) indium crush, (6) glass feedthrough, (7) clamping ring.
        \textit{Right:} Photo of the black outer can (Delrin) attached to the stem, and the inner can (copper) containing the source.
    }
    \label{fig:calib:TaggedSource}
\end{figure}

The source is made from \isotope[45]{Sc} and will be activated via neutron bombardment to an activity of $\sim$200\,Bq prior to the calibration campaign.\footnote{Activation is done at the SLOWPOKE reactor, Royal Military College, Kingston ON. \href{https://www.rmc-cmr.ca/en/chemistry-and-chemical-engineering/slowpoke-2-facility}{https://www.rmc-cmr.ca/}}
Once activated, the source is placed onto a PMT enclosed in a copper housing, which is in turn encapsulated in a Delrin housing.
The copper housing lid is sealed by using indium wire in a crush seal, and the Delrin housing is sealed with FFKM O-rings.\footnote{TRP Polymer Solutions. \href{https://trp.co.uk/products/high-performance-elastomer-O-rings/}{https://trp.co.uk/products/high-performance-elastomer-O-rings/}}
\Cref{fig:calib:TaggedSource} shows the design and picture of the \isotope[46]{Sc} source.
Other isotopes that decay via $\beta$ and $\gamma$ emission may be used with this design by changing the plastic scintillator containing the isotope.\footnote{Saint-Gobain Crystals, BC-408 plastic scintillator. \href{https://www.crystals.saint-gobain.com/products/bc-408-bc-412-bc-416}{https://www.crystals.saint-gobain.com/}}
The PMT detects the interaction of the $\beta$ particle in the plastic scintillator and produces a tag.
A tagged source aids in the identification of $\gamma$s and thus produces a lower background spectrum, allowing a measurement of the high energy tails of the distribution.

\subsubsection{Untagged Radioactive Calibration Sources}

\begin{figure}[htp]
    \centering
    \includegraphics[height=38mm]{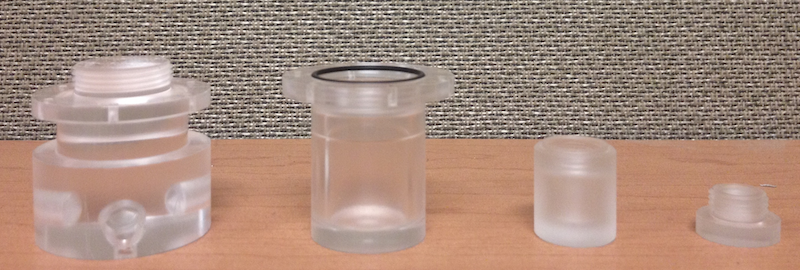}
    \quad
    \includegraphics[height=38mm]{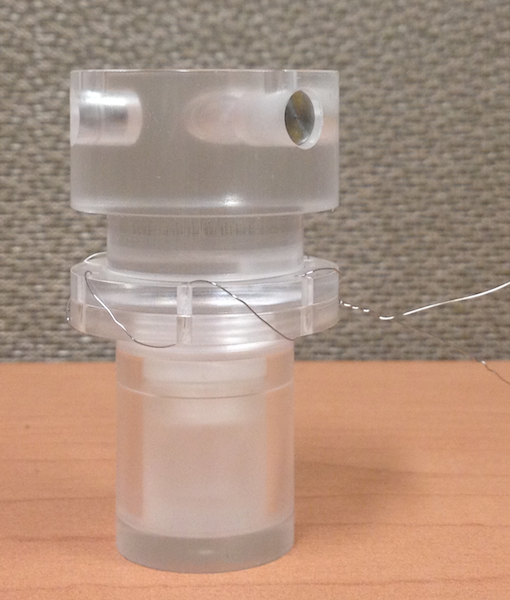}
    \caption{\textit{Left:} Photo of the \isotope[48]{Sc} source components.
        From left to right: the lid of the outer container, the outer container with the FFKM O-ring, the inner container, and the lid of the inner container.
        \textit{Right:} Photo of the \isotope[48]{Sc} source once closed.
        The safety stainless-steel wire prevents the outer container from opening while the source is moved around the detector volume.
        The stainless steel cross dowels are also shown (top lid).
    }
    \label{fig:calib:Sc48bis}
\end{figure}

\isotope[48]{Sc}, \isotope[137]{Cs} and \isotope[57]{Co} are three $\gamma$ sources designed to calibrate the energy scale during the scintillator and tellurium phases.
Untagged sources will be encapsulated in two independent acrylic containers (shown for \isotope[48]{Sc} in \Cref{fig:calib:Sc48bis}).
The UVA acrylic is the same type as was used in the Cherenkov source (see \Cref{sec:cherenkov}).
Each container is closed using a LAB-resistant FFKM O-ring.
The encapsulated source is connected to the umbilical via an acetal stem and the source connector (see \Cref{sec:source_connector}) that acts as a weight.
As a safety measure, stainless steel wire is used to prevent the opening of the outer container during movement of the source in the liquid scintillator, which can loosen the screws.
The energy losses as a result of interactions with the container material have been studied with MC simulations.

\paragraph{Scandium-48}

The decay modes of \isotope[48]{Sc} (T$_{1/2}$ = 43.67\,hours) are shown in \Cref{tab:calib:Sc48_decay}.
The source is produced via bombardment of a high purity titanium cylinder (12.7\,mm height and 6\,mm diameter) with 14\,MeV neutrons from a DT neutron generator.\footnote{HZDR Laboratory in Dresden-Rossendorf (Germany). \href{https://www.hzdr.de}{https://www.hzdr.de}}
At these energies, the expected production cross section for \isotope[48]{Sc} is $\sim$46\,mb for the \isotope[\mathrm{nat}]{Ti}(n,x) reaction.
Because of the short half-life, the source is expected to be produced shortly before the calibration campaign.
The activity of the source during the calibration campaign is expected to be 200--300 Bq.
The encapsulation can be opened to exchange the titanium cylinder for a future calibration campaign once the \isotope[48]{Sc} has decayed.
Effects of other isotopes (\isotope[46]{Sc}, \isotope[47]{Sc}) produced in the source by neutron bombardment have been estimated; the effect was found to be negligible.
The selected thickness of the encapsulation will stop the emitted $\beta$ particles (contamination smaller than 0.4\%), leaving a pure $\gamma$ source.
80--85\% of the $\gamma$s deposit their entire energy in the scintillator, with only a $\sim$0.1\% probability of being absorbed by the source itself.

\paragraph{Cesium-137}
The decay modes of \isotope[137]{Cs} (T$_{1/2}$ = 30.1\,y) are shown in \Cref{tab:calib:Sc48_decay}.
The source is a polyethylene disk (12.7\,mm diameter, 0.2\,mm thickness) produced externally.\footnote{National Metrology Institute of Germany (PTB). \href{https://www.ptb.de}{https://www.ptb.de}}
The polyethylene disk will be surrounded by two stainless steel disks (0.1\,mm thickness), used to stop low energy $\gamma$s or electrons.
About 83.6\% of the high energy $\gamma$s deposit their entire energy in the scintillator, with less than 1\% probability for the $\beta$s emitted to reach the scintillator.
The expected activity to be used during the calibration campaign is 123\,Bq.

\paragraph{Cobalt-57}
The \isotope[57]{Co} source is constructed identically to the \isotope[137]{Cs} source (see above).
The decay modes of \isotope[57]{Co} ($T_{1/2}=271.74$\,d) are shown in \Cref{tab:calib:Sc48_decay}.
The activity of the source during the calibration campaign will be 70--80\,Bq, with contamination from \isotope[60]{Co} expected to be at the level of $(30\pm6)$\,mBq.
About 70\% of the $\gamma$s deposit their entire energy in the scintillator, with less than 1\% of the conversion electrons and low energy $\gamma$s reaching the scintillator.
The $\gamma$ energy spectrum is mainly composed of two peaks --- one at 0.12\,MeV (with a branching ratio of 85.6\%) and one at 0.14\,MeV (10.68\%).
The peaks cannot be distinguished, though the shift in the peak position from 0.12\,MeV (a result of the higher energy $\gamma$s) is estimated to be below 0.5\%.

\begin{table}[htp]
    \centering
    \caption{Decay modes of the various calibration isotopes.}
    \label{tab:calib:Sc48_decay}
    \scalebox{0.9}{ 
    \begin{tabular}{c c c c}
        \hline \hline
        Isotope & Branching ratio [\%] &  $\beta$ max. energy [MeV] & $\gamma$ energies [MeV] \\ \hline
        $^{48}$Sc & 89.98 & 0.66 & (1.04, 1.31, 0.98) \\
        & 10.02 & 0.48 & (1.21, 1.31, 0.98) or (0.17, 1.04, 1.31, 0.98) \\ \hline
        $^{137}$Cs & 94.4 & 0.514 & 0.6617  \\
        & 0.00061 & 0.892 & 0.2835 \\
        & 5.64 & 1.176 & --- \\ \hline
        $^{57}$Co & 85.6 & --- & (0.122, 0.014) \\
        & 10.68 & --- & 0.136 \\
        & 0.17 & --- & multiple \\
        \hline \hline
    \end{tabular}
    }
\end{table}

    \section{Conclusions and Outlook}
\label{conclusions}

The infrastructure of the SNO experiment has been refurbished and upgraded for the SNO+ experiment.
The water systems were upgraded and three purpose-built plants were constructed for the purification, loading, and deployment of active target materials.
Repairs to the cavity floor and PMT electronics were completed.
A hold-down rope-net was engineered and installed, anchoring the acrylic vessel to the cavity floor.
A steady-state nitrogen cover gas system was installed to maintain low background levels and overall detector cleanliness.
Readout electronics and trigger systems were upgraded to handle higher data flow rates.
New calibration sources and systems were developed to understand the detector's timing and energy response, while minimizing the influx of radioactivity.
The resulting detector offers a wide and distinctly different physics program from SNO, with the ultimate goal of searching for neutrinoless double beta (\dbd{0}) decay.

During the SNO+ water phase, valuable detector experience was gained; both physics and calibration data were collected.
World-leading limits on invisible nucleon decay lifetimes were set~\cite{snop_nd}, the \isotope[8]{B} solar neutrino flux was measured with very low backgrounds~\cite{snop_solar}, and the rate of neutrons capturing on water was measured with unprecedented efficiency~\cite{snop_ncapture}.
Additional analyses are underway, including a measurement of muon-induced neutron spallation and a search for reactor antineutrinos in pure water.

The ultra-pure water in the acrylic vessel is currently being replaced with a liquid scintillator developed by the SNO+ collaboration~\cite{snop_scint}.
By doing so, the energy threshold for physics analyses will be lowered to $\mathcal{O}(1)$\,MeV, expanding the range of physics topics that may be studied to include low-energy solar neutrinos, reactor and geo-antineutrinos, supernova neutrinos, and various exotic searches.

The scintillator will be loaded with a tellurium diol compound using a novel diolisation method developed by the SNO+ collaboration.
The final phase will enable world-leading searches for \dbd{0} decay, with a lifetime sensitivity of 2.1\ee{26}\,y expected after five years of data-taking.
An initial concentration of 0.5\% Te with 1.3\,tonnes of \isotope[130]{Te} is planned, though the loading technique is scalable.
A future increase in loading may push the \dbd{0} decay sensitivity above $10^{27}$\,y, probing the inverted mass ordering for the first time.

With background levels under control, data acquisition and calibration systems running, and scintillator deployed within the detector, the SNO+ experiment is on target to search for neutrinoless double beta decay once tellurium is added to the detector.
A \dbd{0} decay discovery would provide evidence for lepton number violation, which could explain the matter-antimatter asymmetry in the universe.

    \section*{Acknowledgements}
\label{sec:acknowledgements}

Capital construction funds for the SNO+ experiment were provided by the Canada Foundation for Innovation (CFI) and matching partners.
This research was supported by:
{\bf Canada}: Natural Sciences and Engineering Research Council, the Canadian Institute for Advanced Research (CIFAR), Queen’s University at Kingston, Ontario Ministry of Research, Innovation and Science, Alberta Science and Research Investments Program, National Research Council, Federal Economic Development Initiative for Northern Ontario (FedNor), Northern Ontario Heritage Fund Corporation, Ontario Early Researcher Awards, Arthur B. McDonald Canadian Astroparticle Physics Research Institute;
{\bf Germany}: the Deutsche Forschungsgemeinschaft;
{\bf Mexico}: DGAPA-UNAM and Consejo Nacional de Ciencia y Tecnolog\'{i}a;
{\bf Portugal}: Funda\c{c}\~{a}o para a Ci\^{e}ncia e a Tecnologia (FCT-Portugal);
{\bf UK}: Science and Technology Facilities Council (STFC), the European Union’s Seventh Framework Programme under the European Research Council (ERC) grant agreement, the Marie Curie grant agreement;
{\bf US}: Department of Energy Office of Nuclear Physics, National Science Foundation, the University of California, Berkeley, Department of Energy National Nuclear Security Administration through the Nuclear Science and Security Consortium.

We would like to thank SNOLAB and its staff for support through underground space, logistical, and technical services.
SNOLAB operations are supported by the Canada Foundation for Innovation and the Province of Ontario, with underground access provided by Vale Canada Limited at the Creighton mine site.

This research was enabled in part by support provided by WestGRID (\url{www.westgrid.ca}) and ComputeCanada (\url{www.computecanada.ca}), in particular computer systems and support from the University of Alberta (\url{www.ualberta.ca}) and from Simon Fraser University (\url{www.sfu.ca}); and by the GridPP Collaboration, in particular computer systems and support from Rutherford Appleton Laboratory~\cite{gridpp_2006,gridpp_2009}.
Additional high-performance computing was provided through the ``Illume'' cluster funded by CFI and Alberta Economic Development and Trade (EDT) and operated by ComputeCanada and the Savio computational cluster resource provided by the Berkeley Research Computing program at the University of California, Berkeley (supported by the UC Berkeley Chancellor, Vice Chancellor for Research, and Chief Information Officer).
Additional long-term storage was provided by the Fermilab Scientific Computing Division.
Fermilab is managed by Fermi Research Alliance, LLC (FRA) under Contract with the U.S. Department of Energy, Office of Science, Office of High Energy Physics.



    \bibliographystyle{bib/JHEP}
    \bibliography{bib/references}

\providecommand{\href}[2]{#2}\begingroup\raggedright\begin{thebibliography}{100}

\bibitem{snonim}
{\scshape SNO} collaboration, J.~Boger et~al., \emph{The {Sudbury Neutrino
  Observatory}},
  \href{http://dx.doi.org/10.1016/S0168-9002(99)01469-2}{\emph{Nucl. Instrum.
  Meth.} {\bfseries A449} (2000) 172--207},
  [\href{https://arxiv.org/abs/nucl-ex/9910016}{{\ttfamily nucl-ex/9910016}}].

\bibitem{cuore-0}
{\scshape CUORE} collaboration, C.~Alduino et~al., \emph{Measurement of the
  two-neutrino double-beta decay half-life of {$^{130}$Te} with the {CUORE-0}
  experiment},
  \href{http://dx.doi.org/10.1140/epjc/s10052-016-4498-6}{\emph{Eur. Phys. J.}
  {\bfseries C77} (2017) 13},
  [\href{https://arxiv.org/abs/1609.01666}{{\ttfamily 1609.01666}}].

\bibitem{snop2015}
{\scshape SNO+} collaboration, S.~Andringa et~al., \emph{Current status and
  future prospects of the {SNO+} experiment},
  \href{http://dx.doi.org/10.1155/2016/6194250}{\emph{Adv. High Energy Phys.}
  {\bfseries 2016} (2016) 6194250},
  [\href{https://arxiv.org/abs/1508.05759}{{\ttfamily 1508.05759}}].

\bibitem{furry}
W.~H. Furry, \emph{{On transition probabilities in double
  beta-disintegration}},
  \href{http://dx.doi.org/10.1103/PhysRev.56.1184}{\emph{Phys. Rev.} {\bfseries
  56} (1939) 1184--1193}.

\bibitem{seesaw1}
P.~Minkowski, \emph{$\mu \to e\gamma$ at a rate of one out of $10^{9}$ muon
  decays?}, \href{http://dx.doi.org/10.1016/0370-2693(77)90435-X}{\emph{Phys.
  Lett.} {\bfseries 67B} (1977) 421--428}.

\bibitem{seesaw2}
M.~Gell-Mann, P.~Ramond and R.~Slansky, \emph{Complex spinors and unified
  theories}, {\emph{Conf. Proc.} {\bfseries C790927} (1979) 315--321},
  [\href{https://arxiv.org/abs/1306.4669}{{\ttfamily 1306.4669}}].

\bibitem{seesaw3}
T.~Yanagida, \emph{Horizontal symmetry and masses of neutrinos},
  \href{http://dx.doi.org/10.1143/PTP.64.1103}{\emph{Prog. Theor. Phys.}
  {\bfseries 64} (1980) 1103}.

\bibitem{seesaw4}
M.~Magg and C.~Wetterich, \emph{Neutrino mass problem and gauge hierarchy},
  \href{http://dx.doi.org/10.1016/0370-2693(80)90825-4}{\emph{Phys. Lett.}
  {\bfseries 94B} (1980) 61--64}.

\bibitem{seesaw5}
R.~N. Mohapatra and G.~Senjanovic, \emph{Neutrino mass and spontaneous parity
  nonconservation},
  \href{http://dx.doi.org/10.1103/PhysRevLett.44.912}{\emph{Phys. Rev. Lett.}
  {\bfseries 44} (1980) 912}.

\bibitem{seesaw6}
R.~N. Mohapatra and G.~Senjanovic, \emph{Neutrino masses and mixings in gauge
  models with spontaneous parity violation},
  \href{http://dx.doi.org/10.1103/PhysRevD.23.165}{\emph{Phys. Rev.} {\bfseries
  D23} (1981) 165}.

\bibitem{leptogenesis}
W.~Buchmüller, R.~D. Peccei and T.~Yanagida, \emph{Leptogenesis as the origin
  of matter},
  \href{http://dx.doi.org/10.1146/annurev.nucl.55.090704.151558}{\emph{Ann.
  Rev. Nucl. Part. Sci.} {\bfseries 55} (2005) 311--355},
  [\href{https://arxiv.org/abs/hep-ph/0502169}{{\ttfamily hep-ph/0502169}}].

\bibitem{lindner_talk}
M.~Lindner, ``{Neutrinoless Double Beta Decay and new Physics}.'' presented at
  ``Beyond the Standard model with Neutrinos and Nuclear Physics'', Brussels,
  Nov 29 -- Dec 1, 2017.
  \url{http://www.solvayinstitutes.be/event/workshop/beyond_2018/slides/M_Lindner.pdf}.

\bibitem{nme1}
H.~Ejiri, \emph{{Nuclear Matrix Elements for $\beta$ and $\beta \beta$ Decays
  and Quenching of the Weak Coupling $g_A$ in QRPA}},
  \href{http://dx.doi.org/10.3389/fphy.2019.00030}{\emph{Front.in Phys.}
  {\bfseries 7} (2019) 30}.

\bibitem{nme2}
J.~Engel and J.~Menéndez, \emph{{Status and Future of Nuclear Matrix Elements
  for Neutrinoless Double-Beta Decay: A Review}},
  \href{http://dx.doi.org/10.1088/1361-6633/aa5bc5}{\emph{Rept. Prog. Phys.}
  {\bfseries 80} (2017) 046301},
  [\href{https://arxiv.org/abs/1610.06548}{{\ttfamily 1610.06548}}].

\bibitem{biller_2017}
S.~Biller and {\relax Sz}.~Manecki, \emph{A new technique to load {$^{130}$Te}
  in liquid scintillator for neutrinoless double beta decay experiments},
  \href{http://dx.doi.org/10.1088/1742-6596/888/1/012084}{\emph{Journal of
  Physics: Conference Series} {\bfseries 888} (2017) 012084}.

\bibitem{pdg2018}
{\scshape Particle Data Group} collaboration, M.~Tanabashi et~al., \emph{Review
  of particle physics},
  \href{http://dx.doi.org/10.1103/PhysRevD.98.030001}{\emph{Phys. Rev.}
  {\bfseries D98} (2018) 030001}.

\bibitem{giuliani2012}
A.~Giuliani and A.~Poves, \emph{Neutrinoless double-beta decay},
  \href{http://dx.doi.org/10.1155/2012/857016}{\emph{Adv. High Energy Phys.}
  {\bfseries 2012} (2012) 857016}.

\bibitem{borexino_CNO}
{\scshape BOREXINO} collaboration, M.~Agostini et~al., \emph{{Experimental
  evidence of neutrinos produced in the CNO fusion cycle in the Sun}},
  \href{http://dx.doi.org/10.1038/s41586-020-2934-0}{\emph{Nature} {\bfseries
  587} (2020) 577--582}, [\href{https://arxiv.org/abs/2006.15115}{{\ttfamily
  2006.15115}}].

\bibitem{tretyak2002}
V.~I. Tretyak and Y.~G. Zdesenko, \emph{Tables of double beta decay data: An
  update}, \href{http://dx.doi.org/10.1006/adnd.2001.0873}{\emph{Atom. Data
  Nucl. Data Tabl.} {\bfseries 80} (2002) 83--116}.

\bibitem{simkovic2010}
F.~Šimkovic, \emph{Double beta decay: A problem of particle, nuclear and
  atomic physics},
  \href{http://dx.doi.org/10.1016/j.ppnp.2009.12.015}{\emph{Prog. Part. Nucl.
  Phys.} {\bfseries 64} (2010) 219--227}.

\bibitem{geoneutrinos}
G.~Fiorentini, M.~Lissia and F.~Mantovani, \emph{{Geo-neutrinos and Earth's
  interior}},
  \href{http://dx.doi.org/10.1016/j.physrep.2007.09.001}{\emph{Phys. Rept.}
  {\bfseries 453} (2007) 117--172},
  [\href{https://arxiv.org/abs/0707.3203}{{\ttfamily 0707.3203}}].

\bibitem{kamland}
P.~Alivisatos et~al., \emph{{KamLAND}: A liquid scintillator anti-neutrino
  detector at the {Kamioka} site},  tech. rep., {Stanford-HEP-98-03,
  Tohoku-RCNS-98-15}, 1998.
\newblock
  \url{http://grattalab3.stanford.edu/neutrino/kamland_design_report.pdf}.

\bibitem{snop_solar}
{\scshape SNO+} collaboration, M.~Anderson et~al., \emph{Measurement of the
  {$^8$B} solar neutrino flux in {SNO+} with very low backgrounds},
  \href{http://dx.doi.org/10.1103/PhysRevD.99.012012}{\emph{Phys. Rev.}
  {\bfseries D99} (2019) 012012},
  [\href{https://arxiv.org/abs/1812.03355}{{\ttfamily 1812.03355}}].

\bibitem{snolab_science}
F.~Duncan, A.~J. Noble and D.~Sinclair, \emph{{The construction and anticipated
  science of SNOLAB}},
  \href{http://dx.doi.org/10.1146/annurev.nucl.012809.104513}{\emph{Ann. Rev.
  Nucl. Part. Sci.} {\bfseries 60} (2010) 163--180}.

\bibitem{wolfenstein_1978}
L.~Wolfenstein, \emph{Neutrino oscillations in matter},
  \href{http://dx.doi.org/10.1103/PhysRevD.17.2369}{\emph{Phys. Rev.}
  {\bfseries D17} (1978) 2369--2374}.

\bibitem{msw_1986}
S.~P. Mikheyev and A.~{\relax Yu}. Smirnov, \emph{Resonance amplification of
  oscillations in matter and spectroscopy of solar neutrinos}, {\emph{Sov. J.
  Nucl. Phys.} {\bfseries 42} (1985) 913--917}.

\bibitem{nsi_2013}
T.~Ohlsson, \emph{Status of non-standard neutrino interactions},
  \href{http://dx.doi.org/10.1088/0034-4885/76/4/044201}{\emph{Rept. Prog.
  Phys.} {\bfseries 76} (2013) 044201},
  [\href{https://arxiv.org/abs/1209.2710}{{\ttfamily 1209.2710}}].

\bibitem{nsi_2015}
O.~G. Miranda and H.~Nunokawa, \emph{Non standard neutrino interactions:
  current status and future prospects},
  \href{http://dx.doi.org/10.1088/1367-2630/17/9/095002}{\emph{New J. Phys.}
  {\bfseries 17} (2015) 095002},
  [\href{https://arxiv.org/abs/1505.06254}{{\ttfamily 1505.06254}}].

\bibitem{solar_metal1}
J.~N. Bahcall, S.~Basu, M.~Pinsonneault and A.~M. Serenelli,
  \emph{Helioseismological implications of recent solar abundance
  determinations}, \href{http://dx.doi.org/10.1086/426070}{\emph{Astrophys. J.}
  {\bfseries 618} (2005) 1049--1056},
  [\href{https://arxiv.org/abs/astro-ph/0407060}{{\ttfamily
  astro-ph/0407060}}].

\bibitem{solar_metal2}
J.~N. Bahcall, A.~M. Serenelli and S.~Basu, \emph{10,000 standard solar models:
  a {Monte Carlo} simulation},
  \href{http://dx.doi.org/10.1086/504043}{\emph{Astrophys. J. Suppl.}
  {\bfseries 165} (2006) 400--431},
  [\href{https://arxiv.org/abs/astro-ph/0511337}{{\ttfamily
  astro-ph/0511337}}].

\bibitem{solar_metal3}
W.~M. Yang and S.~L. Bi, \emph{Solar models with revised abundances and
  opacities}, \href{http://dx.doi.org/10.1086/513694}{\emph{Astrophys. J.}
  {\bfseries 658} (2007) L67--L70},
  [\href{https://arxiv.org/abs/0805.3644}{{\ttfamily 0805.3644}}].

\bibitem{solar_metal4}
S.~Basu and H.~M. Antia, \emph{Helioseismology and solar abundances},
  \href{http://dx.doi.org/10.1016/j.physrep.2007.12.002}{\emph{Phys. Rept.}
  {\bfseries 457} (2008) 217--283},
  [\href{https://arxiv.org/abs/0711.4590}{{\ttfamily 0711.4590}}].

\bibitem{solar_metal5}
A.~Serenelli, S.~Basu, J.~W. Ferguson and M.~Asplund, \emph{New solar
  composition: The problem with solar models revisited},
  \href{http://dx.doi.org/10.1088/0004-637X/705/2/L123}{\emph{Astrophys. J.}
  {\bfseries 705} (2009) L123--L127},
  [\href{https://arxiv.org/abs/0909.2668}{{\ttfamily 0909.2668}}].

\bibitem{solar_metal6}
D.~G. Cerdeño, J.~H. Davis, M.~Fairbairn and A.~C. Vincent, \emph{{CNO}
  neutrino grand prix: The race to solve the solar metallicity problem},
  \href{http://dx.doi.org/10.1088/1475-7516/2018/04/037}{\emph{JCAP} {\bfseries
  1804} (2018) 37}, [\href{https://arxiv.org/abs/1712.06522}{{\ttfamily
  1712.06522}}].

\bibitem{snews}
P.~Antonioli et~al., \emph{{SNEWS}: The {SuperNova Early Warning System}},
  \href{http://dx.doi.org/10.1088/1367-2630/6/1/114}{\emph{New J. Phys.}
  {\bfseries 6} (2004) 114},
  [\href{https://arxiv.org/abs/astro-ph/0406214}{{\ttfamily
  astro-ph/0406214}}].

\bibitem{snop_nd}
{\scshape SNO+} collaboration, M.~Anderson et~al., \emph{Search for invisible
  modes of nucleon decay in water with the {SNO+} detector},
  \href{http://dx.doi.org/10.1103/PhysRevD.99.032008}{\emph{Phys. Rev.}
  {\bfseries D99} (2019) 032008},
  [\href{https://arxiv.org/abs/1812.05552}{{\ttfamily 1812.05552}}].

\bibitem{snolab_facility}
N.~J.~T. Smith, \emph{{The SNOLAB deep underground facility}},
  \href{http://dx.doi.org/10.1140/epjp/i2012-12108-9}{\emph{Eur. Phys. J. Plus}
  {\bfseries 127} (2012) 108}.

\bibitem{snoprd}
{\scshape SNO} collaboration, B.~Aharmim et~al., \emph{Measurement of the
  cosmic ray and neutrino-induced muon flux at the {Sudbury} neutrino
  observatory}, \href{http://dx.doi.org/10.1103/PhysRevD.80.012001}{\emph{Phys.
  Rev.} {\bfseries D80} (2009) 012001},
  [\href{https://arxiv.org/abs/0902.2776}{{\ttfamily 0902.2776}}].

\bibitem{snop_te}
{\scshape SNO+} collaboration, M.~Anderson et~al., \emph{A method to load
  tellurium in liquid scintillator to study neutrinoless double beta decay},
  in preparation.

\bibitem{borexino}
{\scshape Borexino} collaboration, G.~Alimonti et~al., \emph{The {Borexino}
  detector at the {Laboratori Nazionali del Gran Sasso}},
  \href{http://dx.doi.org/10.1016/j.nima.2008.11.076}{\emph{Nucl. Instrum.
  Meth.} {\bfseries A600} (2009) 568--593},
  [\href{https://arxiv.org/abs/0806.2400}{{\ttfamily 0806.2400}}].

\bibitem{kreslo_2011}
I.~Kreslo et~al., \emph{Pulse-shape discrimination of scintillation from alpha
  and beta particles with liquid scintillator and geiger-mode multipixel
  avalanche diodes},
  \href{http://dx.doi.org/10.1088/1748-0221/6/07/P07009}{\emph{JINST}
  {\bfseries 6} (2011) P07009},
  [\href{https://arxiv.org/abs/1105.4432}{{\ttfamily 1105.4432}}].

\bibitem{wan_2015}
B.~Wan et~al., \emph{Digital pulse shape discrimination methods for n-$\gamma$
  separation in an {EJ-301} liquid scintillation detector},
  \href{http://dx.doi.org/10.1088/1674-1137/39/11/116201}{\emph{Chin. Phys.}
  {\bfseries C39} (2015) 116201},
  [\href{https://arxiv.org/abs/1502.01807}{{\ttfamily 1502.01807}}].

\bibitem{buck_2016}
C.~Buck and M.~Yeh, \emph{{Metal-loaded organic scintillators for neutrino
  physics}}, \href{http://dx.doi.org/10.1088/0954-3899/43/9/093001}{\emph{J.
  Phys. G} {\bfseries 43} (2016) 093001},
  [\href{https://arxiv.org/abs/1608.04897}{{\ttfamily 1608.04897}}].

\bibitem{DYB}
{\scshape Daya Bay} collaboration, F.~An et~al., \emph{{The Detector System of
  The Daya Bay Reactor Neutrino Experiment}},
  \href{http://dx.doi.org/10.1016/j.nima.2015.11.144}{\emph{Nucl. Instrum.
  Meth. A} {\bfseries 811} (2016) 133--161},
  [\href{https://arxiv.org/abs/1508.03943}{{\ttfamily 1508.03943}}].

\bibitem{reno}
J.~Park et~al., \emph{{Production and optical properties of Gd-loaded liquid
  scintillator for the RENO neutrino detector}},
  \href{http://dx.doi.org/10.1016/j.nima.2012.12.121}{\emph{Nucl. Instrum.
  Meth. A} {\bfseries 707} (2013) 45--53}.

\bibitem{cosine100}
G.~Adhikari et~al., \emph{{Initial Performance of the COSINE-100 Experiment}},
  \href{http://dx.doi.org/10.1140/epjc/s10052-018-5590-x}{\emph{Eur. Phys. J.
  C} {\bfseries 78} (2018) 107},
  [\href{https://arxiv.org/abs/1710.05299}{{\ttfamily 1710.05299}}].

\bibitem{juno}
D.~Cao et~al., \emph{{Light Absorption Properties of the High Quality Linear
  Alkylbenzene for the JUNO Experiment}},
  \href{http://dx.doi.org/10.1016/j.nima.2019.01.077}{\emph{Nucl. Instrum.
  Meth. A} {\bfseries 927} (2019) 230--235},
  [\href{https://arxiv.org/abs/1801.08363}{{\ttfamily 1801.08363}}].

\bibitem{sabre}
{\scshape SABRE} collaboration, M.~Antonello et~al., \emph{{The SABRE project
  and the SABRE Proof-of-Principle}},
  \href{http://dx.doi.org/10.1140/epjc/s10052-019-6860-y}{\emph{Eur. Phys. J.
  C} {\bfseries 79} (2019) 363},
  [\href{https://arxiv.org/abs/1806.09340}{{\ttfamily 1806.09340}}].

\bibitem{snop_scint}
{\scshape SNO+} collaboration, M.~R. Anderson et~al., \emph{{Development,
  characterisation, and deployment of the SNO+ liquid scintillator}},
  \href{http://dx.doi.org/10.1088/1748-0221/16/05/p05009}{\emph{JINST}
  {\bfseries 16} (2021) P05009},
  [\href{https://arxiv.org/abs/2011.12924}{{\ttfamily 2011.12924}}].

\bibitem{wright_2009}
A.~J. Wright, \emph{Robust Signal Extraction Methods and {Monte Carlo}
  Sensitivity Studies for the {Sudbury} Neutrino Observatory and {SNO+}
  Experiments}.
\newblock PhD thesis, Queen's University, 2009.
\newblock \url{http://hdl.handle.net/1974/5152}.

\bibitem{bartlett_2018}
D.~Bartlett, \emph{Quality assurance testing, and sensitivity studies for the
  {SNO+} experiment},  Master's thesis, Queen's University, 2018.
\newblock \url{http://hdl.handle.net/1974/23990}.

\bibitem{buck_2015}
C.~Buck, B.~Gramlich and S.~Wagner, \emph{{Light propagation and fluorescence
  quantum yields in liquid scintillators}},
  \href{http://dx.doi.org/10.1088/1748-0221/10/09/P09007}{\emph{JINST}
  {\bfseries 10} (2015) P09007},
  [\href{https://arxiv.org/abs/1509.02327}{{\ttfamily 1509.02327}}].

\bibitem{vonKrosigk_2015}
B.~von Krosigk et~al., \emph{Measurement of $\alpha$-particle quenching in
  {LAB} based scintillator in independent small-scale experiments},
  \href{http://dx.doi.org/10.1140/epjc/s10052-016-3959-2}{\emph{Eur. Phys. J.}
  {\bfseries C76} (2016) 109},
  [\href{https://arxiv.org/abs/1510.00458}{{\ttfamily 1510.00458}}].

\bibitem{vonKrosigk_2013}
B.~von Krosigk, L.~Neumann, R.~Nolte, S.~Röttger and K.~Zuber,
  \emph{Measurement of the proton light response of various {LAB} based
  scintillators and its implication for supernova neutrino detection via
  neutrino-proton scattering},
  \href{http://dx.doi.org/10.1140/epjc/s10052-013-2390-1}{\emph{Eur. Phys. J.}
  {\bfseries C73} (2013) 2390},
  [\href{https://arxiv.org/abs/1301.6403}{{\ttfamily 1301.6403}}].

\bibitem{borexino_background}
G.~Zuzel, \emph{Low background techniques applied in the {BOREXINO}
  experiment}, \href{http://dx.doi.org/10.1063/1.4928003}{\emph{AIP Conf.
  Proc.} {\bfseries 1672} (2015) 110001}.

\bibitem{kamland_reactor}
{\scshape KamLAND} collaboration, A.~Suzuki, \emph{Results from {KamLAND}
  reactor neutrino detection},
  \href{http://dx.doi.org/10.1088/0031-8949/2005/T121/004}{\emph{Phys. Scripta}
  {\bfseries T121} (2005) 33--38}.

\bibitem{borexino_bkg2}
{\scshape Borexino} collaboration, M.~Agostini et~al., \emph{{First
  Simultaneous Precision Spectroscopy of $pp$, $^7$Be, and $pep$ Solar
  Neutrinos with Borexino Phase-II}},
  \href{http://dx.doi.org/10.1103/PhysRevD.100.082004}{\emph{Phys. Rev. D}
  {\bfseries 100} (2019) 082004},
  [\href{https://arxiv.org/abs/1707.09279}{{\ttfamily 1707.09279}}].

\bibitem{ford_scintplant}
{\scshape SNO+} collaboration, R.~J. Ford, \emph{{A Scintillator Purification
  Plant and Fluid Handling System for SNO+}},
  \href{http://dx.doi.org/10.1063/1.4927998}{\emph{AIP Conf. Proc.} {\bfseries
  1672} (2015) 080003}, [\href{https://arxiv.org/abs/1506.08746}{{\ttfamily
  1506.08746}}].

\bibitem{scint_borexino}
J.~Benziger et~al., \emph{The scintillator purification system for the
  {Borexino} solar neutrino detector},
  \href{http://dx.doi.org/10.1016/j.nima.2007.12.043}{\emph{Nucl. Instrum.
  Meth.} {\bfseries A587} (2008) 277--291},
  [\href{https://arxiv.org/abs/0709.1503}{{\ttfamily 0709.1503}}].

\bibitem{scint_aip2011}
R.~Ford, M.~Chen, O.~Chkvorets, D.~Hallman and E.~V\'{a}zquez-J\'{a}uregui,
  \emph{{SNO+} scintillator purification and assay},
  \href{http://dx.doi.org/10.1063/1.3579580}{\emph{AIP Conf. Proc.} {\bfseries
  1338} (2011) 183--194}.

\bibitem{scint_neutrino2010}
O.~Chkvorets, R.~Ford, D.~Hallman and E.~V\'{a}zquez-J\'{a}uregui, \emph{Liquid
  scintillator purification and assay {R\&D} at {SNO+}},
  \href{http://dx.doi.org/10.1016/j.nuclphysbps.2012.09.156}{\emph{Nucl. Phys.
  Proc. Suppl.} {\bfseries 229-232} (2012) 519}.

\bibitem{snop_scint_2011}
H.~M. O'Keeffe, E.~O'Sullivan and M.~C. Chen, \emph{Scintillation decay time
  and pulse shape discrimination in oxygenated and deoxygenated solutions of
  linear alkylbenzene for the {SNO+} experiment},
  \href{http://dx.doi.org/10.1016/j.nima.2011.03.027}{\emph{Nucl. Instrum.
  Meth.} {\bfseries A640} (2011) 119--122},
  [\href{https://arxiv.org/abs/1102.0797}{{\ttfamily 1102.0797}}].

\bibitem{scint_nima2009}
{\scshape SNO} collaboration, B.~Aharmim et~al., \emph{High sensitivity
  measurement of {Ra-224} and {Ra-226} in water with an improved hydrous
  titanium oxide technique at the {Sudbury} neutrino observatory},
  \href{http://dx.doi.org/10.1016/j.nima.2009.01.227}{\emph{Nucl. Instrum.
  Meth.} {\bfseries A604} (2009) 531--535},
  [\href{https://arxiv.org/abs/0803.4162}{{\ttfamily 0803.4162}}].

\bibitem{scint_purification}
J.~B. Benziger et~al., \emph{{A scintillator purification system for a large
  scale solar neutrino experiment}},
  \href{http://dx.doi.org/10.1016/S0168-9002(98)00767-0}{\emph{Nucl. Instrum.
  Meth.} {\bfseries A417} (1998) 278--296}.

\bibitem{Lozza_2015}
V.~Lozza and J.~Petzoldt, \emph{{Cosmogenic activation of a natural tellurium
  target}},
  \href{http://dx.doi.org/10.1016/j.astropartphys.2014.06.008}{\emph{Astropart.
  Phys.} {\bfseries 61} (2015) 62--71},
  [\href{https://arxiv.org/abs/1411.5947}{{\ttfamily 1411.5947}}].

\bibitem{hans_2015}
S.~Hans et~al., \emph{Purification of telluric acid for {SNO+} neutrinoless
  double-beta decay search},
  \href{http://dx.doi.org/10.1016/j.nima.2015.05.045}{\emph{Nucl. Instrum.
  Meth.} {\bfseries A795} (2015) 132--139}.

\bibitem{sno_urylon}
E.~D. Hallman, \emph{{SNO} underground spray testing -- urylon polyurethane
  coatings},  tech. rep.,
  \href{https://sno.phy.queensu.ca/str/SNO-STR-93-010.pdf}{SNO-STR-93-010},
  1993.

\bibitem{khaghani_2016}
P.~Khaghani, \emph{Neck sense rope system and leaching studies for {SNO+}},
  Master's thesis, Laurentian University, 2016.
\newblock \url{https://core.ac.uk/download/pdf/222897163.pdf}.

\bibitem{snop_ropes}
A.~Bialek et~al., \emph{A rope-net support system for the liquid scintillator
  detector for the {SNO+} experiment},
  \href{http://dx.doi.org/10.1016/j.nima.2016.04.114}{\emph{Nucl. Instrum.
  Meth.} {\bfseries A827} (2016) 152--160}.

\bibitem{stachiw_acrylics}
J.~Stachiw et~al., \emph{Handbook of Acrylics for Submersibles, Hyperbaric
  Chambers and Aquaria}.
\newblock Best Pub. Co., 2003.

\bibitem{rope_stretch}
L.~Govaert, C.~Bastiaansen and P.~Leblans, \emph{Stress-strain analysis of
  oriented polyethylene},
  \href{http://dx.doi.org/https://doi.org/10.1016/0032-3861(93)90546-M}{\emph{Polymer}
  {\bfseries 34} (1993) 534--540}.

\bibitem{snoprc}
{\scshape SNO} collaboration, B.~Aharmim et~al., \emph{Measurement of the
  $\nu_e$ and total {$^{8}$B} solar neutrino fluxes with the {Sudbury} neutrino
  observatory phase-{III} data set},
  \href{http://dx.doi.org/10.1103/PhysRevC.87.015502}{\emph{Phys. Rev.}
  {\bfseries C87} (2013) 015502},
  [\href{https://arxiv.org/abs/1107.2901}{{\ttfamily 1107.2901}}].

\bibitem{bonventre_2014}
R.~Bonventre, \emph{Neutron Multiplicity in Atmospheric Neutrino Events at the
  Sudbury Neutrino Observatory}.
\newblock PhD thesis, University of Pennsylvania, 2014.
\newblock \url{https://repository.upenn.edu/edissertations/1213}.

\bibitem{snolab_sulfide}
R.~Ford, \emph{{SNOLAB: Review of the facility and experiments}},
  \href{http://dx.doi.org/10.1063/1.3700605}{\emph{AIP Conf. Proc.} {\bfseries
  1441} (2012) 521--524}.

\bibitem{snolab_h2s}
I.~T. Lawson, \emph{Results of the air quality analysis of the underground
  laboratory for {H$_{2}$S} contamination},  tech. rep.,
  \href{https://www.snolab.ca/docushare/dsweb/Get/Document-10353/SNOLAB-STR-2014-003-Air-Quality.pdf}{SNOLAB-STR-2014-003},
  2014.

\bibitem{zebra}
J.~Zoll et~al., \emph{{Overview of the ZEBRA System}},  tech. rep., CERN
  Program Library Long Writeups Q100/Q101, 1995.
\newblock \url{https://cds.cern.ch/record/2296399}.

\bibitem{ORCA}
M.~A. Howe et~al., \emph{{Sudbury} neutrino observatory neutral current
  detector acquisition software overview},
  \href{http://dx.doi.org/10.1109/TNS.2004.829527}{\emph{IEEE Trans. Nucl.
  Sci.} {\bfseries 51} (2004) 878--883}.

\bibitem{braidwood}
T.~Bolton, \emph{The {Braidwood} reactor antineutrino experiment},
  \href{http://dx.doi.org/10.1016/j.nuclphysbps.2005.05.041}{\emph{Nucl. Phys.
  Proc. Suppl.} {\bfseries 149} (2005) 166--169}.

\bibitem{miniclean}
M.~Akashi-Ronquest et~al., \emph{Triplet lifetime in gaseous argon},
  \href{http://dx.doi.org/10.1140/epja/i2019-12867-2}{\emph{Eur. Phys. J.}
  {\bfseries A55} (2019) 176},
  [\href{https://arxiv.org/abs/1903.06706}{{\ttfamily 1903.06706}}].

\bibitem{deap}
{\scshape DEAP-3600} collaboration, P.~A. Amaudruz et~al., \emph{Design and
  construction of the {DEAP-3600} dark matter detector},
  \href{http://dx.doi.org/10.1016/j.astropartphys.2018.09.006}{\emph{Astropart.
  Phys.} {\bfseries 108} (2019) 1--23}.

\bibitem{geant4}
{\scshape GEANT4} collaboration, S.~Agostinelli et~al., \emph{{GEANT4}: A
  simulation toolkit},
  \href{http://dx.doi.org/10.1016/S0168-9002(03)01368-8}{\emph{Nucl. Instrum.
  Meth.} {\bfseries A506} (2003) 250--303}.

\bibitem{geant4-2}
J.~Allison et~al., \emph{Geant4 developments and applications},
  \href{http://dx.doi.org/10.1109/TNS.2006.869826}{\emph{IEEE Trans. Nucl.
  Sci.} {\bfseries 53} (2006) 270}.

\bibitem{glg4sim}
G.~Horton-Smith, ``An introduction to {GLG4sim} features.'' presented on May
  19, 2006.
  \url{http://neutrino.phys.ksu.edu/~GLG4sim/GLG4sim-intro-2006-05-19.pdf}.

\bibitem{root}
R.~Brun and F.~Rademakers, \emph{{ROOT} --- an object oriented data analysis
  framework},
  \href{http://dx.doi.org/10.1016/S0168-9002(97)00048-X}{\emph{Nucl. Instrum.
  Meth.} {\bfseries A389} (1997) 81--86}.

\bibitem{snonim_pmt}
C.~Jillings et~al., \emph{{The photomultiplier tube testing facility for the
  Sudbury Neutrino Observatory}},
  \href{http://dx.doi.org/10.1016/0168-9002(96)00067-8}{\emph{Nucl. Instrum.
  Meth. A} {\bfseries 373} (1996) 421--429}.

\bibitem{OSullivan_2012}
E.~O'Sullivan, H.~S. Wan Chan~Tseung, N.~Tolich, H.~M. O'Keeffe and M.~Chen,
  \emph{{SNO+} liquid scintillator characterization: Timing, quenching, and
  energy scale},
  \href{http://dx.doi.org/10.1016/j.nuclphysbps.2012.09.186}{\emph{Nucl. Phys.
  Proc. Suppl.} {\bfseries 229-232} (2012) 549}.

\bibitem{barnard_2014}
Z.~Carranza-Barnard, \emph{Low radon permeable gloves and laserball simulations
  for {SNO+}},  Master's thesis, Laurentian University, 2014.
\newblock \url{https://core.ac.uk/download/pdf/19908125.pdf}.

\bibitem{mamedov}
{\scshape SuperNEMO} collaboration, F.~Mamedov, P.~Čermák, K.~Smolek and
  I.~Štekl, \emph{Measurement of radon diffusion through shielding foils for
  the {SuperNEMO} experiment},
  \href{http://dx.doi.org/10.1088/1748-0221/6/01/C01068}{\emph{JINST}
  {\bfseries 6} (2011) C01068}.

\bibitem{artradon}
M.~Liu, H.~W. Lee and A.~B. McDonald, \emph{{$^{222}$Rn} emanation into
  vacuum}, \href{http://dx.doi.org/10.1016/0168-9002(93)90948-H}{\emph{Nucl.
  Instrum. Meth.} {\bfseries A329} (1993) 291--298}.

\bibitem{petriw_2012}
Z.~Petriw, \emph{An underwater six-camera array for monitoring and position
  measurements in {SNO+}},  Master's thesis, University of Alberta, 2012.
\newblock \url{https://doi.org/10.7939/R3SQ6H}.

\bibitem{ellie_nuphys_2016}
E.~Falk, J.~Lidgard, M.~I. Stringer and E.~Turner, \emph{Commissioning of
  {ELLIE} for {SNO+}}, {\emph{NuPhys2016} {\bfseries 4} (2017) },
  [\href{https://arxiv.org/abs/1705.00354}{{\ttfamily 1705.00354}}].

\bibitem{snop_calib_jinst}
{\scshape SNO+} collaboration, R.~Alves et~al., \emph{The calibration system
  for the photomultiplier array of the {SNO+} experiment},
  \href{http://dx.doi.org/10.1088/1748-0221/10/03/P03002}{\emph{JINST}
  {\bfseries 10} (2015) P03002},
  [\href{https://arxiv.org/abs/1411.4830}{{\ttfamily 1411.4830}}].

\bibitem{lombardi_2013}
P.~Lombardi, F.~Ortica, G.~Ranucci and A.~Romani, \emph{Decay time and pulse
  shape discrimination of liquid scintillators based on novel solvents},
  \href{http://dx.doi.org/10.1016/j.nima.2012.10.061}{\emph{Nucl. Instrum.
  Meth.} {\bfseries A701} (2013) 133--144}.

\bibitem{MOFFAT2005255}
B.~A. Moffat et~al., \emph{Optical calibration hardware for the {Sudbury}
  neutrino observatory},
  \href{http://dx.doi.org/10.1016/j.nima.2005.08.029}{\emph{Nucl. Instrum.
  Meth.} {\bfseries A554} (2005) 255--265},
  [\href{https://arxiv.org/abs/nucl-ex/0507026}{{\ttfamily nucl-ex/0507026}}].

\bibitem{mirizzi_2016}
A.~Mirizzi et~al., \emph{Supernova neutrinos: Production, oscillations and
  detection}, \href{http://dx.doi.org/10.1393/ncr/i2016-10120-8}{\emph{Riv.
  Nuovo Cim.} {\bfseries 39} (2016) 1--112},
  [\href{https://arxiv.org/abs/1508.00785}{{\ttfamily 1508.00785}}].

\bibitem{huedepohl_2010}
L.~Hüdepohl, B.~Müller, H.~T. Janka, A.~Marek and G.~G. Raffelt,
  \emph{Neutrino signal of electron-capture supernovae from core collapse to
  cooling}, \href{http://dx.doi.org/10.1103/PhysRevLett.104.251101,
  10.1103/PhysRevLett.105.249901}{\emph{Phys. Rev. Lett.} {\bfseries 104}
  (2010) 251101}, [\href{https://arxiv.org/abs/0912.0260}{{\ttfamily
  0912.0260}}].

\bibitem{Tagg:2002}
N.~J. Tagg et~al., \emph{The {Li-8} calibration source for the {Sudbury}
  neutrino observatory},
  \href{http://dx.doi.org/10.1016/S0168-9002(02)00860-4}{\emph{Nucl. Instrum.
  Meth.} {\bfseries A489} (2002) 178--188},
  [\href{https://arxiv.org/abs/nucl-ex/0202024}{{\ttfamily nucl-ex/0202024}}].

\bibitem{Dragowsky:2001ax}
M.~R. Dragowsky et~al., \emph{The {N-16} calibration source for the {Sudbury}
  neutrino observatory},
  \href{http://dx.doi.org/10.1016/S0168-9002(01)02062-9}{\emph{Nucl. Instrum.
  Meth.} {\bfseries A481} (2002) 284--296},
  [\href{https://arxiv.org/abs/nucl-ex/0109011}{{\ttfamily nucl-ex/0109011}}].

\bibitem{loach_2008}
J.~C. Loach, \emph{Measurement of the Flux of {$^{8}$B} Solar Neutrinos at the
  {Sudbury} Neutrino Observatory}.
\newblock PhD thesis, University of Oxford, 2008.
\newblock
  \href{https://ethos.bl.uk/OrderDetails.do?uin=uk.bl.ethos.490107}{uk.bl.ethos.490107}.

\bibitem{TabRad_v5}
M.-M. B\'e et~al., \emph{Table of Radionuclides}, vol.~5 of
  \emph{\href{http://www.bipm.org/utils/common/pdf/monographieRI/Monographie_BIPM-5_Tables_Vol5.pdf}{Monographie
  BIPM-5}}.
\newblock Bureau International des Poids et Mesures, Pavillon de Breteuil,
  F-92310 S\`evres, France, 2010.

\bibitem{Liu_2020}
Y.~Liu, \emph{Neutron Measurements and Reactor Antineutrino Search with the
  SNO+ Detector in the Water Phase}.
\newblock PhD thesis, Queen's University, 2020.
\newblock \url{http://hdl.handle.net/1974/27931}.

\bibitem{snop_ncapture}
{\scshape SNO+} collaboration, M.~Anderson et~al., \emph{{Measurement of
  neutron-proton capture in the SNO+ water phase}},
  \href{http://dx.doi.org/10.1103/PhysRevC.102.014002}{\emph{Phys. Rev. C}
  {\bfseries 102} (2020) 014002},
  [\href{https://arxiv.org/abs/2002.10351}{{\ttfamily 2002.10351}}].

\bibitem{semenec_2017}
I.~Semenec, \emph{Design of a neutron calibration source for the {SNO+}
  experiment},  Master's thesis, Laurentian University, 2017.
\newblock \url{https://core.ac.uk/download/pdf/222897502.pdf}.

\bibitem{neutron_capture}
B.~B. Kinsey and G.~A. Bartholomew, \emph{Neutron capture
  $\ensuremath{\gamma}$-rays from titanium, chromium, iron, nickel, and zinc},
  \href{http://dx.doi.org/10.1103/PhysRev.89.375}{\emph{Phys. Rev.} {\bfseries
  89} (Jan, 1953) 375--385}.

\bibitem{sibley_2016}
L.~A.~B. Sibley, \emph{The {SNO+} liquid scintillator response to low-energy
  electrons and its effect on the experiment’s sensitivity to a future
  neutrinoless double beta decay signal}.
\newblock PhD thesis, University of Alberta, 2016.
\newblock \url{https://doi.org/10.7939/R3WM1445C}.

\bibitem{gridpp_2006}
{\scshape GridPP} collaboration, P.~J.~W. Faulkner et~al., \emph{{GridPP}:
  Development of the {UK} computing grid for particle physics},
  \href{http://dx.doi.org/10.1088/0954-3899/32/1/N01}{\emph{J. Phys.}
  {\bfseries G32} (2006) N1--N20}.

\bibitem{gridpp_2009}
D.~Britton et~al., \emph{{GridPP}: the {UK} grid for particle physics},
  \href{http://dx.doi.org/10.1098/rsta.2009.0036}{\emph{Phil. Trans. Roy. Soc.
  Lond.} {\bfseries A367} (2009) 2447--2457}.

\end{thebibliography}\endgroup

\end{document}